\documentclass[aps,pre,showkeys,a4paper,floatfix]{revtex4}
\usepackage{natbib}
\usepackage{dcolumn}
\usepackage{graphicx}
\usepackage{epsfig}

\begin{document}
\title{Computational Methods for Simulating Quantum Computers
\footnote{Published in: Handbook of Theoretical and Computational Nanotechnology
Vol. 3: Quantum and molecular computing, quantum simulations, Chapter 1, pp. 2 - 48,
M. Rieth and W. Schommers eds.,American Scientific Publisher, Los Angeles (2006).}
}

\author{H. De Raedt}
 \email{deraedthans@gmail.com}
\affiliation{Department of Applied Physics, 
Materials Science Centre, University of Groningen, Nijenborgh 4,
NL-9747 AG Groningen, The Netherlands}
 \homepage{http://www.compphys.org}
\author{K. Michielsen}
 \email{k.michielsen@fz-juelich.de}
\affiliation{Department of Applied Physics, 
Materials Science Centre, University of Groningen, Nijenborgh 4,
NL-9747 AG Groningen, The Netherlands}
\begin{abstract}
This review gives a survey of numerical algorithms and software to simulate quantum computers.
It covers the basic concepts of quantum computation and quantum algorithms
and includes a few examples that illustrate the use of simulation software
for ideal and physical models of quantum computers.
\keywords{Quantum computation, computer simulation, time-integration algorithms}
\end{abstract}
\date{\today}

\maketitle
{{\hsize=17cm\hfuzz=10pt \tableofcontents}}

\def\ORDER#1{\hbox{${\cal O}(#1)$}}
\def\NOBAR#1{#1}
\def\BAR#1{\overline{#1}}
\def\BRA#1{\langle #1 \vert}
\def\KET#1{\vert #1 \rangle}
\def\EXPECT#1{\langle #1 \rangle}
\def\BRACKET#1#2{\langle #1 \vert #2 \rangle}
\def\hbar{{\mathchar'26\mskip-9muh}}
\def\mod{{\mathop{\hbox{mod}}}}
\def\CNOT{{\mathop{\hbox{CNOT}}}}
\def\Tr{{\mathop{\hbox{Tr}}}}
\def\thavg#1{E\left({#1}\right)}
\def\empavg#1{\overline{#1}}
\def\var{{\mathop{\hbox{var}}}}
\def\covar{{\mathop{\hbox{covar}}}}
\def\bPsi{{\mathbf{\Psi}}}
\def\bPhi{{\mathbf{\Phi}}}
\def\bzero{{\mathbf{0}}}

\def\Mn{\hbox{Mn$_{12}$}}
\def\V{\hbox{V$_{15}$}}
\def\DM{Dzyaloshinskii-Moriya }
\def\DMI{Dzyaloshinskii-Moriya interactions }
\def\figsize{\hsize}
\def\smallfigsize{8cm}
\def\Eq#1{(\ref{#1})}

\section{Introduction}\label{intro}

The basic ideas of quantum computation were formulated more than 20 years
ago~\cite{BENI80,FEYN82}.
In the 1990's several algorithms have been discovered~\cite{SIMO94,SHOR94,GROV96,GROV97,SIMO97,SHOR99}
that run much faster on a quantum computer than on a conventional computer,
promising the solution of some problems that are considered to be
intractable for conventional computers.
These discoveries have fueled many new developments, both
theoretically and experimentally.
Quantum information theory is a new interdisciplinary field
of research, building on concepts from theoretical computer science
and theoretical physics~\cite{NIEL00}.
On the experimental front, considerable progress has been
made to engineer and control quantum systems that may
be used for quantum information processing~\cite{%
CIRA95,MONR95,SLEA95,DOMO95,JONE98a,JONE98b,CHUA98a,CHUA98b,
HUGH98,WINE98,NAKA99,NOGU99,
BLIA00,OLIV00,SYPE00,MARX00,KNIL00,CORY00,JONE00,
SYPE01a,SYPE01b,JONE01,TAKE01}.
The feasibility of executing short quantum algorithms
on quantum systems with a few ($<8$ bits) has been demonstrated~\cite{
JONE98a,JONE98b,CHUA98a,CHUA98b,SYPE01a,SYPE01b,JONE01}.
The technological challenges to build a quantum computer
that can perform calculations that would exhaust
the resources of a conventional computer seem tremendous,
and there is no indication that such a machine will
become available in the next few years.

Any physically realizable
quantum computer is a complicated many-body system
that interacts with its environment.
In quantum statistical mechanics and quantum chemistry, it is well known
that simulating an interacting quantum many-body
system becomes exponentially more difficult
as the size of the system grows.
Actually, it is this observation that lead Feynman to the
question what kind of computer would be needed to overcome
this exponential increase~\cite{FEYN82}.
Formally, the answer to this question is ``a quantum computer.''
It seems that only a quantum computer can efficiently simulate itself~\cite{FEYN82,ZALK98}.

Computer simulation has long been accepted as
the third methodology in many branches of science and engineering~\cite{LAND00}.
Conventional computers can be used to simulate
quantum computers that are relatively small (such as 24 qubits)
but are significantly larger than the experimental machines that have been built.
Therefore, it is striking that theoretical ideas about quantum computation
are seldom confronted with numerical experiments that can be carried
out on present-day (super) computers.
Conventional computers can simulate the abstract model
of an ideal quantum computer on which most theoretical work is based, and,
most important,
they can also simulate the physical behavior of quantum computer hardware~\cite{PRES98}.

There are a number of excellent
reviews~\cite{EKER96,BENN97,VEDR98,BRAN98,JONE01,EKER01,BLAC02,GALI02}
and books~\cite{BERM98,KHLO98,NIEL00,MACC00}
that cover the theoretical and/or experimental aspects of
quantum computations, but there is no review that
discusses methods to simulate ideal and
physical models of quantum computers.
The purpose of this review is to fill this void.
It contains enough information and detail to allow advanced students
to develop their own quantum computer simulator.
It also contains an up-to-date account of existing simulation software.
Most of the simulation methods covered in this review
have a much broader scope than the application to quantum computation
might suggest. As a matter of fact, much of the impetus to
develop these methods stems from problems encountered
in the study of quantum dynamics of nanomagnets.

This chapter is organized as follows.
Section II gives a brief account of the basics of quantum computation.
Section III discusses the implemention of simulation models of quantum computers
on a conventional computer and also gives a survey of the numerical algorithms
that are used to simulate these models.
A short review of quantum algorithms is given in Section IV.
Sections V and VI deal with the simulation of ideal and realistic models
of quantum computers, respectively.
In Section VII we give an overview of existing quantum computer simulator
software and also provide examples of how a simulator is used
to perform quantum computation.
A brief summary and outlook are provided in Section VIII.

\section{The Ideal Quantum Computer}

In contrast to a digital computer, in which the state of an elementary storage unit
is specified by the logical values 0 and 1 (= one bit),
the state of an elementary storage unit of a quantum computer, the quantum bit or qubit,
is described by a two-dimensional vector of Euclidean length one.
Denoting two orthogonal basis vectors of the
two-dimensional vector space by $\KET{0}$ and $\KET{1}$, the state $\KET{\Phi}$
of the {\bf qubit} can be written as a linear superposition of the basis states
$\KET{0}$ and $\KET{1}$:

\begin{equation}
\KET{\Phi}=a_0\KET{0} +a_1\KET{1},
\end{equation}
where $a_0$ and $a_1$ are complex numbers such that $|a_0|^2+|a_1|^2=1$.
The appearance of complex numbers suggests
that one qubit can contain an infinite amount of information.
However, the principles of quantum mechanics, do not allow
retrieving all this information~\cite{SCHI68,BAYM74,BALL03}.
The result of inquiring about the state of the qubit, that is a measurement,
yields either the result 0 or 1.
The frequency of obtaining 0 (1) can be estimated
by repeated measurement of the same state of the
qubits and is given by $|a_0|^2$ ($|a_1|^2$).

Just as a conventional computer with only one bit is fairly useless,
a quantum computer should also contain more than one qubit.
Disregarding the enormous technological hurdles that have
to be taken to put more than, say 7, qubits together,
on present-day personal computers we can readily
simulate quantum computers with $L=20$ qubits.
The internal state of a quantum computer with $L$ qubits
is represented by a unit vector in a $D=2^L$ dimensional space
(see Section~\ref{states}).

The ideal quantum computer differs from a conventional (probabilistic) computer
in the sense that the state of the qubits changes according to the rules
of quantum mechanics, that is through rotations of the vectors representing
the state of the qubits~\cite{NIEL00}.
From elementary linear algebra, we know that a rotation of a vector
corresponds to the multiplication of the vector by a unitary matrix.
Thus, the internal state of the quantum computer evolves
in time according to a sequence of unitary transformations.
Any such sequence is called a {\bf quantum algorithm}.
Of course, not every sequence corresponds to a meaningful computation.
If a quantum algorithm cannot exploit the fact that
the intermediate state of the quantum computer is described by a linear
superposition of basis vectors, it will not be faster than its classical
counterpart.

The unitary time evolution of the internal state of the quantum computer
is interrupted at the point where we inquire about the value
of the qubits, that is as soon as we perform a measurement on the qubits.
If we perform the readout operation on a qubit, we get a definite answer, either 0 or 1,
and the information encoded in the superposition is lost.
The process of measurement cannot be described by a unitary transformation~\cite{BALL03}.
Therefore, we do not consider it to be part of the quantum algorithm.
However, to estimate the efficiency of a quantum
computation, both the operation count of the quantum algorithm
and the cost of measuring the outcome of the calculation
have to be taken into consideration~\cite{RAED02b}.

In the quantum computation literature
the convention, is to count each application of a unitary transformation
as one operation on a quantum computer~\cite{NIEL00}.
As the unitary transformation may change all amplitudes simultaneously,
a quantum computer is a massively parallel machine.
To simulate these unitary operations on a conventional computer, we actually
perform matrix-vector multiplications that requires \ORDER{2^{2L}}
(in the worst case) arithmetic operations.
This may cause some confusion at some points, so the reader should
try to keep this in mind.

To summarize: At any point in time, the internal state of a quantum computer
with $L$ qubits is described by a unit vector in a $D=2^L$ dimensional space.
A quantum algorithm that executes on the quantum computer changes the
internal state by rotating the vector. Therefore, each step in the
quantum algorithm corresponds to the multiplication
of the vector by a unitary matrix.
Readout of the result of a quantum algorithm is destructive in
the sense that it destroys the information contained in the superposition.

In the following subsections, we discuss the basic ingredients of quantum computation
in more detail.

\subsection{Spin-1/2 Algebra}

In the simplest form, a qubit is a two-state quantum system,
of which there are many examples in the quantum world~\cite{SCHI68,BAYM74,BALL03}.
Measurements of a component of the spin (= internal angular momentum)
of particles such as electrons, protons, and neutrons
along any direction yield either $\hbar/2$ or $-\hbar/2$.
The convention is to call the state of the spin
that corresponds to the outcome $\hbar/2$ ``spin up''
($\KET{\uparrow}$) and the other state ``spin down''
($\KET{\downarrow}$).
The fact that the internal angular momentum takes only two values
implies that it corresponds to a total angular momentum quantum
number of 1/2, hence the name spin 1/2 ($S=1/2$) particle.

In principle, any physical object that carries a $S=1/2$ degree of freedom
can be used as a physical realization of a qubit.
In general, the wavefunction that describes the spin of these objects
can be written as a linear combination of
the spin-up and spin-down states~\cite{SCHI68,BAYM74,BALL03}:

\begin{equation}
\KET{\Phi}=a_0\KET{\uparrow}+a_1\KET{\downarrow}
,
\label{SPIN1}
\end{equation}
where $a_0$ and $a_1$ are complex numbers.
It is convenient to normalize the length of the vector $\KET{\Phi}$ to one.
Then $|a_0|^2+|a_1|^2=1$.

The three components of the spin-1/2 operator ${\bf S}$ acting on the Hilbert space spanned
by the states $\KET{\uparrow}$ and $\KET{\downarrow}$
are defined by~\cite{SCHI68,BAYM74,BALL03}

\begin{equation}
S^x={1\over 2}
\left(
\begin{array}{cc}
0&1 \\
1&0
\end{array}
\right),\quad
S^y={1\over 2}
\left(
\begin{array}{cc}
\phantom{-}0&-i \\
\phantom{-}i&\phantom{-}0
\end{array}
\right),\quad
S^z={1\over 2}
\left(
\begin{array}{cc}
\phantom{-}1&\phantom{-}0 \\
\phantom{-}0&-1
\end{array}
\right)
,
\label{spinmatrices}
\label{SPIN2}
\end{equation}
in units such that $\hbar=1$.
According to quantum theory, a measurement
of a physical quantity $A=A^\dagger$ of a quantum system in a state $\KET{\Phi}$
yields a real number, the {\bf expectation value}~\cite{SCHI68,BAYM74,BALL03}

\begin{equation}
\EXPECT{A}=\BRACKET{\Phi}{A|\Phi}/\BRACKET{\Phi}{\Phi}.
\label{EXPECT}
\end{equation}
From Eqs.~\Eq{SPIN1} and \Eq{SPIN2} it follows that

\begin{equation}
\EXPECT{S^x}=\frac{a_0a_1^*+a_0^*a_1}{2}\,\quad,
\EXPECT{S^y}=\frac{ia_0a_1^*-ia_0^*a_1}{2}\,\quad,
\EXPECT{S^z}=\frac{a_0a_0^*-a_1^*a_1}{2}\,\quad
,
\label{SPIN3}
\end{equation}
and we conclude that the representations Eqs.~\Eq{SPIN1} and \Eq{SPIN2},
have been chosen such that
$\KET{\uparrow}$ and $\KET{\downarrow}$ are eigenstates of $S^z$ with
eigenvalues $+1/2$ and $-1/2$, respectively.
This is the convention adopted in the physics literature.

In the context of quantum computation, it is convenient to define~\cite{NIEL00}

\begin{equation}
\KET{0}=\KET{\uparrow}=
\left(
\begin{array}{c}
1 \\
0
\end{array}
\right)
\quad,\quad \KET{1}=\KET{\downarrow}=
\left(
\begin{array}{c}
0 \\
1
\end{array}
\right)
.
\label{SPIN4}
\end{equation}
Then we have
\begin{equation}
\KET{\Phi}=a_0\KET{0}+a_1\KET{1}
,
\label{SPIN5}
\end{equation}
and the expectation values of the three components of the qubits are
defined as

\begin{equation}
\EXPECT{Q^x}=\frac{1-a_0a_1^*-a_0^*a_1}{2}\quad,\quad
\EXPECT{Q^y}=\frac{1-ia_0a_1^*+ia_0^*a_1}{2}\quad,\quad
\EXPECT{Q^z}=\frac{1-a_0a_0^*+a_1^*a_1}{2}\quad,\quad
\label{SPIN6}
\end{equation}
such that $0\le \EXPECT{Q^\alpha}\le 1$ for $\alpha=x,y,z$.
In the following, when we say that the state of a qubit is 0(1), we actually mean that
$\EXPECT{Q^z}=0(1)$ and $\EXPECT{Q^x}=\EXPECT{Q^y}=1/2$.

A change of the state of a qubit corresponds to a rotation of the spin.
In general, a rotation of the spin
by an angle $\phi$ about an axis $\beta$ can be written as ($\hbar =1$)

\begin{equation}
S^{\alpha}(\phi , \beta)=e^{i\phi S^{\beta}}S^{\alpha}e^{-i\phi S^{\beta}}
=S^{\alpha}\cos\phi +\epsilon_{\alpha\beta\gamma}S^{\gamma}\sin\phi,
\label{SPIN7}
\end{equation}
where use has been made of the commutation rules of
the components of the angular momentum operator ${\bf S}$~\cite{SCHI68,BAYM74,BALL03}:

\begin{equation}
[S^{\alpha},S^{\beta}]=i\epsilon_{\alpha\beta\gamma}S^{\gamma}
,
\label{SPIN8}
\end{equation}
where
$\epsilon_{\alpha\beta\gamma}$ is the totally asymmetric unit tensor ($\epsilon_{xyz}=\epsilon_{yzx}=\epsilon_{zxy}=1$,
$\epsilon_{\alpha\beta\gamma}=-\epsilon_{\beta\alpha\gamma}=-\epsilon_{\gamma\beta\alpha}=-\epsilon_{\alpha\gamma\beta}$,
$\epsilon_{\alpha\alpha\gamma} =0$) and the summation convention is used.
In addition to Eq.~\Eq{SPIN8} the $S=1/2$ operators \Eq{SPIN2} also satisfy
the relations~\cite{SCHI68,BAYM74,BALL03}

\begin{eqnarray}
S^{x}S^{x}&=&S^{y}S^{y}=S^{z}S^{z}=\frac{1}{4},
\label{SPIN8a}
\\
S^{x}S^{y}&=&\frac{i}{2}S^{z},\quad S^{x}S^{z}=-\frac{i}{2}S^{y},\quad
S^{y}S^{z}=\frac{i}{2}S^{x}
,\quad
S^{y}S^{x}
=
-\frac{i}{2}S^{z},\quad S^{z}S^{x}=\frac{i}{2}S^{y},\quad
S^{z}S^{y}=-\frac{i}{2}S^{x}
,
\label{SPIN9}
\end{eqnarray}
which are often very useful to simplify products of $S=1/2$ matrices.
From Eq.~\Eq{SPIN9} it follows that $2S^x$, $2S^y$, and $2S^z$
are unitary operators. Hence each of them represents
a genuine quantum computer operation.

Rotations of the spin by $\pi /2$ about the $x$-axis and $y$-axis
are often used as elementary quantum computer operations.
In matrix notation, they are given by

\begin{equation}
{X}\equiv e^{i\pi S^x/2}={1\over\sqrt{2}}
\left(
\begin{array}{cc}
1&i \\
i&1
\end{array}
\right)
,\quad
{Y}\equiv e^{i\pi S^y/2}={1\over\sqrt{2}}
\left(
\begin{array}{cc}
\phantom{-}1&\phantom{-}1 \\ -1&\phantom{-}1\\
\end{array}
\right)
,
\label{SPIN10}
\end{equation}
respectively.
$\NOBAR X$ and $\NOBAR Y$ represent operations on single qubits.
The matrix expressions for the inverse of the rotations
$\NOBAR X$ and $\NOBAR Y$, denoted by $\BAR X$
and $\BAR Y$, respectively, are obtained by taking the Hermitian conjugates
of the matrices in \Eq{SPIN10}.

With our convention, $\BRA {0}\BAR Y S^x\NOBAR Y\KET{0}=-1/2$ so that a positive angle corresponds
to a rotation in the clockwise direction.
In general, a rotation of the state about a vector ${\bf v}$ corresponds
to the matrix

\begin{equation}
e^{i{\bf v}\cdot{\bf S}}=\openone\cos\frac{v}{2}+\frac{2i{\bf v}\cdot{\bf S}}{v}\sin\frac{v}{2}
,
\label{SPIN11}
\end{equation}
where $\openone$ denotes the unit matrix and $v=\sqrt{v_x^2+v_y^2+v_z^2}$ is the length of the
vector ${\bf v}$.
\subsection{State Representation}\label{states}

A quantum computer must have more than one qubit.
Let us denote by $L$ the number of $S=1/2$ objects of the corresponding physical system.
In general, the quantum state of a system of $L$ spins is defined by
the linear superposition of the direct-product states of the $L$ single-spin states:

\begin{eqnarray}
\KET{\Phi}=
& &a({\uparrow\uparrow\ldots\uparrow}) \KET{\uparrow\uparrow\ldots\uparrow}
+a({\downarrow\uparrow\ldots\uparrow}) \KET{\downarrow\uparrow\ldots\uparrow}
+\ldots 
+a({\uparrow\downarrow\ldots\downarrow}) \KET{\uparrow\downarrow\ldots\downarrow}
+a({\downarrow\downarrow\ldots\downarrow}) \KET{\downarrow\downarrow\ldots\downarrow}
.
\label{STAT0}
\end{eqnarray}
As before, the coefficients (amplitudes)
$a({\uparrow\uparrow\ldots\uparrow}),\ldots,a({\downarrow\downarrow\ldots\downarrow})$
are complex numbers, and it is convenient to normalize the vector
$\KET{\phi}$ by the rescaling

\begin{eqnarray}
\sum_{\sigma_1\ldots\sigma_L=\uparrow,\downarrow}|a(\sigma_1\ldots\sigma_L)|^2=1
,
\label{STAT1}
\end{eqnarray}
such that $\BRACKET{\Phi}{\Phi}=1$.

In physics, it is customary to let the first label correspond to the first spin,
the second to the second spin, and so on.
Thus, the second term in Eq.~\Eq{STAT0}
is the contribution of the state with spin 1 down and all other spins up.
In computer science, it is more natural (and logically equivalent)
to think of qubit (spin) 1 as the least significant bit of the integer index that runs from
zero (all spins up) to $2^L-1$ (all spins down).
This is also the convention that is adopted in the literature on quantum computation:
spin up (down) corresponds to a qubit in state 0 (1).
Thus, in quantum-computation notation Eq.~\Eq{STAT0} reads

\begin{eqnarray}
\KET{\Phi}&=&a({0\ldots00}) \KET{0\ldots00}
+a({0\ldots01}) \KET{0\ldots01}
+\ldots
+a({1\ldots10}) \KET{1\ldots10}
+a({1\ldots11}) \KET{1\ldots11},
\nonumber \\
&=&a_0 \KET{0}
+a_1 \KET{1}
+\ldots
+a_{2^L-2} \KET{2^L-2}
+a_{2^L-1} \KET{2^L-1}
.
\label{STAT2}
\end{eqnarray}
For later reference, it is useful to write down explicitly
some examples of translating from physics to quantum computer notation
and vice versa:

\begin{eqnarray}
a({\uparrow\uparrow\uparrow\ldots\uparrow}) \KET{\uparrow\uparrow\uparrow\ldots\uparrow}
&=&
a({0\ldots000}) \KET{0\ldots000}
=a_0 \KET{0},
\nonumber \\
a({\downarrow\uparrow\uparrow\ldots\uparrow}) \KET{\downarrow\uparrow\uparrow\ldots\uparrow}
&=&
a({0\ldots001}) \KET{0\ldots001}
=a_1 \KET{1},
\nonumber \\
a({\uparrow\downarrow\uparrow\ldots\uparrow}) \KET{\uparrow\downarrow\uparrow\ldots\uparrow}
&=&
a({0\ldots010}) \KET{0\ldots010}
=a_2 \KET{2},
\nonumber \\
a({\uparrow\downarrow\downarrow\ldots\downarrow}) \KET{\uparrow\downarrow\downarrow\ldots\downarrow}
&=&
a({1\ldots110}) \KET{1\ldots110}
=a_{2^L-2} \KET{2^L-2},
\nonumber \\
a({\downarrow\downarrow\downarrow\ldots\downarrow}) \KET{\downarrow\downarrow\downarrow\ldots\downarrow}
&=&
a({1\ldots111})  \KET{{1\ldots111}}
=a_{2^L-1}  \KET{2^L-1}
.
\label{STAT3}
\end{eqnarray}

Obviously, there is a one-to-one correspondence between the
last line of Eq.~\Eq{STAT2} and the way the amplitudes $a_i$
are stored in the memory of a conventional computer.
From representation \Eq{STAT2}, it is easy to
estimate the amount of computer memory that is needed
to simulate a quantum spin system of $L$ spins on a conventional digital computer.
The dimension $D$ of the Hilbert space (that is the number of amplitudes $a_i$)
spanned by the $L$ spin-1/2 states is $D=2^{L}$.
For applications that require highly optimized code, it is
often more efficient to store the real and imaginary part
of the amplitudes $a_i$ in separate arrays.
Furthermore, even for simple looking problems, it is
advisable to use 13 - 15 digit floating-point arithmetic (corresponding
to 8 bytes for a real number).
Thus, to represent a state of the quantum system of $L$ qubits
in a conventional, digital computer, we need a least $2^{L+4}$ bytes.
For example, for $L=20$ ($L=24$)
we need at least 16 (256) Mb of memory to store a single arbitrary state $\KET{\Phi}$.
Not surprisingly, the amount of memory that is required to simulate a quantum
system with $L$ spins increases exponentially with the number of spins $L$.

Operations on the $j$th spin are denoted by simply attaching the label $j$
to the three components of the spin operator.
For example, the operation that flips the second spin (or qubit)
is given by

\begin{eqnarray}
\KET{\Phi'}&=&S_2^x\KET{\Phi}\nonumber \\
&=&
a({\uparrow\uparrow\uparrow\ldots\uparrow}) \KET{\uparrow\downarrow\uparrow\ldots\uparrow}
+a({\downarrow\uparrow\uparrow\ldots\uparrow}) \KET{\downarrow\downarrow\uparrow\ldots\uparrow}
+\ldots
 \nonumber \\
&&+a({\uparrow\downarrow\downarrow\ldots\downarrow}) \KET{\uparrow\uparrow\downarrow\ldots\downarrow}
+a({\downarrow\downarrow\downarrow\ldots\downarrow}) \KET{\downarrow\uparrow\downarrow\ldots\downarrow}
 \nonumber \\
&=&
a(0\ldots000) \KET{0\ldots010}+a(0\ldots001)\KET{0\ldots011}
+\ldots
 \nonumber \\
&&+a(1\ldots110)\KET{1\ldots100}+a(1\ldots111)\KET{1\ldots101}
.
\label{STAT4}
\end{eqnarray}
It is important to note that although the spin operator $S^x_j$ only
flips the $j$th spin, to compute $\KET{\Phi'}$ on a conventional computer
we have to update {\sl all} the $2^L$ amplitudes.
In general, on a conventional computer {\sl each} operation on the wavefunction
involves at least $2^L$ arithmetic operations
(here and in the sequel, if we count operations on a conventional
computer, we make no distiction between operations such as add, multiply,
or get from and put to memory).
The fact that both the amount of memory and
arithmetic operation count increase exponentially with the number of qubits
is the major bottleneck for simulating quantum systems (including quantum computers)
on a conventional computer.
However, at the time of writing, the number of qubits that can be simulated
far exceeds the number of qubits of experimentally realizable systems.

\subsection{Universal Computer}\label{universal}

It is easy to see that any (Boolean) logic circuit of a conventional computer can
be built by an appropriate combination of NAND-gates~\cite{CORM94}.
Although in practice it is often convenient to think in terms
of logic circuits that can perform more complex tasks than a NAND operation, conceptually
it is important to know that NAND-gates are all that is needed
to build a universal computing device.
Also, a quantum computer can be constructed from a set of basic gates (= unitary
transformations), but instead of one gate we need
several~\cite{LLOY93,DIVI95,DIVI95a,EKER96,VEDR98,NIEL00}.
The minimal set of gates is not unique, but,
for ideal quantum computers, it is mainly a matter
of taste to choose a particular set. However, in the real world,
the kind of operations we can perform on a specific physical system
is limited, and this will bias the choice of the set of basic gates.
Nevertheless, as long as the chosen set of basic gates
allows us to synthesize {\sl any} unitary tranformation on the state
$\KET{\Phi}$ of the quantum computer, this set can be used for universal
quantum computation~\cite{NIEL00,MYER97}.

In analogy with Boolean logic circuits, simplicity is an important
guideline to determine the set of elementary unitary gates of
a quantum computer.
Some basic results from linear algebra are very helpful in this respect.
It is well known that any vector of $N$ elements can be transformed to the vector $(1,0,\ldots,0)$
by at most $N-1$ plane rotations~\cite{WILK65,GOLU96}.
Each plane rotation is represented by a $2\times2$ unitary matrix
that operates on the elements $(1,j)$ for $j=2,\ldots,N$.
Application of this procedure to the $N$, mutually orthogonal,
column vectors of a $N\times N$ unitary matrix $U$ immediately
leads to the conclusion that $U$ can be written as a
product of at most $N(N-1)/2$ plane rotations.
Thus, we can synthesize {\sl any}
$N\times N$ unitary matrix by multiplying unitary matrices each of which
involves only two of the $N$ basis states~\cite{BARE95}.

As the order in which we apply the plane rotations
is arbitrary, the decomposition is not unique.
Furthermore, on a quantum computer $N=2^L$ so
that for a given quantum algorithm (or equivalently, a given unitary matrix $U$)
we should be able to find a decomposition that
takes much less than $2^{L-1}(2^L-1)$ plane rotations.
Otherwise the implementation of the quantum algorithm
is not efficient, and there is no point of even
contemplating the use of a quantum computer to solve the problem.
Finding the optimal decomposition of a $N\times N$ unitary matrix $U$
in terms of plane rotations is a difficult computational problem,
in particular because it is known that for some $U$
there is no other alternative than to decompose
it into $N(N-1)/2$ plane rotations~\cite{NIEL00}.
It has been shown that any of the $2^{L-1}(2^L-1)$ plane rotations
can be constructed by a combination of single qubit gates
and the so-called CNOT gate that operates on two qubits~\cite{DIVI95a,NIEL00}. Therefore
these single qubit operations and the CNOT gate constitute a set of gates
that can be used to construct a universal quantum computer.

Another basic result of linear algebra is that
any nonsingular matrix can be written as the matrix exponential
of a Hermitian matrix~\cite{BELL97}.
As a unitary matrix is nonsingular, we can write $U=e^{-itH}$,
where we already anticipated that, for the case at hand,
the Hermitian matrix $H$ corresponds to the Hamiltonian that models the qubits.
The next question is then to ask for the class of model
Hamiltonians that can be used for universal computation.
Again, simplicity is an important criterion to
determine useful models, but, as the Hamiltonian represents a physical
system, there may be some additional constraints on the form of
the interactions among qubits and the like. We discuss these
aspects in more detail in Section~\ref{physical}, where we consider
models of physically realizable quantum computers.

It is a remarkable fact that the simplest quantum many-body
system, the Ising spin model, can be used for universal quantum
computation~\cite{LLOY93}.
The Ising model

\begin{equation}
H(t)=
-\sum_{i,j=1}^L J_{i,j}^z(t) S_i^z S_j^z
-\sum_{i=1}^L\sum_{\alpha=x,y,z} h_{i}^\alpha(t) S_i^\alpha
,
 \label{Ising}
 \label{UNIV0}
\end{equation}
is often used as a physical model of an
ideal universal quantum computer~\cite{LLOY93,BERM94,NIEL00}.
The first sum in Eq.~\Eq{UNIV0}
runs over all pairs $P\le (L-1)L/2$ of spins that interact with each other.
For instance, if the spins interact only with their nearest neighbors,
we have $P=L/2$ and $J_{i,j}^z(t)=0$ except when $i$ and $j$ refer to spins
that are neighbors. In the case of dipolar interaction, $P=(L-1)L/2$.
Note that Eq.~\Eq{UNIV0} implicitly assumes that we
have complete control over all interaction parameters $J_{i,j}^z(t)$
and external fields $h_{i}^\alpha(t)$.
In Sections \ref{singlequbit} and \ref{twoqubit}
we explain how a universal set of gates, the single qubit and the CNOT gates,
can be implemented on the quantum computer defined by the Ising model \Eq{UNIV0}.

The more general form of the Hamiltonian of a physical model of a
quantum computer reads

\begin{equation}
H(t)=
-\sum_{i,j=1}^L\sum_{\alpha=x,y,z} J_{i,j}^\alpha(t) S_i^\alpha S_j^\alpha
-\sum_{i=1}^L\sum_{\alpha=x,y,z} h_{i}^\alpha(t) S_i^\alpha
.
 \label{hamiltonian}
\end{equation}
The exchange parameters $J_{i,j}^\alpha(t)$ determine the strength
of the interactions between the $\alpha$-components of spins $i$ and $j$.

Hamiltonian \Eq{hamiltonian} is sufficiently generic to represent most models for
candidates of physical realizations of quantum computer hardware.
The spin-spin term in Eq.~\Eq{hamiltonian} is sufficiently
general to describe the most common types of interactions such
as Ising, anisotropic Heisenberg, and dipolar coupling between the spins.
Furthermore, if we also use spin-1/2 degrees of freedom to represent
the environment then, on this level of description,
the interaction between the quantum computer and its environment is included in model
\Eq{hamiltonian}, too.
In other words, the Hamiltonian \Eq{hamiltonian} is sufficiently generic to cover
most cases of current interest.

In the context of quantum computation and experiments in general, the quantum
dynamics of model \Eq{hamiltonian} is controlled and/or probed by static and/or
time-dependent external fields $h_{i}^\alpha(t)$.
Depending on the physical system, also the exchange parameters $J_{i,j}^\alpha(t)$
can be controlled by external fields.

\subsection{Time Evolution of Quantum Computers}

A quantum algorithm is a sequence of unitary transformations
that change the state vector of the quantum computer.
Physically, this change corresponds to the time evolution of a quantum system.
In general, the quantum system has non-negative
probabilities $p_0, p_1,\ldots,p_{2^L-1}$ for being in the states
$\KET{0},\ldots,\KET{2^L-1}$.
Then, the state of the quantum system is described by
the density matrix~\cite{SCHI68,BALL03}

\begin{equation}
\rho(t)=\sum_{i=1}^{2^L-1}\KET{i}p_i(t)\BRA{i}
,
\label{TDSE-1}
\end{equation}
where $p_i(t=0)=p_i$ and the expectation values of physical quantities
are given by $\EXPECT{A(t)}=\Tr \rho(t) A$.
The time evolution of the density matrix \Eq{TDSE-1} trivially follows
from the time evolution of each of the pure states $\KET{i}$.
Therefore, it is sufficient to focus on methods to compute the time
evolution of a pure quantum state.
In practice, a calculation of the time evolution of the density matrix takes
$2^L$ times more CPU time than the same calculation for a single pure state.

The time evolution of a pure state of the quantum system is given by the solution
of the time-dependent Schr\"odinger equation~\cite{SCHI68,BAYM74,BALL03}

\begin{equation}
i{\partial \over\partial t} \KET{\Phi(t)}= H(t) \KET{\Phi(t)}
.
\label{TDSE}
\label{TDSE0}
\end{equation}
The solution of Eq.~\Eq{TDSE0} can be written as~\cite{SCHI68,BAYM74,BALL03}

\begin{equation}
\KET{\Phi(t+\tau)} = U(t+\tau,t)\KET{\Phi(t)}
 = \exp_+\left(-i \int_{t}^{t+\tau} H(u) du\right)\KET{\Phi(t)}
,
\label{TDSE1}
\end{equation}
where $\tau$ denotes the time step.
The propagator $U(t+\tau,t)=\exp_+\left(-i \int_{t}^{t+\tau} H(u) du\right)$ is a unitary
matrix that transforms (that is, rotates) the state $\KET{\Phi(t})$ into the state
$\KET{\Phi(t+\tau)}$.
The time-ordered matrix exponential
$\exp_+\left(-i \int_{t}^{t+\tau} H(u) du\right)$
is formally defined by

\begin{eqnarray}
\exp_+\left(-i \int_{t}^{t+\tau} H(u) du\right)=1&+&(-i) \int_{t}^{t+\tau} du_1\,H(u_1)
+(-i)^2 \int_{t}^{t+\tau} du_1 \int_{t}^{u_1} du_2 H(u_1)H(u_2)\nonumber \\
&+&(-i)^3 \int_{t}^{t+\tau} du_1 \int_{t}^{u_1} du_2 \int_{t}^{u_2} du_3 H(u_1)H(u_2)H(u_3)+\ldots
.
\label{TDSE2}
\end{eqnarray}
For computational purposes, a naive truncation of the series \Eq{TDSE2}
would be fairly useless because it would yield a nonunitary approximation
to $U(t+\tau,t)$ and this would violate one of the basic axioms of quantum theory.

In practice, the time dependence of the external fields can always
be regarded as piece-wise constant in time.
Hence we divide the interval $[t,t+\tau]$ into $n$
smaller intervals $[t_k,t_{k+1}]$
where $t_k=t+\sum_{j=0}^k \tau_j$, $\tau_k$ denotes
the $k$-th time interval for which $H(t)$
is constant and $\tau=\sum_{j=0}^n \tau_j$ ($\tau_0=0$).
Then we can write~\cite{SUZU93}

\begin{equation}
U(t+\tau,t)=U(t_{n},t_{n-1})U(t_{n-1},t_{n-2})\ldots U(t_{1},t_{0})
,
\label{TDSE3}
\end{equation}
where

\begin{equation}
U(t_{j},t_{j-1})=e^{-i\tau_j H(t_{j-1}+\tau_j/2)}
.
\label{TDSE4}
\end{equation}
Note that we have not made any assumption about the
size of the time intervals $[t_k,t_{k+1}]$ (with respect to the relevant energy scales).
These formal manipulations show that a numerical solution of the time-dependent Schr\"odinger equation for
the time dependent model \Eq{hamiltonian} boils down to the calculation of matrix exponentials
of the form \Eq{TDSE4}. For $t_{j-1}\le t <t_{j}$, the Hamiltonian that appears in
\Eq{TDSE4} does not change with time. Hence we can use the shorthand $U(t)=e^{-itH}$,
keeping in mind that in actual applications, both $t$ and $H$ may depend on
the particular time interval $[t_{j-1},t_{j}]$.

In general, there are two closely related,
strategies to construct algorithms to solve equations such as
the time-dependent Schr\"odinger equation~\cite{SMITH85}.
The traditional approach is to discretize (with an increasing level of sophistication)
the derivative with respect to time.
A fundamental difficulty with this approach is that it is hard
to construct algorithms that preserve the essential character of
quantum dynamical time evolution, namely the fact that it is a {\sl unitary
transformation} of the state of the quantum system.
In particular, in the context of quantum computation it is more natural to consider
algorithms that yield a unitary time evolution by construction.
In this chapter we review only methods that are based
on the second strategy~\cite{SMITH85,RAED87},
namely, to approximate the formally exact solution, that is
the matrix exponential $U(t)=e^{-itH}$ by some unitary time evolution matrix
that is also unitary (to machine precision) $\widetilde U(t)$.

If the approximation $\widetilde U(t)$ is itself an orthogonal transformation,
then $\Vert\widetilde U(t)\Vert=1$ where
$\Vert X\Vert$ denotes the 2-norm of a vector or matrix
$X$~\cite{WILK65,PARL81,GOLU96,BELL97}.
This implies that $\Vert\widetilde\bPhi(t)\Vert=\Vert\widetilde{U}(t)\bPhi(0)\Vert =
 \Vert\bPhi(0)\Vert$,
for an arbitrary initial condition $\bPhi(0)$ and for all times $t$.
Hence the time integration algorithm defined by $\widetilde U(t)$
is unconditionally stable by construction~\cite{SMITH85,RAED87}.
In other words, the numerical stability of the algorithm does not depend
on the time step $t$ that is used.
In Sections~\ref{numerical} we review and compare different algorithms to compute
$U(t)=e^{-itH}$.

\subsection{Single-qubit Operations}\label{singlequbit}

The simplest, single-qubit operation on qubit $j$ changes the phase
of the amplitudes, depending on whether qubit $j$ is 0 or 1.
In terms of spins, the Hamiltonian that performs this operation is

\begin{equation}
H=-h_{j}^z S_j^z
,
 \label{SING0}
\end{equation}
and is clearly a special case of the ideal quantum computer model \Eq{UNIV0}.
Taking $j=1$ as an example, the unitary transformation

\begin{equation}
U=e^{-itH}=e^{ith_{1}^z S_1^z}
,
 \label{SING1}
\end{equation}
changes the state \Eq{STAT0} into

\begin{eqnarray}
\KET{\Phi'}=U\KET{\Phi}
&=&e^{i\phi}a({\uparrow\uparrow\ldots\uparrow}) \KET{\uparrow\uparrow\ldots\uparrow}
+e^{-i\phi}a({\downarrow\uparrow\ldots\uparrow}) \KET{\downarrow\uparrow\ldots\uparrow}
+\ldots \nonumber \\&&
+e^{i\phi}a({\uparrow\downarrow\ldots\downarrow}) \KET{\uparrow\downarrow\ldots\downarrow}
+e^{-i\phi}a({\downarrow\downarrow\ldots\downarrow}) \KET{\downarrow\downarrow\ldots\downarrow},
\nonumber \\
&=&e^{i\phi}
\left[
a({\uparrow\uparrow\ldots\uparrow}) \KET{\uparrow\uparrow\ldots\uparrow}
+e^{-2i\phi}a({\downarrow\uparrow\ldots\uparrow}) \KET{\downarrow\uparrow\ldots\uparrow}
+\ldots
\right.
\nonumber \\
&&+
\left.
a({\uparrow\downarrow\ldots\downarrow}) \KET{\uparrow\downarrow\ldots\downarrow}
+e^{-2i\phi}a({\downarrow\downarrow\ldots\downarrow}) \KET{\downarrow\downarrow\ldots\downarrow}
\right],
\nonumber \\
&=&e^{i\phi}
\left[
a_0 \KET{0}
+e^{-2i\phi}a_1 \KET{1}
+\ldots
+a_{2^L-2} \KET{2^L-2}
+e^{-2i\phi}a_{2^L-1} \KET{2^L-1}
\right]
,
\label{SING2}
\end{eqnarray}
where the phase shift $\phi$ is given by $\phi=th_{1}^z/2$,
and we used the fact that in the (spin up, spin down)-representation,
$S^z_1$ is a diagonal matrix with eigenvalues (1/2,-1/2).
From Eq.~\Eq{SING2} it is clear that by a proper choice of the time $t$ and
external field $h_{1}^z$ we can perform any phase shift operation.
To obtain the last line of Eq.~\Eq{SING2}, a global phase factor has
been extracted from all the amplitudes. This manipulation does not
alter the outcome of the calculation as it eventually drops out
in the calculation of expectation values~\cite{SCHI68,BAYM74,BALL03}.
We will often use this trick to simplify the expressions,
dropping global phase factors whenever possible.

The phase shift operation $Z_j=e^{i\phi S^z_j}$ with the $\pi/2$ rotations
\Eq{SPIN10} allows us to compose any single qubit operation.
For example, using the algebraic properties of the spin operators, it follows that

\begin{eqnarray}
e^{i\phi S^x}&=&e^{-i\pi S^y/2}e^{i\phi S^z}e^{i\pi S^y/2},
\label{SING3}
\end{eqnarray}
showing that we can readily construct a $\pi/2$ rotation about
the $z$-axis by combining rotations about the $x$ and $y$-axis.

Some simple examples may help to understand the various operations.
For a two-qubit quantum computer we have $\KET{a}=a_0\KET{00}+a_1\KET{01}+a_2\KET{10}+a_3\KET{11}$
and

\begin{equation}
X_1
\left(
\begin{array}{c}
a_0\\ a_1\\ a_2\\ a_3
\end{array}
\right)
\equiv{1\over\sqrt{2}}
\left(
\begin{array}{cccc}
1&i&0&0 \\
i&1&0&0 \\
0&0&1&i \\
0&0&i&1
\end{array}
\right)
\left(
\begin{array}{c}
a_0\\ a_1\\ a_2\\ a_3
\end{array}
\right)
,
\label{SING4}
\end{equation}
where $X_1$ denotes a rotation of qubit 1 by $\pi /2$ about the $x$-axis.
For example ${\NOBAR X_1}\KET{10}=(\KET{10}+i\KET{11})/\sqrt{2}$ and
${\BAR X_1}\KET{10}=(\KET{10}-i\KET{11})/\sqrt{2}$.
Using the same labeling of states as in \Eq{SING4}, we have

\begin{equation}
Y_2
\equiv{1\over\sqrt{2}}
\left(
\begin{array}{cccc}
\phantom{-} 1&\phantom{-} 0&\phantom{-} 1&\phantom{-} 0\\  
\phantom{-} 0&\phantom{-} 1&\phantom{-} 0&\phantom{-} 1\\  
- 1&\phantom{-} 0&\phantom{-} 1&\phantom{-} 0\\  
\phantom{-} 0&-1&\phantom{-} 0&\phantom{-} 1    
\end{array}
\right)
,
\label{SING5}
\end{equation}
for example, ${\NOBAR Y_2}\KET{10}=(\KET{00}+\KET{10})/\sqrt{2}$ and
${\BAR Y_2}\KET{10}=(-\KET{00}+\KET{11})/\sqrt{2}$.

For later use we introduce the symbol

\begin{equation}
R_j(\phi)=e^{i\phi/2}e^{-i\phi S_j^z}=
\left(
\begin{array}{cc}
\phantom{-} 1&\phantom{-} 0\\
\phantom{-} 0&\phantom{-} e^{i\phi}
\end{array}
\right)
,
\label{SING6}
\end{equation}
to represent the {\bf single-qubit phase-shift} operation by a phase $\phi$ on qubit $j$.
A symbolic representation of $R_j(\phi =\pi /k)$ is given in Fig.~\ref{fig:Gates}a.

\subsection{Two-qubit Operations}\label{twoqubit}

\begin{table}[t]
\caption{Input and output states and the corresponding expectation
values ($Q_1^z$,$Q_2^z$) of the qubits for the CNOT operation.}
  \begin{center}
    \begin{ruledtabular}
    \begin{tabular}{cccccc}
Input state&$Q_2^z$&$Q_1^z$&Output state&$Q_2^z$&$Q_1^z$\\
\hline
\noalign{\vskip 4pt}
$\KET{00}$&0&0&$\KET{00}$&0&0\\
$\KET{01}$&1&0&$\KET{11}$&1&1\\
$\KET{10}$&0&1&$\KET{10}$&1&0\\
$\KET{11}$&1&1&$\KET{01}$&0&1\\
    \end{tabular}
    \end{ruledtabular}
    \label{tab:CNOTgate}
  \end{center}
\end{table}

\begin{figure*}[ht]
\begin{center}
\includegraphics[width=10cm]{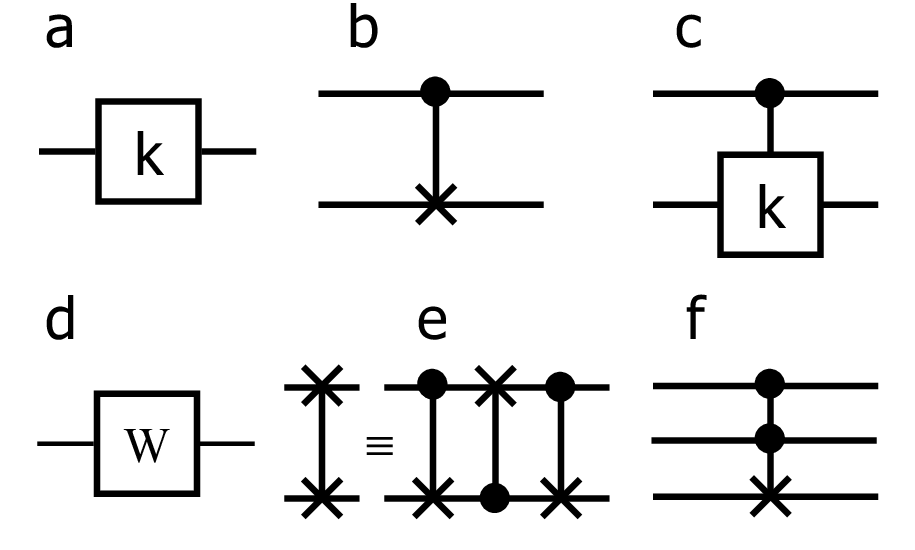}
\end{center}
\caption{
Graphical representation of some of the basic gates used in quantum computation;
(a) single qubit phase shift $R_j(\phi =\pi /k)$ (see Eq.~\ref{SING6});
(b) CNOT gate (see Eq.~\ref{COMM3});
(c) controlled phase shift by $R_{ji}(\phi =\pi/k)$ (see Eq.~\ref{COMM4});
(d) Walsh-Hadamard gate (see Eq.~\ref{HADA1});
(e) SWAP gate (see Eq.~\ref{SWAP1});
(f) Toffoli gate (see Eq.~\ref{TOFF2}).
The horizontal lines denote the qubits involved in the quantum operations. The dots and crosses denote the
control and target bits, respectively.}
\label{fig:Gates}
\end{figure*}

As explained previously, the single qubit operations and the CNOT gate
constitute a universal set of gates~\cite{NIEL00}.
The CNOT gate is defined by its action on the computational basis states,
as shown in Table~\ref{tab:CNOTgate}.
To simplify the notation, we consider only the two relevant qubits
and call them qubit 1 and 2.
The CNOT gate flips the second qubit if the state of the first qubit is $\KET{1}$,
that is, the first qubit acts as a control qubit for the second one.
As shown in Fig.~\ref{fig:Gates}b, the CNOT gate can be symbolically represented by a vertical line connecting
a dot (control bit) and a cross (target bit).
On the ideal quantum computer model \Eq{UNIV0},
the CNOT gate can be implemented by a combination of single-qubit operations
and a controlled phase shift operation.
In matrix notation, the CNOT operation (see Table~\ref{tab:CNOTgate})
can be written as

\begin{equation}
\CNOT
\left(
\begin{array}{c}
a_0\\ a_1\\ a_2\\ a_3
\end{array}
\right)
\equiv
C_{21}
\left(
\begin{array}{c}
a_0\\ a_1\\ a_2\\ a_3
\end{array}
\right)
=
\left(
\begin{array}{cccc}
1&0&0&0 \\
0&0&0&1 \\
0&0&1&0 \\
0&1&0&0
\end{array}
\right)
\left(
\begin{array}{c}
a_0\\ a_1\\ a_2\\ a_3
\end{array}
\right)
.
\label{COMM1}
\end{equation}

Our goal is to show that up to an irrelevant global
phase factor, $\CNOT={\BAR Y_2}I_{21}(\pi){\NOBAR Y_2}$ where

\begin{equation}
I_{21}(\phi)=e^{i\phi(2S_1^zS_2^z -S_1^z-S_2^z)/2}=
\left(
\begin{array}{cccc}
\phantom{-} e^{-i\phi/4} &\phantom{-}0&\phantom{-}0&\phantom{-}0 \\
\phantom{-}0&\phantom{-} e^{-i\phi/4} &\phantom{-}0&\phantom{-}0 \\
\phantom{-}0&\phantom{-}0&\phantom{-} e^{-i\phi/4} &\phantom{-}0 \\
\phantom{-}0&\phantom{-}0&\phantom{-}0& e^{3i\phi/4}
\end{array}
\right).
\label{COMM2}
\end{equation}
It is easy to see that

\begin{equation}
I_{21}(\phi)=e^{-it(-JS_1^zS_2^z -hS_1^z-hS_2^z)},
\label{COMM2a}
\end{equation}
where $h=-J/2$ and $Jt=\phi$.
Eq.~\Eq{COMM2a} shows that physically the phase-shift operation \Eq{COMM2}
corresponds to the time evolution of a quantum spin system described by the Ising model.

Using the properties of the $S=1/2$ matrices, it is easy to show that

\begin{equation}
\CNOT=
\BAR Y_2
\left(
\begin{array}{cccc}
\phantom{-}1&\phantom{-}0&\phantom{-}0&\phantom{-}0 \\
\phantom{-}0&\phantom{-}1&\phantom{-}0&\phantom{-}0 \\
\phantom{-}0&\phantom{-}0&\phantom{-}1&\phantom{-}0 \\
\phantom{-}0&\phantom{-}0&\phantom{-}0&-1
\end{array}
\right) \NOBAR Y_2=
e^{i\pi/4}\BAR Y_2 I_{21}(\pi)\NOBAR Y_2
\label{COMM3}
.
\end{equation}
This completes the construction of the set of universal gates that
can be implemented on the ideal Ising-model quantum computer \Eq{UNIV0}.

The CNOT operation contains a special case of the {\bf controlled
phase shift operation} $R_{ji}(\phi)$. The latter appears in several of the
quantum algorithms that are discussed later
and therefore it is appropriate to introduce it here.
The controlled phase shift operation on qubits 1 and 2 reads

\begin{eqnarray}
R_{21}(\phi)&=&e^{i\phi/4}I_{21}(\phi)=
\left(
\begin{array}{cccc}
\phantom{-}1&\phantom{-}0&\phantom{-}0&\phantom{-}0 \\
\phantom{-}0&\phantom{-}1&\phantom{-}0&\phantom{-}0 \\
\phantom{-}0&\phantom{-}0&\phantom{-}1&\phantom{-}0 \\
\phantom{-}0&\phantom{-}0&\phantom{-}0&e^{i\phi}
\end{array}
\right)
\label{COMM4}
.
\end{eqnarray}
Graphically, the controlled phase shift $R_{ij}(\phi=\pi/k)$
is represented by a vertical line connecting a dot (control bit)
and a box denoting a single qubit phase shift by $\pi /k$ (see Fig.~\ref{fig:Gates}c).

\setlength{\unitlength}{1cm}

\section{Numerical Methods}\label{numerical}

We start by discussing aspects that are common to all numerical algorithms discussed
in this review such as techniques to let (combinations of) spin-1/2 operators act on a state.
Then we review four different algorithms to compute the time evolution
$\widetilde U(t)\KET{\Phi}=e^{-itH}\KET{\Phi}$.
We start by a brief exposition of the full-diagonalization approach.
Then we describe the Chebyshev polynomial algorithm,
the short-iterative Lanczos procedure, and
several algorithms based on Suzuki-Trotter product formulas.

In numerical work where the use of floating-point
arithmetic (e.g. 13-15 digit precision) is necessary,
there always will be errors due to the finite precision, rounding and the like.
In this chapter we use the term {\bf numerically exact} to indicate that
the results are correct up to an error that vanishes
with the number of digits that are used to perform the calculation.
For example, the value of $\arccos(-1)=\pi$ is numerically exact.

\subsection{General Aspects}\label{numa}

As explained previously, the state of a spin-1/2 system with $P$ pairs of
$L$ interacting spins or, equivalently, the
state of a $L$-qubit quantum computer, is represented by a complex-valued
vector of length $D=2^L$.
In quantum mechanics matrices operating on a vector of length $D$ have dimension $D\times D$.
Matrix elements of the Hamiltonian \Eq{hamiltonian} can be complex numbers.
Hence we need $2^{2L+4}$ bytes to store the full matrix.
For instance, in the case of a 10-qubit quantum
computer we need 16 Mb (16 kb) of memory to store the Hamiltonian (state)
which poses no problem for present-day PCs.
However, to store the Hamiltonian (state) of the $L=20$ system,
a computer should have 16 Tb (16Mb) of memory.
This simple example clearly shows that
any simulation scheme that requires storage for the matrix
representing the Hamiltonian will be of (very) limited use.
Thus, it is necessary to consider techniques to avoid the use of
large matrices. Fortunately, for the spin systems considered in this
review it is rather straightforward to organize the calculations
such that we need only storage for \ORDER{PL} matrices of dimension $2\times2$
and $4\times4$.

We first consider the operations $\KET{\Phi'}=S_j^\alpha\KET{\Phi}$.
More specifically, let us denote

\begin{eqnarray}
\KET{\Phi}=
& &a({\uparrow\uparrow\ldots\uparrow}) \KET{\uparrow\uparrow\ldots\uparrow}
+a({\downarrow\uparrow\ldots\uparrow}) \KET{\downarrow\uparrow\ldots\uparrow}
+\ldots 
+a({\uparrow\downarrow\ldots\downarrow}) \KET{\uparrow\downarrow\ldots\downarrow}
+a({\downarrow\downarrow\ldots\downarrow}) \KET{\downarrow\downarrow\ldots\downarrow}
,
\label{NUM0}
\end{eqnarray}
and

\begin{eqnarray}
\KET{\Phi'}&=&S_j^\alpha\KET{\Phi}\nonumber \\
&=&a'({\uparrow\uparrow\ldots\uparrow}) \KET{\uparrow\uparrow\ldots\uparrow}
+a'({\downarrow\uparrow\ldots\uparrow}) \KET{\downarrow\uparrow\ldots\uparrow}
+\ldots 
+a'({\uparrow\downarrow\ldots\downarrow}) \KET{\uparrow\downarrow\ldots\downarrow}
+a'({\downarrow\downarrow\ldots\downarrow}) \KET{\downarrow\downarrow\ldots\downarrow}
.
\label{NUM1}
\end{eqnarray}
From Eq.~\Eq{spinmatrices}, it follows that $S_j^z$ is a diagonal matrix
in the representation that we use to store the vector $\KET{\Phi}$.
Thus we obtain $\KET{\Phi'}$ by reversing the sign
of all coefficients of $\KET{\Phi}$ for which the $j$th bit of their index is one.
In terms of amplitudes, we have

\begin{eqnarray}
a'({*\ldots*0*\ldots*})&=&+\frac{1}{2}\, a({*\ldots*0*\ldots*})\nonumber \\
a'({*\ldots*1*\ldots*})&=&-\frac{1}{2}\, a({*\ldots*1*\ldots*}),
\label{NUM2}
\end{eqnarray}
where we use the asterisk to indicate that the bits on the corresponding position
are the same.
From Eq.~\Eq{spinmatrices}, it follows that $S_j^x$ interchanges states up and down.
It immediately follows that the elements of $\KET{\Phi'}$ are obtained
by swapping pairs of elements of $\KET{\Phi}$.
We have

\begin{eqnarray}
a'({*\ldots*0*\ldots*})&=&+\frac{1}{2}\, a({*\ldots*1*\ldots*})\nonumber \\
a'({*\ldots*1*\ldots*})&=&+\frac{1}{2}\, a({*\ldots*0*\ldots*}).
\label{NUM3}
\end{eqnarray}
In Eq.~\Eq{NUM3}, the bit strings on the left- and right-hand side are identical
except for the $j$th bit.
In the case of $\KET{\Phi'}=S_j^y\KET{\Phi}$, a similar argument shows that

\begin{eqnarray}
a'({*\ldots*0*\ldots*})&=&-\frac{i}{2}\, a({*\ldots*1*\ldots*})
\nonumber \\
a'({*\ldots*1*\ldots*})&=&+\frac{i}{2}\, a({*\ldots*0*\ldots*}).
\label{NUM4}
\end{eqnarray}
Note that all these operations can be done {\sl in place},
that is, without using another vector of length $2^L$.
The three operations discussed here are sufficient to construct
any algorithm to simulate the quantum dynamics of model \Eq{hamiltonian}.
However, for reasons of computational efficiency, it is
advantageous to extend the repertoire of elementary operations a little.

The two-spin operation $\KET{\Phi''}=S_j^\alpha S_k^\alpha\KET{\Phi}$
can be implemented in at least three different ways.
In the discussion that follows, we exclude the trivial case where $j=k$.
One obvious method would be to write it as two single-spin operations
of the kind discussed previously:
$\KET{\Phi'}=S_k^\alpha \KET{\Phi}$ followed by $\KET{\Phi''}=S_j^\alpha \KET{\Phi'}$.
For a single pair $(j,k)$, this is an efficient method.
However, if there are many
pairs, as in the case of Hamiltonian \Eq{hamiltonian},
we have either to recalculate $S_k^\alpha \KET{\Phi}$ several times,
or we have to allocate extra storage to hold $S_k^\alpha \KET{\Phi}$ for $k=1,\ldots,L$.
In both cases, we use extra, costly resources.

In the second approach, we work out analytically how the amplitudes
change if we apply $S_j^\alpha S_k^\alpha$ to $\KET{\Phi}$.
If $\alpha=z$, we have to reverse the sign of the amplitude only
if the $j$th and $k$th bit of the amplitude index are different.
We have

\begin{eqnarray}
a''({*\ldots*0*\ldots*0*\ldots*})&=&+\frac{1}{4}\, a({*\ldots*0*\ldots*0*\ldots*})\nonumber \\
a''({*\ldots*1*\ldots*0*\ldots*})&=&-\frac{1}{4}\, a({*\ldots*1*\ldots*0*\ldots*})\nonumber \\
a''({*\ldots*0*\ldots*1*\ldots*})&=&-\frac{1}{4}\, a({*\ldots*0*\ldots*1*\ldots*})\nonumber \\
a''({*\ldots*1*\ldots*1*\ldots*})&=&+\frac{1}{4}\, a({*\ldots*1*\ldots*1*\ldots*}).
\label{NUM5}
\end{eqnarray}
If $\alpha=x$, we interchange quadruples of amplitudes according to the
following, rather obvious, rules:

\begin{eqnarray}
a''({*\ldots*0*\ldots*0*\ldots*})&=&\frac{1}{4}\, a({*\ldots*1*\ldots*1*\ldots*})\nonumber \\
a''({*\ldots*1*\ldots*0*\ldots*})&=&\frac{1}{4}\, a({*\ldots*0*\ldots*1*\ldots*})\nonumber \\
a''({*\ldots*0*\ldots*1*\ldots*})&=&\frac{1}{4}\, a({*\ldots*1*\ldots*0*\ldots*})\nonumber \\
a''({*\ldots*1*\ldots*1*\ldots*})&=&\frac{1}{4}\, a({*\ldots*0*\ldots*0*\ldots*}).
\label{NUM6}
\end{eqnarray}
We leave the case $\alpha=y$ as an exercise for the reader.
Clearly, this approach does not require additional storage or calculations to
compute intermediate results.

The third method relies on the observation that for $\alpha=x,y$ and $j\not=k$

\begin{eqnarray}
S_j^\alpha S_k^\alpha \KET{\Phi}=R_j^\alpha R_k^\alpha S_j^z  S_k^z (R_j^\alpha)^\dagger
(R_k^\alpha)^\dagger \KET{\Phi},
\label{NUM7}
\end{eqnarray}
where $R_j^\alpha$ is a rotation that transforms $S_j^z$ into $S_j^\alpha$
and that the calculation of $S_j^z  S_k^z \KET{\Phi}$ can be done very efficiently.
From Eq.~\Eq{SPIN7} it follows that
${\NOBAR Y_j}S_j^x{\BAR Y_j}=S_j^z$ and ${\BAR X_j}S_j^y{\NOBAR X_j}=S_j^z$.
Hence it follows that $R_j^x={\BAR Y_j}$ and $R_j^y={\NOBAR X_j}$.
From Eq.~\Eq{NUM7}, it is clear that we have to determine the rules only to compute
$\KET{\Phi'}= (R_j^\alpha)^\dagger \KET{\Phi}$ for $\alpha=x,y$.
These rules follow directly from the matrix representations \Eq{SPIN10} of
${\NOBAR X_j}$ and ${\NOBAR Y_j}$.
For instance, in the case of $\KET{\Phi'}= (R_j^x)^\dagger \KET{\Phi}$, we have

\begin{eqnarray}
a'({*\ldots*0*\ldots*})&=&+\frac{1}{\sqrt{2}}[a({*\ldots*0*\ldots*})-a({*\ldots*1*\ldots*})]
\nonumber \\
a'({*\ldots*1*\ldots*})&=&+\frac{1}{\sqrt{2}}[a({*\ldots*0*\ldots*})+a({*\ldots*1*\ldots*})],
\label{NUM8}
\end{eqnarray}
and similar expressions hold for the other cases.

All the single- and two-spin operations discussed here
can be carried out in \ORDER{2^L} floating-point operations.
The full diagonalization method, the
Chebyshev and short-iterative Lanczos algorithm to be discussed later
require a procedure that computes $\KET{\Phi'}=H\KET{\Phi}$.
Given the parameters of $H$ [see Eq.~\Eq{hamiltonian}], it is
straightforward to combine the single- and two-spin operations that were described
to perform the operation $\KET{\Phi'}=H\KET{\Phi}$.

The single- and two-qubit operations described before
suffice to implement any unitary transformation on the state vector.
As a matter of fact, the simple rules described are all that we need to
simulate ideal quantum computers on a conventional computer.
In Section~\ref{idealqc} we discuss this topic in more detail.
For now, we continue with reviewing numerical techniques for
computing the time evolution $e^{-itH}$ of a physical system
described by a Hamiltonian $H$.

\subsection{Full Diagonalization Approach}

As $H$ is a $D\times D$ Hermitian matrix, it has a complete set of eigenvectors
and real-valued eigenvalues~\cite{WILK65,GOLU96,PARL81,BELL97}.
Let us denote the diagonal matrix of all these eigenvalues by $\Lambda$
and the unitary matrix of all these eigenvectors by $V$.
Then we have $V^{\dagger}HV=\Lambda$ and $U(t)=e^{-itH}=Ve^{-it\Lambda}V^{\dagger}$.
In other words, once we know $\Lambda$ and $V$, $U(t)$ is obtained by
simple matrix multiplication.
Thus, the most straightforward method to compute $U(t)=e^{-itH}=Ve^{-it\Lambda}V^{\dagger}$
is to use standard linear-algebra algorithms to diagonalize the matrix $H$.
An appealing feature of this approach is that the most complicated
part of the algorithm, the diagonalization of the matrix $H$,
is relegated to well-developed linear algebra packages.
Memory and CPU time of the standard
diagonalization algorithms scale as $D^2$
and $D^3$, respectively~\cite{WILK65,GOLU96,PARL81}.

The matrix elements of $H$ can be computed by repeated use of the
$\KET{\Phi'}=H\KET{\Phi}$ procedure.
If we set $\KET{\Phi}=(1,0,\ldots,0)^T$, then
$\KET{\Phi'}=H\KET{\Phi}$ gives us the first column of the matrix $H$.
The same calculation for $\KET{\Phi}=(0,1,\ldots,0)^T$ yields the second
column, and so on.

The application of the full-diagonalization approach is limited by
the memory and CPU resources it requires.
The former is often the most limiting factor (see the prior discussion),
in spite of the fact that full diagonalization takes
\ORDER{2^{3L}} floating-point operations.
In practice, solving the time-dependent Schr\"odinger equation for a system of 12-13 spins for many pairs ($H$,$t$)
(which is necessary if there are time-dependent fields)
is many orders of magnitude more expensive in terms of computational resources
than solving the same problem by the methods discussed later.
Nevertheless, any toolbox for simulating quantum spin dynamics should include
code that is based on the full diagonalization approach.
For all but the most trivial problems, it is an essential tool
to validate the correctness of other algorithms that solve the time-dependent Schr\"odinger equation.

\subsection{Chebyshev Polynomial Algorithm}

The basic idea of this approach is to make use of a numerically exact
polynomial approximation to the matrix exponential
$U(t)=e^{-itH}$~\cite{TALE84,TALE89,LEFO91,IITA97,SILV97,LOH00,DOBR03}.
The first step in the algorithm is to ``normalize'' the matrix $H$ such that
its (unknown) eigenvalues lie in the interval $[-1,1]$.
In general, the eigenvalues $E_j$ of the Hermitian matrix $H$ are real numbers
in the interval $[-\Vert H\Vert,\Vert H\Vert]$ where
$\Vert H \Vert\equiv\max_{j}|E_j|$
is the 2-norm of the matrix $H$~\cite{WILK65,GOLU96}.
Unless we already solved the eigenvalue problem of $H$,
we usually do not know the exact value of $\Vert H\Vert$.
For the Hamiltonian \Eq{hamiltonian},
it is easy to compute an upper bound to $\Vert H\Vert$ by repeated use of the triangle inequality

\begin{equation}
\Vert X + Y \Vert \le \Vert X\Vert + \Vert Y\Vert,
\label{triangle}
\end{equation}
and the elementary bounds $\Vert S_j^\alpha\Vert=1/2$ and $\Vert S_j^\alpha S_k^\alpha\Vert=1/4$.
For fixed $t$, we can write $H=H(t)$, and we find~\cite{DOBR03}

\begin{equation}
\Vert H \Vert \le
\Vert H \Vert_b \equiv\frac{1}{4}\sum_{i,j=1}^L\sum_{\alpha=x,y,z} | J_{i,j}^\alpha |
+\frac{1}{2}\sum_{i=1}^L\sum_{\alpha=x,y,z} | h_{i}^\alpha|.
\label{upperbound1}
\end{equation}
Maximum efficiency of the Chebyshev polynomial algorithm
is obtained if we know the exact value $\Vert H \Vert$ (see later).
Thus, if we have more specific information about the model parameters
$J_{i,j}^\alpha$ and $h_{i}^\alpha$,
it should be used to improve the bound on $\Vert H \Vert$.
By construction, the eigenvalues of $\widehat H\equiv H/\Vert H\Vert_b$
all lie in the interval $[-1,1]$.

Expanding the initial state $\Phi(0)$ in the (unknown) eigenvectors $\KET{E_j}$
of $H$, we have

\begin{equation}
\KET{\Phi(t)}=e^{-itH}\KET{\Phi(0)}=
e^{-i\widehat t\widehat H}\KET{\Phi(0)} =\sum_j e^{-i\widehat t\widehat E_j} \KET{E_j}\BRACKET{E_j}{\Phi(0)},
\label{phit}
\end{equation}
where $\widehat t=t\Vert H\Vert_b$ and
the $\widehat E_j=E_j/\Vert H\Vert_b$ denote the (unknown) eigenvalues of $\widehat H$
(in practice, there is no need to know the eigenvalues and eigenvectors of
$H$ explicitly).
Next, recall that for $|x|\le 1$ we have~\cite{ABRA64}

\begin{equation}
e^{-izx}=J_0(z) + 2\sum_{k=1}^{\infty} (-i)^{k} J_{k}(z) T_{k}(x)
,
\label{expzx}
\end{equation}
where $J_k(z)$ is the Bessel function of integer order $k$,
and $T_k(x)=\cos[k\arccos(x)]$ is the Chebyshev polynomial
of the first kind~\cite{ABRA64}.
As $-1\le \widehat E_j \le 1$, we can use Eq.~\Eq{expzx} to write
Eq.~\Eq{phit} as
\begin{equation}
\KET{\Phi(t)}
=\left[J_0(-\widehat t)\openone + 2\sum_{k=1}^{\infty} J_{k}(-\widehat t)\widehat{T}_{k}(\widehat H)\right]
\KET{\Phi(0)}
,
\label{CHEB0}
\end{equation}
where $\widehat T_{k}(\widehat H)=i^k T_k(\widehat H)=
i^k\sum_j \KET{E_j}T_k(\widehat E_j)\BRA{E_j}$ is a
matrix-valued, modified Chebyshev polynomial that is defined by the recursion
relations
\begin{equation}
\widehat{T}_{0}(\widehat H)\KET{\Psi}=\KET{\Psi}\,,\quad
\widehat{T}_{1}(\widehat H)\KET{\Psi}=i\widehat H \KET{\Psi}\,,\quad
\label{CHEB1}
\end{equation}
and
\begin{equation}
\widehat{T}_{k+1}(\widehat H)\KET{\Psi}=
2i\widehat H\widehat{T}_{k}(\widehat H)\KET{\Psi}+\widehat{T}_{k-1}(\widehat H)\KET{\Psi}\,,
\label{CHEB2}
\end{equation}
for $k\ge1$.

For practical purposes the sum in Eq.(\ref{CHEB0}) has to be truncated.
We compute the approximation $\KET{\widetilde\Phi(t)}=\widetilde U(t)\KET{\Phi(0)}$
by keeping the first $K+1$ contributions:

\begin{equation}
\KET{\widetilde\Phi(t)}
=\left[J_0(-\widehat t)\openone + 2\sum_{k=1}^{K} J_{k}(-\widehat t)\widehat{T}_{k}(\widehat H)\right]
\KET{\Phi(0)}.
\label{CHEB3}
\end{equation}
As $|J_k(-\widehat z)|<1$ and
$\Vert\widehat T_{k}(\widehat H)\Vert=\Vert T_{k}(\widehat H)\Vert\le1$,
all contributions in Eq.~\Eq{CHEB3} are of the same order of magnitude.
In contrast, the truncated Taylor series
$e^{-itH}\approx\sum_{k=1}^{K}(-itH)^k/k!$ is a sum of many small and large numbers,
and a numerical calculation of the sum is known to suffer from severe
instabilities~\cite{MOLE78}.

\begin{figure}[t]
\begin{center}
\includegraphics[width=10cm]{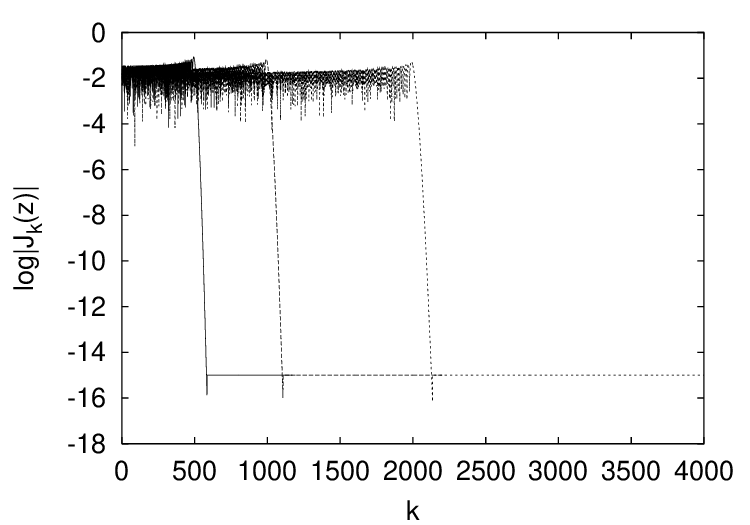}
\caption{Dependence of the Bessel function $J_k(z)$ on the order $k$;
solid line: $z=500$;
dashed line: $z=1000$;
dotted line: $z=2000$.}
\label{Jnz}
\end{center}
\end{figure}

Using the downward recursion relation of the Bessel functions,
it is straightforward to calculate the first $K$ Bessel functions to machine precision
in \ORDER{K} arithmetic operations~\cite{ABRA64,PRES86}.
In practice, a calculation of the first 20,000 Bessel functions takes less than 1 second
on a Pentium III 600 MHz mobile processor, using 14 - 15 digit arithmetic.
To get a feeling for the value of $K$ at which the Chebyshev polynomial expansion
might be truncated, it is instructive to plot $J_k(z)$ as a function of $k$.
From Fig.\ref{Jnz} it is clear that $|J_k(z)|$ vanishes (very) rapidly
if $k$ becomes larger than $z$. In fact, $|J_k(z)|\le z^k/2^k k!$ for real $z$ hence
$|J_k(z)|$ vanishes exponentially fast as $k$ increases~\cite{ABRA64}.
Thus we can fix the number $K$ by requiring that $|J_{k}(z)|>\kappa$ for all $k\le K$.
Here $\kappa$ is a small number that determines the accuracy of the approximation.
From Fig.\ref{Jnz}, it follows that $K\approx z$ and that
$|J_k(z=2000)|<10^{-10}$ for all $k>z+100$.

Once we have found the smallest $K$ such that $|J_{k}(-\widehat t)|>\kappa$
for all $k\le K$, there is no point of taking more than $K$ terms in the expansion.
Since $\Vert\widehat T_{k}(\widehat H)\Vert\le1$,
Fig.\ref{Jnz} suggests that including such contributions would only add to the noise,
and numerical tests show that this is indeed the case.
However, taking less than $K$ terms has considerable negative impact on
the accuracy of the results. Hence in practice the choice of $K$ is rather limited
(such as $K\approx z+200$ if $z=2000$ and machine precision is required).
Most important, for fixed $\kappa$, $K$ increases linearly with $t \Vert H \Vert_b$.
Therefore, if we can replace the bound $\Vert H \Vert_b$ by a sharper one,
the value of $K$ can be reduced accordingly and the calculation will take less CPU time.

In a strict sense, the truncated Chebyshev polynomial in Eq.~\Eq{CHEB3} is not a unitary matrix.
However, the error $\Vert \Phi(t)-\widehat\Phi(t)\Vert$ between the exact state $\Phi(t)$
and the truncated Chebyshev polynomial result $\widehat\Phi(t)$
vanishes (exponentially) fast as $K$ increases.
Therefore, for practical purposes the truncated Chebyshev polynomial
can be viewed as an extremely stable
time-integration algorithm because it yields an approximation to the
exact time evolution operator $U(t)=e^{-itH}$ that is numerically exact.
Under these conditions, the Chebyshev algorithm may safely be used repeatedly
to perform multiple time steps with a (very) large fixed time step~\cite{DOBR03}.

According to Eq.(\ref{CHEB3}), performing one time step amounts to repeatedly using
recursion relation (\ref{CHEB2}) to obtain
$\widehat T_{k}(\widehat H) \Phi(0)$ for $k=2,\ldots,K$,
then multiply the elements of this vector by $J_{k}(-\widehat t)$ and add
all contributions. This procedure requires storage for two vectors of the same
length as $\Phi$ and a procedure that returns $H\KET{\Phi}$.
Both memory and CPU time of this method are \ORDER{D=2^L}.
CPU time increases linearly with the time step $t$.

\subsection{Short-Iterative Lanczos Algorithm}

The short-iterative Lanczos algorithm~\cite{LEFO91,PARK86,MANT91,JACK94,JACK00}
belongs to the class of reduced-order methods that
are based on the approximation of a matrix by its projection onto a (much) smaller subspace.
The short-iterative Lanczos algorithm is based on the approximation

\begin{equation}
e^{-it H}\KET{\Psi}\approx \widetilde U_N(t)=e^{-it P_NHP_N}\KET{\Psi},
\label{SIL1}
\end{equation}
where $P_N$ is the projector on the $N$-dimensional subspace spanned by the vectors
$\{\Psi,H\Psi,\ldots,H^{N-1}\Psi\}$.
As $e^{-it P_NHP_N}$ is unitary the method is unconditionally stable.

There are two major steps in the short-iterative Lanczos algorithm.
First we calculate $P_N H P_N\Psi$ by generating
the orthogonal Lanczos vectors in the usual manner~\cite{WILK65,GOLU96}:

\begin{eqnarray}
\KET{\Psi^\prime_0}&=&\KET{\Psi},\quad
\nonumber \\
\KET{\Psi_0}&=&\KET{\Psi^\prime_0}/\sqrt{\BRACKET{\Psi^\prime_0}{\Psi^\prime_0}},
\nonumber \\
\KET{\Psi^\prime_1}&=&H\KET{\Psi_0}-\KET{\Psi_0}\BRACKET{\Psi_0}{H|\Psi_0},\quad
\nonumber \\
\KET{\Psi_1}&=&\KET{\Psi^\prime_1}/\sqrt{\BRACKET{\Psi^\prime_1}{\Psi^\prime_1}},
\label{SIL2}
\end{eqnarray}
and
\begin{eqnarray}
\KET{\Psi^\prime_k}&=&H\KET{\Psi_{k-1}}
-\KET{\Psi_{k-1}}\BRACKET{\Psi_{k-1}}{H|\Psi_{k-1}}
-\KET{\Psi_{k-2}}\BRACKET{\Psi_{k-2}}{H|\Psi_{k-2}},
\nonumber \\
\KET{\Psi_k}&=&\KET{\Psi^\prime_k}/\sqrt{\BRACKET{\Psi^\prime_k}{\Psi^\prime_k}},
\label{SIL3}
\end{eqnarray}
for $1\le k\le N$.
By construction, we have $\BRACKET{\Psi_k}{\Psi_{k'}}=\delta_{k,k'}$, and
the projection with $P_N=\sum_{k=1}^N \KET{\Psi_k}\BRA{\Psi_k}$
yields a $N\times N$ tri-diagonal matrix $P_N H P_N$.
The second step in the short-iterative Lanczos algorithm
is to diagonalize this tridiagonal matrix and use the resulting
eigenvalues and eigenvectors to compute $e^{-it P_N H P_N}\KET{\Psi}$.
The latter is done in exactly the same manner
as in the case of the full diagonalization method.

The short-iterative Lanczos algorithm requires storage
for all the eigenvectors and/or the $N$ Lanczos vectors $\KET{\Psi_k}$
of the $N\times N$ matrix $P_N H P_N$.
Thus, for this method to compete with the full diagonalization
method, we require $N\ll D=2^L$.
The accuracy of this algorithm depends both on the order $N$ and the state
$\KET{\Psi}$~\cite{LEFO91,PARK86,MANT91}.
With infinite-precision arithmetic,
$e^{-it H}\KET{\Psi}=\lim_{N\rightarrow\infty} e^{-it P_NHP_N}\KET{\Psi}$, but in
practice, the loss of orthogonality during the Lanczos procedure limits the
order $N$ and the time step $t$ that can be used without
introducing spurious eigenvalues~\cite{WILK65,GOLU96}.
The low-order short-iterative Lanczos algorithm
may work well if $\KET{\Psi}$ contains contributions from
eigenstates of $H$ that are close in energy.
However, if $\KET{\Psi}$ contains contributions from many
eigenstates of $H$ with very different energies,
it is unlikely that all these eigenvalues will be
present in $P_NHP_N$, and the approximation $\widetilde U_N(t)$
may be rather poor unless $N$ is sufficiently large.
Memory and CPU time (for one time step) of the short-iterative Lanczos algorithm scale
as $D$ and $N^2D$, respectively.
In general, $N$ increases with $t$ in a nontrivial, problem-dependent manner
that seems difficult to determine in advance.

\subsection{Suzuki-Trotter Product-Formula Algorithms}\label{SUZTROT}

A systematic approach to construct unitary approximations to unitary matrix
exponentials is to make use of the Lie-Trotter-Suzuki product-formula~\cite{TROT59,SUZU77}
\begin{equation}
U(t)=e^{-itH}=e^{-it(H_1+\ldots+H_K)}=
\lim_{m\rightarrow\infty}
\left(\prod_{k=1}^K e^{-it{H}_k/m}\right)^m,
\label{SUZ0}
\end{equation}
and generalizations thereof~\cite{SUZU85,RAED83,SUZU91}.
Expression \Eq{SUZ0} suggests that
\begin{equation}
\widetilde U_1(t)=e^{-it{H}_1}\ldots e^{-it{H}_K},
\label{SUZ1}
\end{equation}
might be a good approximation to $U(t)$ if $t$ is sufficiently small.
From Eq.~\Eq{SUZ1}, it follows that the Taylor series of
$U(t)$ and $\widetilde U_1(t)$ are identical up to first order in $t$.
We call $\widetilde U_1(t)$ a first-order approximation to $U(t)$.
If all the $H_i$ in Eq.~\Eq{SUZ1}
are Hermitian, then $\widetilde U_1(t)$ is unitary by construction,
and a numerical algorithm based on \Eq{SUZ1} will be unconditionally stable.
For unitary matrices $U(t)$ and $\widetilde U_1(t)$, it can be shown that~\cite{RAED87}
\begin{equation}
\|U(t)-\widetilde U_1(t)\|\leq\frac{t^2}{2}\sum_{i<j}\|[{H}_i,{H}_j]\|,
\label{SUZ2}
\end{equation}
suggesting that $\widetilde U_1(t)$ may be a good
aproximation if we advance the state of the quantum system
by small time steps $t$ such that $t\Vert H\Vert\ll1$.
Note that this is the situation of interest if there are
external time-dependent fields.

The Suzuki-Trotter product-formula approach provides a simple, systematic framework to
improve the accuracy of the approximation to $U(\tau)$
with marginal programming effort and without changing its
fundamental properties.
For example, the matrix
\begin{equation}
\widetilde U_2(t)={\widetilde U_1^\dagger(-t/2)}\widetilde U_1(t/2)=
e^{-it{H}_K/2}\ldots e^{-it{H}_1/2}e^{-it{H}_1/2}\ldots e^{-it{H}_K/2},
\label{SUZ3}
\end{equation}
is a second-order approximation to $\widetilde U(t)$~\cite{SUZU85,SUZU91,RAED83}.
If $\widetilde U_1(t)$ is unitary, so is $\widetilde U_2(t)$.
For unitary $\widetilde U_2(t)$, we have~\cite{RAED87}

\begin{equation}
\Vert U(t)-\widetilde U_2(t)\Vert \leq c_2 t^2,
\label{SUZ4}
\end{equation}
where $c_2$ is a positive constant~\cite{RAED87}.

Higher-order approximations based on $\widetilde U_2(t)$ [or $\widetilde U_1(t)$]
can be constructed by using Suzuki's fractal decomposition~\cite{SUZU85,SUZU91}.
A particularly useful fourth-order approximation is given by~\cite{SUZU85,SUZU91}
\begin{equation}
\widetilde U_4(t)=\widetilde U_2(at)\widetilde U_2(at)\widetilde U_2((1-4a)t)\widetilde U_2(at)\widetilde U_2(at),
\label{SUZ5}
\end{equation}
where $a=1/(4-4^{1/3})$.
As before, if $\widetilde U_2(t)$ is unitary, so is $\widetilde U_4(t)$ and we have

\begin{equation}
\Vert U(t)-\widetilde U_4(t)\Vert \leq c_4 t^4,
\label{SUZ6}
\end{equation}
where $c_4$ is a positive constant.
The rigorous error bounds \Eq{SUZ2}, \Eq{SUZ4} and \Eq{SUZ6}
suggest that for sufficiently small $t$, the numerical error
$\Vert U(t)\Psi -\widetilde U_n(t)\Psi\Vert$ vanishes as $t^n$.
If this behavior is not observed, there is a fair chance that
there is at least one mistake in
the computer program that implements $\widetilde U_n(t)$.
For a fixed accuracy, memory and CPU time of an $\widetilde U_n(t)$ algorithm scale as
\ORDER{D} and \ORDER{t^{(1+1/n)}D}, respectively.
The approximations \Eq{SUZ3} and \Eq{SUZ5}
have proven to be very useful for a wide range of different applications
~\cite{SUZU77,FEIT82,RAED83,RAED87,HANS89,KREC98,KOBA94,ROUH95,KAWA96,OHTS97,SHAD97,TRAN98,MICH98,RAED00,KOLE00}.

In practice, an efficient implementation of the first-order scheme is all
that is needed to build higher-order algorithms \Eq{SUZ3} and \Eq{SUZ5}.
The crucial step that we have to consider now is how to choose the Hermitian $H_i$s
such that the matrix exponentials $\exp(-it H_1)$, ..., $\exp(-it H_K)$
can be calculated efficiently.

If there are external time-dependent fields, as in the case of a physical quantum computer,
it is expedient to decompose the Hamiltonian into single-spin and two-spin contributions.
The rational behind this is that in most cases of interest (NMR,...)
the external fields change on a much shorter time scale than the other parameters
in the Hamiltonian.
As the spin operators with different qubit labels commute, we have

\begin{equation}
\exp\left\{-it\left[-\sum_{j=1}^L\sum_{\alpha=x,y,z} h_{j}^\alpha S_j^\alpha\right]\right\}
=\prod_{j=1}^L \exp\left[ it\sum_{\alpha=x,y,z} h_{j}^\alpha S_j^\alpha\right]
.
 \label{SUZ7}
\end{equation}
The $j$th factor of the product in \Eq{SUZ7} rotates spin $j$
about the vector ${\bf h}_{j}=(h_j^x,h_j^y,h_j^z)$.
Each factor can be worked out analytically, yielding

\begin{equation}
e^{it{\bf S}_j\cdot{\bf h}_{j}}
=\left(
\begin{array}{cc}
\cos \frac{th_j}{2}
+\frac{ih_{j}^z}{h_j} \sin \frac{th_j}{2}&
\frac{ih_{j}^x + h_{j}^y}{h_j} \sin \frac{th_j}{2}\\
\frac{ih_{j}^x - h_{j}^y}{h_j} \sin \frac{th_j}{2}&
\cos \frac{th_j}{2}
-\frac{ih_{j}^z}{h_j} \sin \frac{th_j}{2}\\
\end{array},
\right),
 \label{SUZ8}
\end{equation}%
where $h_j=\Vert{\bf h}_{j}\Vert$ is the length of the vector ${\bf h}_{j}$.
From Eq.~\Eq{SUZ7} and Eq.~\Eq{SUZ8}, we conclude that the time evolution due to
single-spin terms can be calculated by
performing \ORDER{LD} operations involving $2\times2$ matrices.
Because all matrix elements are non-zero, these single-spin operations
are only marginally more complicated than the ones discussed in Section~\ref{numa}.

We now consider two different decompositions of the two-spin terms
that can be implemented very efficiently:
The original pair-product split-up~\cite{SUZU77,VRIE93}
in which $H_j$ contains all contributions
of a particular pair of spins, and an XYZ decomposition in which we break
up the Hamiltonian according to the $x$, $y$, and $z$ components of the spin operators~\cite{RAED00}.

The pair-product decomposition is defined by~\cite{SUZU77,VRIE93}

\begin{equation}
\exp\left\{-it\left[
-\sum_{j,k=1}^L\sum_{\alpha=x,y,z} J_{j,k}^\alpha S_j^\alpha S_k^\alpha
\right]\right\}
=\prod_{j,k=1}^L
\exp\left[it(J_{j,k}^x S_j^x S_k^x+J_{j,k}^y S_j^y S_k^y+J_{j,k}^z S_j^z S_k^z)\right]
,
 \label{SUZ9}
\end{equation}
where the order of the factors in the product over $(j,k)$ is arbitrary.
This freedom may be exploited to increase the execution speed
by vectorization and/or parallelization techniques.
Also in this case the expression of each factor can be worked out
analytically and reads

\begin{equation}
e^{it(J_{j,k}^x S_j^x S_k^x+J_{j,k}^y S_j^y S_k^y+J_{j,k}^z S_j^z S_k^z)}
=\left(
\begin{array}{cccc}
e^{iat}\cos bt  &0&0&ie^{iat}\sin bt \\
0&e^{-iat}\cos ct &ie^{-iat}\sin ct \\
0&ie^{-iat}\sin ct &e^{-iat}\cos ct \\
ie^{iat}\sin bt &0&0&e^{iat}\cos bt
\end{array}
\right)
,
 \label{SUZ10}
\end{equation}
where $a=J_{j,k}^z/4$, $b=(J_{j,k}^x-J_{j,k}^y)/4$, and $c=(J_{j,k}^x+J_{j,k}^y)/4$.
From Eq.~\Eq{SUZ9} and Eq.~\Eq{SUZ10}, it follows that the time evolution due
to the spin-spin coupling terms can be calculated by
performing \ORDER{PD} operations involving $4\times4$ matrices.
These $4\times4$ matrix operations are marginally more complicated
than the ones discussed in Section~\ref{numa} and therefore we omit further details.

The XYZ decomposition is defined by~\cite{RAED00}

\begin{equation}
\exp\left\{-it\left[
-\sum_{j,k=1}^L\sum_{\alpha=x,y,z} J_{j,k}^\alpha S_j^\alpha S_k^\alpha
\right]\right\}
=\prod_{\alpha=x,y,z} e^{-it H^\alpha}
,
 \label{SUZ11}
\end{equation}
where there is no particular order of the factors in Eq.~\Eq{SUZ11} and
\begin{equation}
H^\alpha=-\sum_{j,k=1}^L J_{j,k}^\alpha S_j^\alpha S_k^\alpha
.
 \label{SUZ12}
\end{equation}
The computational basis states are eigenstates of the $S_j^z$ operators.
Thus, in this representation $e^{-it H^z}$ is diagonal by construction,
and it changes the input state by altering the phase of each of the basis vectors.
As $H^z$ is a sum of pair interactions, it is trivial to implement this operation
as a sequence of multiplications by $4\times4$ diagonal matrices.
Still working in the same representation,
an efficient algorithm that implements the action of $e^{-it H^x}$ ($e^{-it H^y}$)
by using the $Y_j$ ($X_j$) rotations can be constructed as follows.
Writing $Y=\prod_{j=1}^L Y_j$ we have

\begin{equation}
e^{-it H^x}={\BAR Y} Ye^{-itH^x}{\BAR Y} Y
={\BAR Y} \exp\left(it\sum_{j,k=1}^L J_{j,k}^x S_j^z S_k^z\right)Y
.
 \label{SUZ13}
\end{equation}
From Eq.~\Eq{SUZ13}, it is clear that the action of $e^{-it H^x}$
can be computed by applying the $L$ single-spin rotations $Y$,
a sequence of multiplications by $4\times4$ diagonal matrices
containing phase shifts, and the $L$ inverse rotations ${\BAR Y}$.
A similar procedure is used to compute the action of
$e^{-itH^y}$.
We have only to replace $Y$ ($J_{j,k}^x$) by ${\BAR X}$ ($J_{j,k}^y$).
The total arithmetic operation count of the XYZ decomposition
\Eq{SUZ11} is \ORDER{LD}.
If there is enough memory on the conventional computer to
store all the values of $\sum_{j,k=1}^L J_{j,k}^\alpha S_j^z S_k^z$
for $\alpha=x,y,z$ and $S_j^z=\pm1/2$, then the XYZ decomposition runs in \ORDER{D}.

\subsection{Comments}\label{comments}

All the algorithms just discussed satisfy
the basic requirement (unitarity to machine precision)
of a valid quantum mechanical time evolution.
A general analysis of the strong and weak points of these algorithms
is rather difficult: experience shows that the conclusions
may change significantly with the particular application.
Therefore, we can only consider specific examples and
compare algorithms in terms of accuracy, memory, and CPU requirements,
and that is what we will do in this subsection.

As an example, we consider a model of two spins (${\mathbf S}_1,{\mathbf S}_2$)
interacting with a ``bath'' of $L-2$ spins (${\mathbf I}_n$)
described by the Hamiltonian~\cite{DOBR03a}
\begin{equation}
H=J_0({\mathbf S}_1+{\mathbf S}_2)^2 + \sum_{n=1}^{L-2} J_n \mathbf{I}_n\cdot({\mathbf S}_1+{\mathbf S}_2).
\label{NU0}
\end{equation}
This model has been used to study decoherence in a two-spin system
due to coupling with a bath of spins~\cite{DOBR03a}.
The Heisenberg exchange interactions $\{J_n\}$ are assumed to be random, uncorrelated,
and small compared with $J_0$ ($|J_n|\ll |J_0|$).
Initially, the two spins ${\mathbf S}_1$ and ${\mathbf S}_2$
are in the state with one spin up and the other spin down.
The initial state of the spins $\{{\mathbf I}_n\}$ is assumed to be random.
Obviously, Eq.~\Eq{NU0} is a special case of the generic Hamiltonian \Eq{hamiltonian}.

\begin{figure}[t]
\begin{center}
\setlength{\unitlength}{1cm}
\begin{picture}(8,6)
\put(-5,0.){\includegraphics[width=9cm]{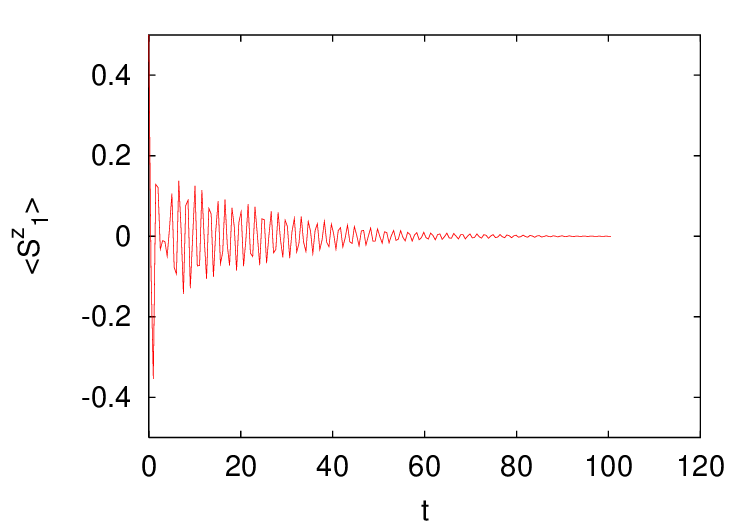}}
\put(4,0){\includegraphics[width=9cm]{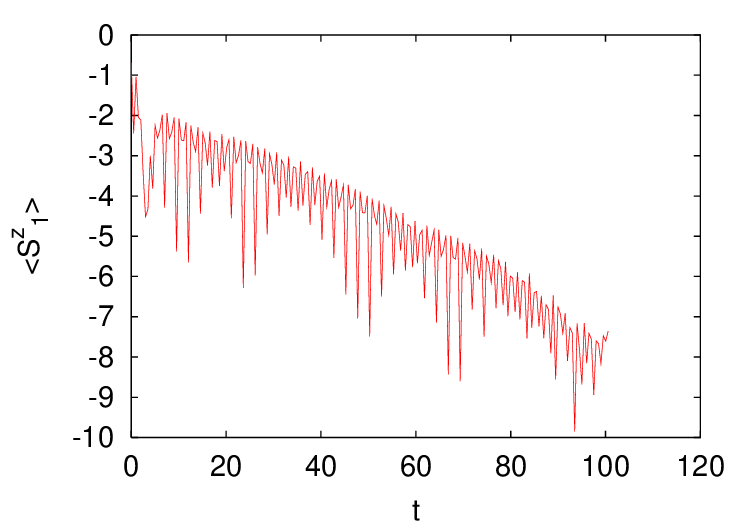}}
\end{picture}
\caption{%
Left: Magnetization $\EXPECT{S_1^z(t)}$ as a function of time as obtained by numerical
simulation of two spins interacting with a bath of $22$ spins.
The parameters of model (\ref{NU0}) are $J_0=8$ and the $J_n$ are
uncorrelated random numbers that satisfy $0<J_n<0.4$.
Right:
$\log|\EXPECT{S_1^z(t)}|$ as a function of time showing
exponential decay at long times (note that the intrinsic time scale of
${S_1^z(t)}$ is set by $J_0=8$).}
\label{fignu0}
\end{center}
\end{figure}

In Fig.~\ref{fignu0}, we show simulation results of the
time evolution of the magnetization $\EXPECT{S_1^z(t)}$
over a large time span, for a spin-bath containing 22 spins ($L=24$).
On a fine time scale (compared with the scale used in Fig.\ref{fignu0}),
the first and second spin oscillate rapidly with a period of order $1/J_0$
(results not shown)~\cite{DOBR03a}.
Initially, the amplitude of the magnetization oscillations rapidly
decays to zero, then increases again, and then decays to zero very slowly~\cite{DOBR03a}.
This is the generic behavior of the magnetization in this class of models~\cite{DOBR03a}.

\begin{table*}[t]
\begin{center}
\caption{Comparison of algorithms to solve the time-dependent Schr\"odinger equation
for model \Eq{NU0}. All calculations use a time step $\tau/2\pi=0.01$.
$L$ is the total number of spins, $m$ the number of time steps,
and $N$ denotes the number of iterations in the short-iterative Lanczos algorithm.
An asterisk in a column indicates that the state obtained by that
particular method is used as reference to compare with states
obtained by other algorithms.
The calculations for $L=22$ where carried out on a Cray SV1 system.
All other calculations were performed on an IBM T40 Notebook
with an Intel Centrino (1.3GHz) processor and 512 MB memory.
}
\begin{ruledtabular}
\begin{tabular}{p{2.5cm}ccccccc}
& Exact  & Chebyshev & Iterative & Suzuki  & Suzuki & Suzuki& Suzuki\\
& Diagonalization & Polynomial&  Lanczos  & Pair(2) & Pair(4)& XYZ(2)& XYZ(4)\\
\hline
\noalign{\vskip 4pt}
\hline \noalign{\vskip 2pt}
$L=10$, $m=400$\\
Error (N=5) &  * & 0.34E-12 & 0.17E-05 & 0.23E-03 & 0.75E-08 & 0.14E+00 & 0.53E-04  \\
CPU time [s]& 72 & 0.5      & 3.5      & 0.8      & 3.6      & 0.6      & 2.6       \\
\hline \noalign{\vskip 2pt}
$L=12$, $m=400$\\
Error (N=5) &  - & *        & 0.27E-05 & 0.27E-03 & 0.80E-08 & 0.14E+00 & 0.55E-04  \\
CPU time [s]&  - & 2.1      & 17.3     & 3.7      & 18.0     & 2.5      & 12.1      \\
\hline \noalign{\vskip 2pt}
$L=12$, $m=400$\\
Error (N=10)&  - & *        & 0.81E-13 & 0.27E-03 & 0.80E-08 & 0.14E+00 & 0.55E-04  \\
CPU time [s]&  - & 2.1      & 36.2     & 3.7      & 18.0     & 2.5      & 12.1      \\
\hline \noalign{\vskip 2pt}
$L=18$, $m=40$\\
Error (N=5) &  - & *       & 0.97E-06 & 0.90E-04 & 0.12E-07 & 0.21E-01 & 0.94E-0.5\\
CPU time [s]&  - & 82      & 316      & 74.9     & 312.5    & 52.3     & 239.6    \\
\hline \noalign{\vskip 2pt}
$L=22$, $m=8$\\
Error (N=5) &  - & *        & 0.40E-06 & 0.35E-04 & 0.21E-07  & 0.57E-02 & 0.39E-05\\
CPU time [s]&  - & 826      & 1817     & 402      & 1510      &  412     & 1549    \\
\hline \noalign{\vskip 2pt}
\end{tabular}
\end{ruledtabular}
\label{tabnu1}
\end{center}
\end{table*}

We use the system \Eq{NU0} to compare the different time-integration algorithms.
In Table \ref{tabnu1}, we show the error on the final state and the CPU time it took
to solve the time-dependent Schr\"odinger equation.
The error is defined as $\Vert \Psi_{\hbox{*}}(m) - \Psi_{\hbox{X}}(m)\Vert$
where * denotes the algorithm (exact diagonalization or Chebyshev polynomial method)
that is used to generate the
reference solution and X is one of the other algorithms.
It is clear that the short iterative Lanczos method is not competitive for this
type of time-dependent Schr\"odinger problem.
The fourth-order pair-approximation is close but is still less efficient than
the Chebyshev algorithm. The other Suzuki product-formula algorithms are clearly
not competitive. The reason that the pair-approximation is performing fairly well in this
case is related to the form of the Hamiltonian (\ref{NU0}).
The present results support our earlier finding~\cite{DOBR03} that
the numerical simulation of decoherence in spin systems is most efficiently done
in a two-step process: The Chebyshev algorithm can be used to make a big leap in time,
followed by a Suzuki product-formula algorithm
calculation to study the time dependence on a more detailed level.

In Table \ref{tabnu0}, we present a necessarily rough overall assessment
of the virtues and limitations of the algorithms discussed in this review.
From a general perspective, to increase the confidence in numerical simulation results,
it is always good to have several different algorithms to perform the same task.

\begin{table*}[t]
\begin{center}
\caption{Overall comparison of different integration methods
to solve the  time-dependent Schr\"odinger equation.
$D=2^L$ is the dimension of the Hilbert space,
$T$ is the time interval of integration,
$\tau$ is the time step,
and $N$ denotes the number of iterations in the short-iterative Lanczos algorithm.
An entry MP indicates that the result is numerically exact to machine precision.}
\begin{ruledtabular}
\begin{tabular}{p{2cm}ccccccc}
& Exact           & Chebyshev & Iterative & Suzuki  & Suzuki & Suzuki& Suzuki\\
& Diagonalization & Polynomial&  Lanczos  & Pair(2) & Pair(4)& XYZ(2)& XYZ(4)\\
\hline
\noalign{\vskip 4pt}
Memory&$\ORDER{D^2}$&$\ORDER{D}$&$\ORDER{ND}$&$\ORDER{D}$&$\ORDER{D}$&$\ORDER{D}$&$\ORDER{D}$\\
CPU&$\ORDER{D^3}$&$\ORDER{DT}$&$\ORDER{N^2DT/\tau}$&$\ORDER{DT^{3/2}}$&$\ORDER{DT^{3/4}}$&$\ORDER{DT^{3/2}}$&$\ORDER{DT^{3/2}}$\\
\hline \noalign{\vskip 2pt}
Unitary&YES& MP &YES&YES&YES&YES&YES\\
\hline \noalign{\vskip 2pt}
$H(t)$ changes

very often&Inefficient&Slow&Ok&Ok&Ok&Ok&Ok\\
\hline \noalign{\vskip 2pt}
Overall    &Essential  &Very efficient&Limited: $N$&Best choice if  &Less accurate  &Good choice if  &Less efficient\\
usefulness &for testing&for large $T$ &not known   &$H(t)$ changes  &but competitive&$H(t)$ changes  &than Suzuki\\
           &           &              &in advance  &rapidly with $t$&with Chebyshev &rapidly with $t$&Pair(4)\\
\end{tabular}
\end{ruledtabular}
\label{tabnu0}
\end{center}
\end{table*}

\section{Quantum Algorithms}\label{qas}
A quantum algorithm is a sequence of unitary operations that changes
the internal state of the quantum computer~\cite{NIEL00}.
Quantum algorithms vary in complexity. In the first subsection, we
discuss the elementary gates of a quantum computer. Two of these gates
--~ namely the CNOT gate and the controlled phase shift --~ have already been discussed in Section~\ref{twoqubit}.
These quantum gates are rather simple quantum algorithms that
can be used to build quantum networks
of more complicated quantum algorithms such as Grover's database search algorithm
and Shor's factoring algorithm
These more complicated quantum algorithms are discussed in Subsections \ref{qas}.B
- \ref{qas}.H.
In this Section we discuss only the salient features of the quantum algorithms.
Much more detailed information can be found in, for example, references
\cite{BARE95,VEDR96,VEDR98,CLEV98,BECK96,NIEL00}.

\subsection{Elementary Gates}
\subsubsection{Hadamard Gate}

The Hadamard gate is a single-qubit gate and is often used to prepare
the state of uniform superposition~\cite{NIEL00}.
The Hadamard operation on qubit $j$ is defined by

\begin{equation}
W_j={1\over\sqrt{2}}
\left(
\begin{array}{cc}
\phantom{-}1&\phantom{-}1 \\
\phantom{-}1&-1
\end{array}
\right).
\label{HADA0}
\end{equation}
For example
\begin{equation}
W_j\KET{0}=
{1\over\sqrt{2}}(\KET{0} +\KET{1}).
\label{HADA1}
\end{equation}
In terms of the elementary rotations $X$ and $Y$,
the Hadamard operation on qubit $j$ reads

\begin{equation}
W_j=-i\NOBAR X_j^2\BAR Y_j=-i\NOBAR Y_j \NOBAR X_j^2 = i\BAR X_j^2\BAR Y_j= i\NOBAR Y_j\BAR X_j^2.
\label{HADA2}
\end{equation}
The Hadamard operation can be generalized to an arbitrary number of qubits~\cite{NIEL00}.
The generalized operation is known as the Hadamard transform or as the Walsh-Hadamard transform.
This operation consists of $L$ Hadamard gates acting in parallel on $L$ qubits.
For example, for $L=2$, $W_2W_1\KET{00}=(\KET{00}+\KET{01}+\KET{10}+\KET{11})/2$.
The Walsh-Hadamard transform produces a uniform superposition of all basis states.
A symbolic representation of the Walsh-Hadamard gate is given in Fig.~\ref{fig:Gates}d.

\subsubsection{Swap Gate}

The swap gate interchanges two qubits and is useful in, for example, quantum
algorithms that perform Fourier transformations~\cite{NIEL00}.
In matrix notation, the SWAP operation reads

\begin{equation}
\hbox{SWAP}
\left(
\begin{array}{c}
a_0\\ a_1\\ a_2\\ a_3
\end{array}
\right)
\equiv
\left(
\begin{array}{cccc}
1&0&0&0 \\
0&0&1&0 \\
0&1&0&0 \\
0&0&0&1
\end{array}
\right)
\left(
\begin{array}{c}
a_0\\ a_1\\ a_2\\ a_3
\end{array}
\right)
=
\left(
\begin{array}{cccc}
1&0&0&0 \\
0&1&0&0 \\
0&0&0&1 \\
0&0&1&0
\end{array}
\right)
\left(
\begin{array}{cccc}
1&0&0&0 \\
0&0&0&1 \\
0&0&1&0 \\
0&1&0&0
\end{array}
\right)
\left(
\begin{array}{cccc}
1&0&0&0 \\
0&1&0&0 \\
0&0&0&1 \\
0&0&1&0
\end{array}
\right)
\left(
\begin{array}{c}
a_0\\ a_1\\ a_2\\ a_3
\end{array}
\right)
\label{SWAP1}
,
\end{equation}
showing that the SWAP operation can be decomposed into three CNOT operations.
A graphical representation of the swap gate is shown in Fig.~\ref{fig:Gates}e).

\subsubsection{Toffoli Gate}

The Toffoli gate is a generalization of the CNOT gate
in the sense that it has two control qubits and one target qubit~\cite{BARE95,NIEL00}.
The target qubit flips if and only if the two control qubits $Q_1^z=Q_2^z=1$
(see Table~\ref{tab:Toffoligate_work}).
Symbolically the Toffoli gate is represented by a vertical line
connecting two dots (control bits) and
one cross (target bit), as shown in Fig.~\ref{fig:Gates}f.
There are many ways to decompose the Toffoli gate in one- and two-qubit
operations~\cite{NIEL00}. We discuss two examples.

\begin{figure*}[ht]
\begin{center}
\includegraphics[width=10cm]{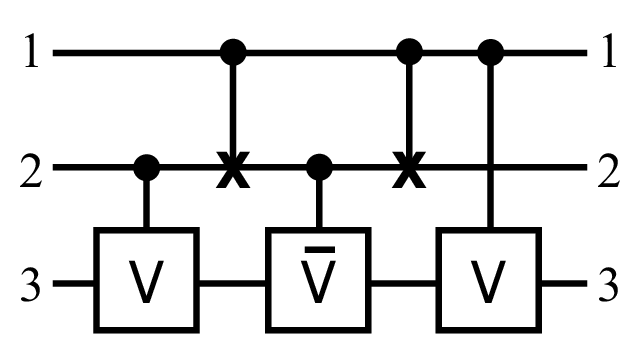}
\end{center}
\caption{
Quantum network for the Toffoli gate using CNOT gates and controlled phase shifts, which are part
of the operations ${\NOBAR V}$ and ${\BAR V}$.}
\label{fig:Toffoli1}
\end{figure*}

A quantum network for the first implementation is given in
Fig.~\ref{fig:Toffoli1}~\cite{BARE95,NIEL00}.
It consists of two CNOT gates and three controlled phase shifts.

\setlength{\unitlength}{1cm}

\setlength{\unitlength}{1cm}
\begin{figure*}[ht]
\begin{center}
\includegraphics[width=13cm]{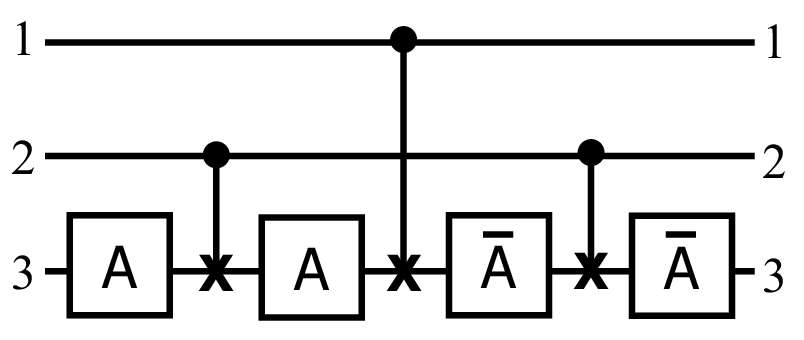}
\end{center}
\caption{Quantum network for the Toffoli gate using CNOT gates and
single-qubit operation $A$.
$A$ (${\BAR A}$) is a rotation by $\pi/4$ ($-\pi/4$) about the $y$-axis.}
\label{fig:Toffoli2}
\end{figure*}

\begin{table}[ht]
\caption{Internal operation of the quantum network, Fig.~\ref{fig:Toffoli1}
representing the Toffoli gate.}
\begin{center}
\begin{ruledtabular}
\begin{tabular}{cc|ccccc|c}
$Q_1^z$ & $Q_2^z$ & $V$ & CNOT & ${\BAR V}$ & CNOT & $V$ & Result\\
\hline
\noalign{\vskip 4pt}
0 & 0 &  &  &  &  &  &  $Q_3^z$ \\
0 & 1 & $V$ &  & ${\BAR V}$ &  &  & ${\NOBAR V}{\BAR V}Q_3^z=Q_3^z$ \\
1 & 0 &  & 1 & ${\BAR V}$ & 0 & $V$ & ${\BAR V}{\NOBAR V}Q_3^z=Q_3^z$ \\
1 & 1 & $V$ & 0 &  & 1 & $V$ & $V^2Q_3^z=1-Q_3^z$
\end{tabular}
\end{ruledtabular}
\label{tab:Toffoligate_work}
\end{center}
\end{table}
The internal operation of this quantum network is shown
in Table~\ref{tab:Toffoligate_work}.
The first two columns give all possible combinations of $Q_1^z$ and $Q_2^z$.
The next five columns schematically show the corresponding operations
as we move through the network (see Fig.~\ref{fig:Toffoli1}) from left to right.
The columns labeled ``CNOT'' show only the value of the second qubit because
the first qubit acts as a control bit and does not change.
The last column summarizes the operation on the third qubit.
From Table~\ref{tab:Toffoligate_work}
it immediately follows that operation $V$ has to be constructed
such that $V^2$ flips $Q_3$
and that ${\BAR V}{\NOBAR V}$ is equal to the identity matrix.
These conditions are fulfilled by taking

\begin{equation}
V
={e^{-i\pi /4}\over\sqrt{2}}
\left(
\begin{array}{cc}
1&i \\
i&1
\end{array}
\right)
.
\end{equation}
Taking into account that $V$ should change only the target qubit (2)
if the control qubit (1) is one, we find

\begin{equation}
\label{eq:V12}
V_{21}
\left(
\begin{array}{c}
a_0\\ a_1\\ a_2\\ a_3
\end{array}
\right)
\equiv
\left(
\begin{array}{cccc}
1&0&0&0 \\
0&e^{-i\pi /4}/\sqrt{2}&0&ie^{-i\pi /4}/\sqrt{2} \\
0&0&1&0 \\
0&ie^{-i\pi /4}/\sqrt{2}&0&e^{-i\pi /4}/\sqrt{2}
\end{array}
\right)
\left(
\begin{array}{l}
a_0\\ a_1\\ a_2\\ a_3
\end{array}
\right)
.
\end{equation}
The matrix $V_{12}$ can be brought into a similar form as
the controlled phase shift:

\begin{eqnarray}
\label{eq:U12}
V_{21}
&=&\BAR Y_2
\left(
\begin{array}{cccc}
1&0&0&0 \\
0&1&0&0 \\
0&0&1&0 \\
0&0&0&e^{-i\pi /2}
\end{array}
\right) \NOBAR Y_2\equiv
\BAR Y_2 {\BAR R_{21}(\pi /2)}\NOBAR Y_2
.
\end{eqnarray}
The controlled phase shift $R_{21}(\pi/2)$ can be implemented on
the Ising model quantum computer (\ref{Ising}) by putting
$h=-J/2$ and $\tau J=-\pi/2$.

It is easy to see that the quantum network shown in Fig.~\ref{fig:Toffoli1}
corresponds to the following sequence of operations:

\begin{equation}
\label{toffolilong}
\hbox{TOFFOLI}={\BAR Y_3}{\BAR R_{31}}{\NOBAR Y_3}
{\BAR Y_2}      I_{21} {\NOBAR Y_2}
{\BAR Y_3}{\NOBAR R_{32}}{\NOBAR Y_3}
{\BAR Y_2}I_{21}{\NOBAR Y_2}
{\BAR Y_3}{\BAR R_{32}}{\NOBAR Y_3},
\label{TOFF1}
\end{equation}
where $I_{21}\equiv I_{21}(\pi)$, $R_{ji}\equiv R_{ji}(\pi /2)$, and
${\BAR R_{ji}}\equiv R_{ji}(-\pi /2)$.
The sequence \Eq{TOFF1} can be shortened by observing that
${\BAR Y_2}I_{21} {\NOBAR Y_2}{\BAR Y_3}
={\BAR Y_3}{\BAR Y_2}I_{21} {\NOBAR Y_2}$.
This leads to

\begin{equation}
\hbox{TOFFOLI}={\BAR Y_3}{\BAR R_{31}}{\BAR Y_2}I_{21}{\NOBAR Y_2}{\NOBAR R_{32}}
{\BAR Y_2}I_{21}{\NOBAR Y_2}{\BAR R_{32}}{\NOBAR Y_3}.
\label{TOFF2}
\end{equation}

\begin{table*}[t]
\caption{Internal operation of the quantum network, Fig.~\ref{fig:Toffoli2}.
$A$ is a single-qubit rotation about the $y$-axis by $\pi/4$.
The gate $N$ flips the qubit.}
\begin{center}
\begin{ruledtabular}
\begin{tabular}{cc|ccccccc|c}
$Q_1^z$ & $Q_2^z$ & ${\NOBAR A}$ & CNOT & ${\NOBAR A}$ & CNOT & ${\BAR A}$ & CNOT &
${\BAR A}$ & Desired result\\
\hline
\noalign{\vskip 4pt}
0 & 0 & ${\NOBAR A}$ &    & ${\NOBAR A}$ &    & ${\BAR A}$ &    & ${\BAR A}$ & ${\BAR A}{\BAR A}{\NOBAR A}{\NOBAR A}Q_3^z=Q_3^z$ \\
0 & 1 & ${\NOBAR A}$ & $N$ & ${\NOBAR A}$ &    & ${\BAR A}$ & $N$ & ${\BAR A}$ & ${\BAR A}N{\BAR A}{\NOBAR A}N{\NOBAR A}Q_3^z=Q_3^z$ \\
1 & 0 & ${\NOBAR A}$ &    & ${\NOBAR A}$ & $N$ & ${\BAR A}$ &    & ${\BAR A}$ & ${\BAR A}{\BAR A}N{\NOBAR A}{\NOBAR A}Q_3^z=Q_3^z$ \\
1 & 1 & ${\NOBAR A}$ & $N$ & ${\NOBAR A}$ & $N$ & ${\BAR A}$ & $N$ & ${\BAR A}$ & ${\BAR A}N{\BAR A}N{\NOBAR A}N{\NOBAR A}Q_3^z=NQ_3^z$
\end{tabular}
\end{ruledtabular}
\label{tab:Toffoligate_work2}
\end{center}
\end{table*}

In Fig.~\ref{fig:Toffoli2}, we show a quantum network that performs
the same operation as the Toffoli gate up to
some known phase factors~\cite{BARE95,NIEL00}.
The internal operation of this network is summarized
in Table~\ref{tab:Toffoligate_work2}.
The first two columns give all possible combinations of the control qubits
$Q_1^z$ and $Q_2^z$.
The next seven columns schematically show the individual operations
as we move through the network from left to right
(see Fig.~\ref{fig:Toffoli2}).
Entries in the colums labeled ``CNOT'' having a value ``N''
indicate that the target qubit (qubit 3) flips.
The last column shows the full operation on the third qubit.
From Table~\ref{tab:Toffoligate_work2},
it follows that the first and second row
of the last column are trivially satisfied.
The third row implies that $N=1$, which is impossible.
Nevertheless, if we choose $A=e^{i\pi S^y/4}$
we find that the condition in the fourth row is satisfied and
that ${\BAR A}{\BAR A}N{\NOBAR A}{\NOBAR A}=
2{\BAR Y}S^x{\NOBAR Y}=-2S^z$.
This means that the target qubit will acquire an extra
phase factor ($-1$) if and only if the first control qubit
is one and the second control qubit is zero.
Thus, the quantum network of Fig.~\ref{fig:Toffoli2}
does not implement a Toffoli gate but performs
an operation that is very similar, up to a phase factor
that depends on the state of the control qubits~\cite{NIEL00}.

\subsection{Quantum Fourier Transform}

The quantum Fourier transform (QFT) is an essential subroutine in, for example, Shor's
factoring algorithm, the order finding algorithm and phase estimation
in general~\cite{NIEL00,EKER96}.
The QFT is not a new kind of Fourier transform: it is a particular
application of the standard discrete Fourier transform.
The latter is defined by ${\bf x}=U{\bf y}$
where ${\bf x}=(x_0,\ldots,x_{N-1})$, ${\bf y}=(y_0,\ldots,y_{N-1})$
and $U_{j,k}=e^{2\pi i j k/N}/\sqrt{N}$. Clearly $U$ is a unitary matrix.
In the context of quantum computation, the vectors ${\bf x}$ and ${\bf y}$ are
just two different representations of the basis vectors that
span the Hilbert space of dimension $N$.
The QFT is defined as

\begin{equation}
\sum_{j=0}^{N-1} a_j \KET{x_j}=
\frac{1}{\sqrt{N}}\sum_{j,k=0}^{N-1} e^{2\pi i j k/N}a_j \KET{y_k}
=\sum_{k=0}^{N-1} b_k \KET{y_k}
,
\label{qft0}
\end{equation}
where the amplitudes $b_k$ are the discrete Fourier transform of the amplitudes $a_j$ and
$N=2^L$ where $L$ is the number of qubits.
Readers more familiar with traditional quantum mechanics recognize
the similarity with the coordinate and momentum representation.
Exploiting the massive parallelism of the ideal quantum computer, the QFT can be carried
out in $\ORDER{\log_2 N}=\ORDER{L}$ quantum operations~\cite{NIEL00,EKER96}.

\begin{figure*}[ht]
\includegraphics[width=\figsize]{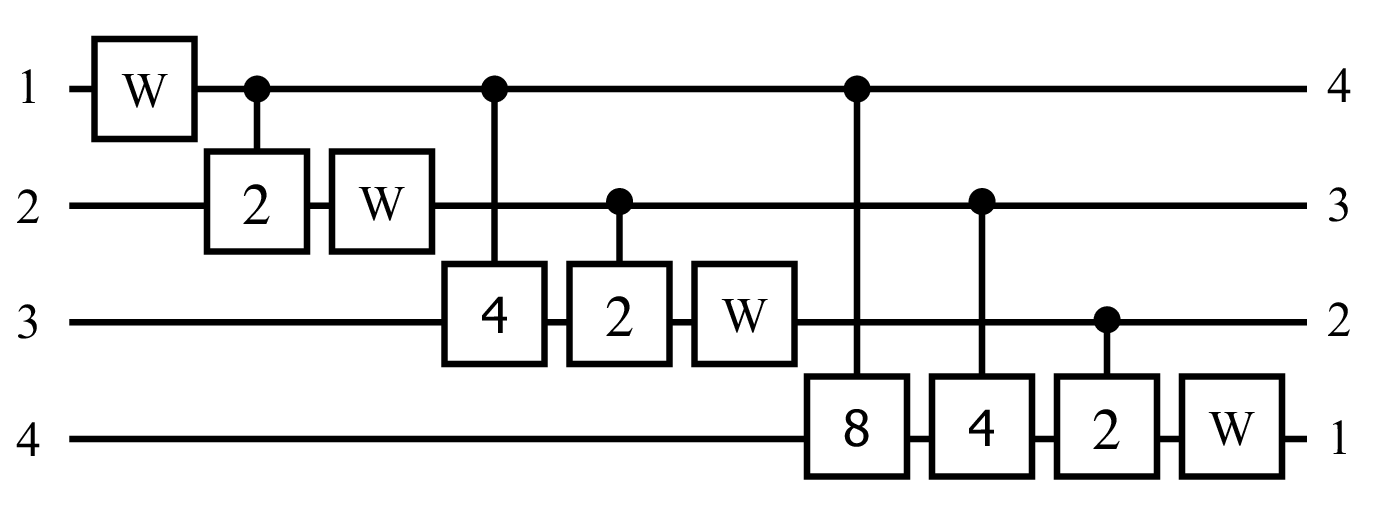}
\caption{
Quantum network that performs a four-qubit quantum Fourier transform.
W denotes the Walsh-Hadamard transform.
The operations ``2'',``4'', and ``8'' perform controlled phase shifts
with angles $\pi/2$, $\pi/4$, and $\pi/8$, respectively.}
\label{qftnetwork}
\label{qftfig}
\end{figure*}

How a quantum network can be derived for the QFT is explained in, for example,
Ref.~\cite{EKER96,NIEL00}.
In Fig.~\ref{qftnetwork}, we show a quantum network that performs a four-qubit QFT~\cite{EKER96}.
The blocks labeled $W$ perform a Walsh-Hadamard transform, and the other blocks perform a
controlled phase shift by the angle indicated.
Not shown in the network is the series of SWAP-gates that interchange
the output qubits (1,4) and (2,3)~\cite{NIEL00,EKER96}, hence the different labeling of input and output lines in Fig.~\ref{qftnetwork}.
For the applications that we discuss later, these interchanges merely add to the operation
count and can therefore be omitted.
To see how the network carries out the QFT
(and also for the derivation of the network),
the states $\KET{x_j}$ and $\KET{y_j}$ in Eq.~\Eq{qft0}
have to be written in binary representation, for example,
$\KET{x_0}=\KET{0000}$,
$\KET{x_1}=\KET{0001}$, and so on.

\setlength{\unitlength}{1cm}

\subsection{Finding the Period of a Periodic Function}\label{periodfunction}

Assume that we are given the function $f(n)=f(n+M)$ for $n=0,\ldots,N-1$.
On a quantum computer, we can determine the period $M$ as follows.
We use one register of qubits to store $n$ and another one to store $f(n)$.
As a first step, the quantum computer is put in the state of uniform superposition over all $n$.
The state of the quantum computer can be written as

\begin{eqnarray}
\frac{1}{\sqrt{N}}\sum_{n=0}^{N-1} \KET{n}\KET{f(n)}
&=&
\frac{1}{\sqrt{N}}\left\{
\sum_{n=0}^{M-1} \KET{n}\KET{f(n)}
+\sum_{n=M}^{2M-1} \KET{n}\KET{f(n)}
+\dots\right\}
\nonumber \\
&=&
\frac{1}{\sqrt{N}}
\sum_{n=0}^{M-1}
\left(\KET{n}+\KET{n+M}+\ldots\right) \KET{f(n)}
,
\end{eqnarray}
where, in the last step, we used the periodicity of $f(n)$.
Using the Fourier representation of $\KET{n}$ we obtain

\begin{eqnarray}
\frac{1}{\sqrt{N}}\sum_{n=0}^{N-1} \KET{n}\KET{f(n)}
&=&
\frac{1}{N}
\sum_{k=0}^{N-1}
\sum_{n=0}^{M-1}
e^{2\pi i k n/N}
\left(1+e^{2\pi i k M/N}+e^{4\pi i k M/N}+\ldots+e^{2\pi i k M(L-1)/N} \right)
\KET{k}\KET{f(n)}\nonumber\\
&+&
\frac{1}{N}
\sum_{k=0}^{N-1}
\sum_{n=0}^{L-1}
e^{2\pi i k n/N}e^{2\pi i k ML/N}
\KET{k}\KET{f(n)}
,
\end{eqnarray}
where $L=\lfloor N/M \rfloor$ denotes the largest integer $L$ such that $ML\le N$.
In simple terms, $L$ is the number of times the period $M$ fits into the
interval $[0,N-1]$.
The probability $p_q(M)$ to observe the quantum computer in the state $\KET{q}$ is given by
the expectation value of the (projection) operator $Q=\KET{q}\BRA{q}$.
With the restriction on $f(n)$ that $f(n)=f(n')$ implies $n=n'$,
we find

\begin{eqnarray}
\EXPECT{Q}=p_q(M)
&=&
\frac{M}{N^2}
\left(\frac{\sin(\pi qML/N)}{\sin(\pi qM/N)}\right)^2
+\frac{N-ML}{N^2}
\frac{\sin(\pi qM(2L+1)/N)}{\sin(\pi qM/N)}
.
\end{eqnarray}
The results for $p_q(M)$ in the case $N=8$ (3 qubits) are given in Table~\ref{qfttab}.
From Table~\ref{qfttab} it follows directly that the expectation values
of the qubits are
($Q_1^z=Q_2^z=Q_3^z=0$) if the period $M=1$,
($Q_1^z=Q_2^z=0$, $Q_3^z=0.5$) if the period $M=2$,
($Q_1^z=0.5$, $Q_2^z=0.375$, $Q_3^z=0.34375$) if the period $M=3$,
and
($Q_1^z=0$, $Q_2^z=Q_3^z=0.5$) if the period $M=4$.
Thus, in this simple case the periodicity of $f(n)$ can be unambiguously determined
from the expectation values of the individual qubits.

\begin{table}[ht]
\caption{
Probability $p_q(M)$ to observe the state $\KET{q}$
after performing the quantum Fourier transform on the perodic function $f(n)=f(n+M)$
for $n=0,\dots,7$.}
\begin{center}
\begin{ruledtabular}
\begin{tabular}{ccccc}
$q$  & $p_q(M=1)$ & $p_q(M=2)$ & $p_q(M=3)$ & $p_q(M=4)$ \\
\hline
\noalign{\vskip 4pt}
 0  &  1  &  0.5  &  0.34375  & 0.25\\
 1  &  0  &  0.0  &  0.01451  & 0.00\\
 2  &  0  &  0.0  &  0.06250  & 0.25\\
 3  &  0  &  0.0  &  0.23549  & 0.00\\
 4  &  0  &  0.5  &  0.31250  & 0.25\\
 5  &  0  &  0.0  &  0.23549  & 0.00\\
 6  &  0  &  0.0  &  0.06250  & 0.25\\
 7  &  0  &  0.0  &  0.01451  & 0.00\\
\end{tabular}
\end{ruledtabular}
\label{qfttab}
\end{center}
\end{table}

\subsection{Grover's Database Search Algorithm}

Next we consider Grover's database search algorithm
to find the needle in a haystack~\cite{GROV96,GROV97}.
On a conventional computer, finding an item out
of $N$ elements requires \ORDER{N} queries~\cite{CORM94}.
Grover has shown that a quantum computer can find the item using
only \ORDER{\sqrt{N}} attempts~\cite{GROV96,GROV97,ZALK00}.
Assuming a uniform probability distribution for the needle,
for $N=4$ the average number of queries required by a conventional
algorithm is 9/4~\cite{CHUA98a,CORM94}.
With Grover's quantum algorithm, the correct answer
for this database with four items can be found in
a single query~\cite{JONE98a,CHUA98a}.
Grover's algorithm for the four-item database can be implemented on a two-qubit quantum computer.

The key ingredient of Grover's algorithm is an
operation that replaces each amplitude
of the basis states in the superposition by
two times the average amplitude minus
the amplitude itself.
This operation is called ``inversion about the mean'' and
amplifies the amplitude of the basis state that represents the
searched-for item~\cite{GROV96,GROV97}.
To see how this works, it is useful to consider an example.
Consider a database containing
four items and functions $g_j(x)$, $j=0,\ldots,3$ that upon query of the database
returns minus one if $x=j$ and plus one if $x\not=j$.
Let us assume that the item to search for corresponds
to, for example, 2 ($g_2(0)=g_2(1)=g_2(3)=1$ and $g_2(2)=-1$).
Using the binary representation of integers,
the quantum computer is in the state (up to an irrelevant phase factor as usual)
\begin{equation}
\label{eq:database}
\KET{\Phi}={1\over 2}(\KET{00}+\KET{01}-\KET{10}+\KET{11}).
\end{equation}

The operator $B$ that inverts states like (\ref{eq:database}) about their means reads
\begin{equation}
\label{eq:D}
B
\left(
\begin{array}{c}
a_0\\ a_1\\ a_2\\ a_3
\end{array}
\right)
={1\over2}
\left(
\begin{array}{cccc}
-1&\phantom{-}1&\phantom{-}1&\phantom{-}1 \\
\phantom{-}1&-1&\phantom{-}1&\phantom{-}1 \\
\phantom{-}1&\phantom{-}1&-1&\phantom{-}1 \\
\phantom{-}1&\phantom{-}1&\phantom{-}1&-1
\end{array}
\right)
\left(
\begin{array}{c}
a_0\\ a_1\\ a_2\\ a_3
\end{array}
\right)
.
\end{equation}
Applying $B$ to $\KET{\Phi}$ results in $B\KET{\Phi}=\KET{10}$,
that is, the correct answer.
In general, for more than two qubits,
more than one application of $B$ is required to get the correct answer~\cite{GROV96,GROV97}.
In this sense the two-qubit case is somewhat special.

Implementation of this example on a two-qubit quantum computer requires
a representation in terms of elementary rotations
of the preparation and query steps and of the operation of inversion about the mean.
Initially, we bring the quantum computer in the state $\KET{00}$ and then transform $\KET{00}$ to the state
(\ref{eq:database}) by a two-step process. First we use the Walsh-Hadamard transform to bring
the quantum computer in the uniform superposition state: $W_2W_1\KET{00}=(\KET{00}+\KET{01}+\KET{10}+\KET{11})/2$,
where $W_j$ is given by Eq.~\Eq{HADA0}. Next we apply
a transformation $F_2$ that corresponds to the application of $g_2(x)$ to the uniform superposition state
\begin{equation}
\label{eq:F2}
F_2
\left(
\begin{array}{c}
a_0\\ a_1\\ a_2\\ a_3
\end{array}
\right)
={1\over2}
\left(
\begin{array}{cccc}
\phantom{-}1&\phantom{-}0&\phantom{-}0&\phantom{-}0 \\
\phantom{-}0&\phantom{-}1&\phantom{-}0&\phantom{-}0 \\
\phantom{-}0&\phantom{-}0&-1&\phantom{-}0 \\
\phantom{-}0&\phantom{-}0&\phantom{-}0&\phantom{-}1
\end{array}
\right)
\left(
\begin{array}{l}
a_0\\ a_1\\ a_2\\ a_3
\end{array}
\right)
.
\end{equation}
This transformation can be implemented by first letting the system evolve in time
\begin{equation}
\label{eq:time}
GW_2W_1\KET{00}= {e^{-i\pi S_1^z S_2^z}\over 2}
(\KET{00}+\KET{01}+\KET{10}+\KET{11})
=
{1\over 2}(e^{-i\pi/4}\KET{00}+e^{+i\pi/4}\KET{01}
+e^{+i\pi/4}\KET{10}+e^{-i\pi/4}\KET{11}),
\end{equation}
where the two-qubit operation $G$ is defined by
\begin{equation}
\label{eq:G12}
G
\left(
\begin{array}{c}
a_0\\ a_1\\ a_2\\ a_3
\end{array}
\right)
=
\left(
\begin{array}{cccc}
e^{-i\pi /4}&0&0&0 \\
0&e^{+i\pi /4}&0&0 \\
0&0&e^{+i\pi /4}&0 \\
0&0&0&e^{-i\pi /4}
\end{array}
\right)
\left(
\begin{array}{c}
a_0\\ a_1\\ a_2\\ a_3
\end{array}
\right)
.
\end{equation}
Then we apply a sequence of single-spin rotations
to change the four phase factors such that we get the desired
state.
The two sequences
$\NOBAR Y_j \NOBAR X_j \BAR Y_j$ and $\NOBAR Y_j\BAR X_j\BAR Y_j$
operating on qubit j are particularly useful for this purpose since
\begin{equation}
\NOBAR Y_j\NOBAR X_j\BAR Y_j\KET{0}=e^{+i\pi /4}\KET{0},\
\NOBAR Y_j\NOBAR X_j\BAR Y_j\KET{1}=e^{-i\pi /4}\KET{1},\
\NOBAR Y_j\BAR X_j\BAR Y_j\KET{0}=e^{-i\pi /4}\KET{0},\
\NOBAR Y_j\BAR X_j\BAR Y_j\KET{1}=e^{+i\pi /4}\KET{1}.
\end{equation}
We find
\begin{eqnarray}
\label{eq:sequence}
&&{\NOBAR Y_1} {\NOBAR X_1} {\BAR Y_1} {\NOBAR Y_2}{\BAR X_2}{\BAR Y_2}
\left[
{1\over 2}(e^{-i\pi/4}\KET{00}+e^{+i\pi/4}\KET{01}
+e^{+i\pi/4}\KET{10}+e^{-i\pi/4}\KET{11})\right]
\nonumber \\
&&=
{1\over 2}(e^{-i\pi/4}\KET{00}+e^{-i\pi/4}\KET{01}
+e^{+3i\pi/4}\KET{10}+e^{-i\pi/4}\KET{11})
=
{e^{-i\pi/4}\over 2}(\KET{00}+\KET{01}
-\KET{10}+\KET{11}).
\end{eqnarray}
Thus we can construct the sequence $F_j$ that
transforms the uniform superposition state
to the state that corresponds to $g_j(x)$:
\begin{equation}
\label{eq:Fs}
F_0=\NOBAR Y_1 \BAR X_1 \BAR Y_1 \NOBAR Y_2 \BAR X_2\BAR Y_2 G,\
F_1=\NOBAR Y_1 \BAR X_1 \BAR Y_1 \NOBAR Y_2 \NOBAR X_2\BAR Y_2 G,\
F_2=\NOBAR Y_1 \NOBAR X_1\BAR Y_1 \NOBAR Y_2 \BAR X_2\BAR Y_2 G,\
F_3=\NOBAR Y_1 \NOBAR X_1 \BAR Y_1 \NOBAR Y_2 \NOBAR X_2\BAR Y_2 G.
\end{equation}

Finally, we need to express the operation of inversion about the mean
--~ that is, the matrix $B$ [see (\ref{eq:D})] --~ by a sequence of
elementary operations. It is not difficult to see that
$B$ can be written as
\begin{equation}
B=W_1 W_2
\left(
\begin{array}{cccc}
\phantom{-}1&\phantom{-}0&\phantom{-}0&\phantom{-}0 \\
\phantom{-}0&-1&\phantom{-}0&\phantom{-}0 \\
\phantom{-}0&\phantom{-}0&-1&\phantom{-}0 \\
\phantom{-}0&\phantom{-}0&\phantom{-}0&-1
\end{array}
\right)
W_1 W_2
\equiv W_1 W_2 P W_1 W_2.
\end{equation}
The same approach that was used to implement $g_2(x)$ also works
for $P$ ($=-F_0$) and yields
\begin{equation}
P=-Y_1{\BAR X}_1{\BAR Y}_1 Y_2 {\BAR X}_2{\BAR Y}_2 G.
\end{equation}
The complete sequence $U_j$ operating on $\KET{00}$ reads
\begin{equation}
U_j=W_1 W_2 P W_1 W_2 F_j W_2 W_1.
\end{equation}
Each sequence $U_j$ can be shortened by
observing that in some cases a rotation is followed by its inverse.
Making use of the alternative representations
of the Walsh-Hadamard transform $W_i$ [see Eq.~\Eq{HADA2}],
the sequence for, for instance, $j=1$ can be written as

\begin{equation}
W_1 W_2 F_1=
\NOBAR X_1 \NOBAR X_1\BAR Y_1 \BAR X_2 \BAR X_2 \BAR Y_2
\NOBAR Y_1 \BAR X_1 \BAR Y_1 \NOBAR Y_2\NOBAR X_2\BAR Y_2 G
=
\NOBAR X_1 \BAR Y_1 \BAR X_2 \BAR Y_2 G.
\end{equation}
The optimized sequences $U_j$ read
\begin{eqnarray}
\label{eq:Groverseq}
U_0&=&-\NOBAR X_1\BAR Y_1 \NOBAR X_2\BAR Y_2 G
\NOBAR X_1 \BAR Y_1 \NOBAR X_2 \BAR Y_2 G\BAR X_2 \BAR X_2 \BAR Y_2 \BAR X_1 \BAR X_1 \BAR Y_1 ,\nonumber \\
U_1&=&-\NOBAR X_1\BAR Y_1 \NOBAR X_2\BAR Y_2 G
\NOBAR X_1 \BAR Y_1 \BAR X_2 \BAR Y_2 G\BAR X_2 \BAR X_2 \BAR Y_2 \BAR X_1 \BAR X_1 \BAR Y_1 ,\nonumber \\
U_2&=&-\NOBAR X_1\BAR Y_1 \NOBAR X_2\BAR Y_2 G
\BAR X_1 \BAR Y_1 \NOBAR X_2 \BAR Y_2 G\BAR X_2 \BAR X_2 \BAR Y_2 \BAR X_1 \BAR X_1 \BAR Y_1 ,\nonumber \\
U_3&=&-\NOBAR X_1\BAR Y_1 \NOBAR X_2\BAR Y_2 G
\BAR X_1 \BAR Y_1 \BAR X_2 \BAR Y_2 G\BAR X_2 \BAR X_2 \BAR Y_2 \BAR X_1 \BAR X_1 \BAR Y_1 .
\label{GROV0}
\end{eqnarray}
Note that the quantum algorithms (\ref{eq:Groverseq}) are by no means unique:
various alternative expressions can be written down by using
the algebraic properties of the $X$s and $Y$s.
Straightforward calculations show that $U_j\KET{0}=\KET{j}$.
Hence the application of a sequence (\ref{eq:Groverseq})
to the initial state $\KET{0}$ brings the quantum computer
in the state that corresponds to the searched-for item.

\subsection{Finding the Order of a Permutation}

We consider the problem of finding the order of a permutation.
An experimental realization of this quantum algorithm on a five-qubit NMR quantum computer
for the case of a permutation of four items is described in Ref.~\cite{SYPE00}.
The problem is defined as follows:
Given a permutation $P$ of the integers $\{0,1, ..., N-1\}$
and an integer $0 \le y \le N-1$, the order $r(y)$ is the smallest integer
for which $P^r(y) y =y$.
Thus, the purpose of the quantum algorithm is to determine $r(y)$,
for a given $y$ and permutation $P$.
The theory in this section closely follows Ref.~\cite{SYPE00}. We therefore also consider the case $N=4$,
as an example.

The basic idea of the quantum algorithm to find the order of a permutation is to
exploit the quantum parallelism to compute $P(y)$, $P^2(y)$, $P^3(y)$, and $P^4(y)$
simultaneously and filter out the power $r(y)$ that yields $P^r(y) y =y$.
Denoting $f(n)=P^n$, finding $r(y)$ is the same as finding the period of
$f(n)$, a problem that is solved by invoking the QFT (see Section~\ref{periodfunction}).

First we consider the problem of generating a permutation of $N=4$ items.
We need two qubits to specify one of the four items.
Using the binary representation of the integers
$\{0,1,2,3\}$, it is easy to see that the CNOT operation $C_{21}$ (the right subscript
denoting the control bit) corresponds to
the permutation (in cycle notation)
$P=(0)(2)(13)$, that interchanges items 1 and 3.
Likewise, $C_{12}$ generates
the permutation $P=(0)(1)(23)$
and $C_{12} C_{21} C_{12}$ is equivalent to
the permutation $P=(0)(3)(12)$.
The remaining interchanges of two items can be generated by a combination
of CNOT gates and NOT operations.
Denoting the NOT operation on the $j$th qubit by $N_j$,
we find that $C_{21}N_2=N_2C_{21}$ yields $P=(02)(1)(3)$,
$C_{12}N_1=N_1C_{12}$ yields $P=(2)(3)(01)$,
and
$C_{12}N_2C_{21}C_{12}$ yields $P=(1)(2)(03)$.
Using these elementary interchanges, we can construct any permutation.

\begin{figure*}[ht]
\begin{center}
\includegraphics[width=\figsize]{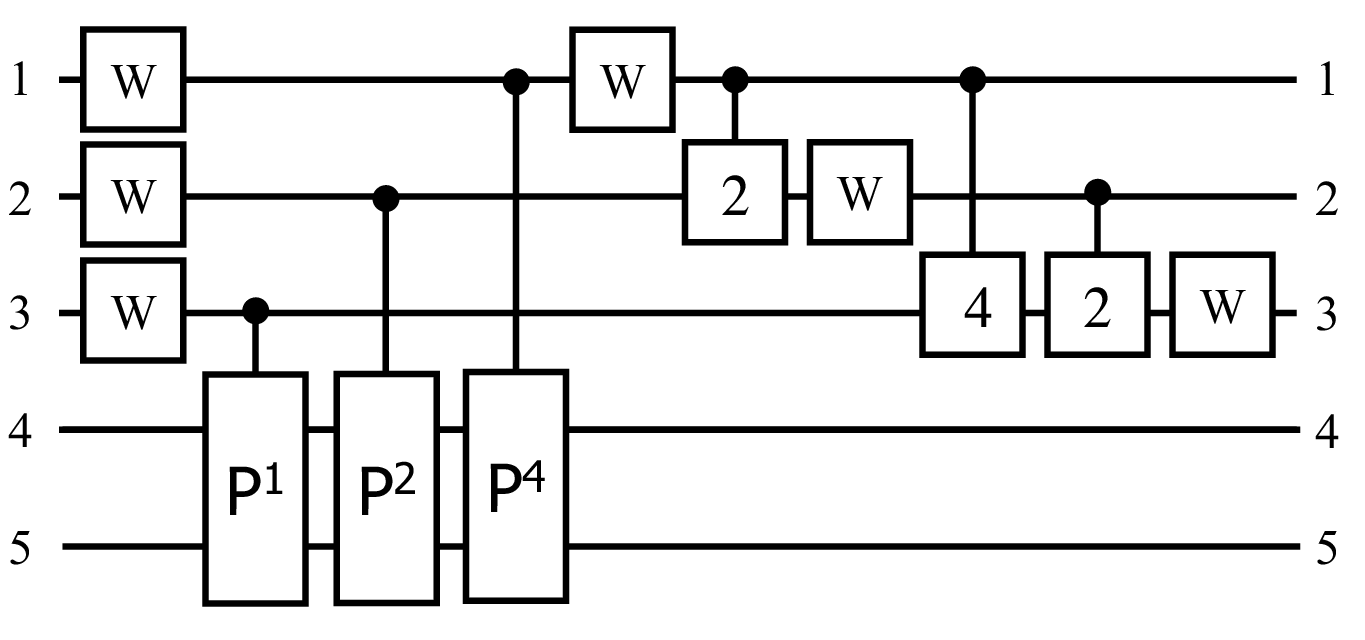}
\end{center}
\caption{Quantum network of a quantum algorithm to find the order of a permutation of four items.
W and P$^k$ denote the Walsh-Hadamard transform and $k$ applications of the permutation P, respectively.
The operations ``2'' and ``4'' perform controlled phase shifts
with angles $\pi/2$ and $\pi/4$, respectively.}
\label{ordernetwork}
\end{figure*}

The quantum network that carries out the quantum algorithm to find the order of
a permutation $P$ of four items is shown in Fig.~\ref{ordernetwork}.
There is a one-to-one mapping from this network onto the quantum algorithm to find the period
of the function $f(n)$ (see Section~\ref{periodfunction}).
The first three qubits hold the integer $n$; qubits 4 and 5 hold $y=2y_1+y_0$.
The three Walsh-Hadamard operations
change the initial state $\KET{000}\KET{y_1y_0}$ into $\KET{uuu}\KET{y_1y_0}$ where we use
the label ``$u$'' to denote the uniform superposition.
Then, we apply the permutations $P^0y, P^1y,\ldots,P^7y$.
In the actual implementation, the sequence of
CNOT and NOT operations that implement the permutation $P$
need to be replaced by Toffoli and CNOT gates, respectively,
because (the power of) $P$ is applied to the fourth and fifth
qubit (representing the integer $y$), conditional on the state of the first three qubits.
Finally, we perform a QFT on the first three qubits and consider the
value of the first three qubits $Q_1^z$, $Q_2^z$, and $Q_3^z$.
As explained before, in this simple case we can extract the period of the
function, and hence the order of $P$ acting on $y$, from the
values of $Q_1^z$, $Q_2^z$, and $Q_3^z$.

\subsection{Number Factoring: Shor's Algorithm}

As another application of the QFT,
we consider the problem of factoring integers, that is, Shor's algorithm.
For the case $N=15$, an experimental realization
of this quantum algorithm on a seven-qubit NMR quantum computer
is given in Ref.~\cite{SYPE01a}.
The theory in this section closely follows Ref.~\cite{SYPE01b}.

The theory behind Shor's algorithm has been discussed at great length elsewhere~\cite{NIEL00,SHOR99,EKER96}.
Therefore, we recall only the basic elements of Shor's algorithm.
Shor's algorithm is based on the fact that the factors $p$ and $q$ of an integer
$N=pq$ can be deduced from the period $M$ of the function $f(j)=a^j \mod N$
for $j=0,\ldots,2^n-1$ where $N\le 2^n$.
Here $a<N$ is a random number that has no common factors with $N$.
Once $M$ has been determined, at least one factor of $N$ can be found by computing
the greatest common divisor (g.c.d.) of $N$ and $a^{M/2}\pm 1$.

Compared with the example of the previous section, the new aspect
is the modular exponentiation $a^j \mod N$.
For $N=15$ this calculation is almost trivial.
Using the binary represention of $j$, we can write
$a^j \mod N= a^{2^{n-1}j_{n-1}}\ldots a^{2j_{1}} a^{j_{0}}\mod N
= (a^{2^{n-1}j_{n-1}}\mod N) \ldots (a^{2j_{1}}\mod N) (a^{j_{0}}\mod N) \mod N$,
showing that we only need to implement $(a^{2^{k}j_{k}}\mod N)$.
For $N=15$ the allowed values for $a$ are $a=2,4,7,8,11,13,14$.
If we pick $a=2,7,8,13$ then $a^{2^k} \mod N=1$ for all $k>1$
while for the remaining cases we have $a^{2^k} \mod N=1$ for all $k>0$.
Hence for all $a$ we need to ``calculate'' $1\mod N$ and $a\mod N$ and for $a=2,7,8,13$
we in addition have to calculate $a^2\mod N$.
Thus, for $N=15$, only two (not four) qubits are sufficient
to obtain the period of $f(j)=a^j \mod N$~\cite{SYPE01b}.
As a matter of fact, this analysis provides enough information
to deduce the factors of $N=15$ using Shor's procedure
so that no further computation is necessary.
Nontrivial quantum operations
are required if we decide to use three (or more) qubits to determine
the period of $f(j)=a^j \mod N$~\cite{SYPE01b}.
Following Ref.~\cite{SYPE01b}, we consider a seven-qubit quantum computer
with four qubits to hold $f(j)$ and three qubits to perform the QFT.

In essence, the quantum network for the Shor algorithm
is the same as the one shown in Fig.~\ref{ordernetwork} (and therefore not shown)
with the permutations (two qubits) replaced by modular exponentiation (four qubits).
The quantum networks to compute $a^j \mod 15$ for $j=0,\ldots,7$ and a fixed
input $a$ are easy to construct.
For example, consider the case $a=11=\KET{1011}$.
If $j$ is odd, then $11^j \mod 15=11$ and the network should leave $\KET{1011}$ unchanged.
Otherwise, $11^j \mod 15=1$, and hence it should return $\KET{0001}$
(in this case $M=2$ and $\hbox{g.c.d.}(N,a^{M/2}\pm 1)=
\{ \hbox{g.c.d.}(15,10),\hbox{g.c.d.}(15,12)\}=\{5,3\}$, showing
that there is no need to perform a quantum computation).
The network for this operation consists of two CNOT gates that have
as control qubit, the same least-significant qubit of the three qubits
that are input to QFT.
The sequence of CNOT and Toffoli gates that performs similar operations
for the other cases can be found in the same manner.

\subsection{A Three-Input Adder}\label{adder}

\begin{figure*}[t]
\begin{center}
\includegraphics[width=\figsize]{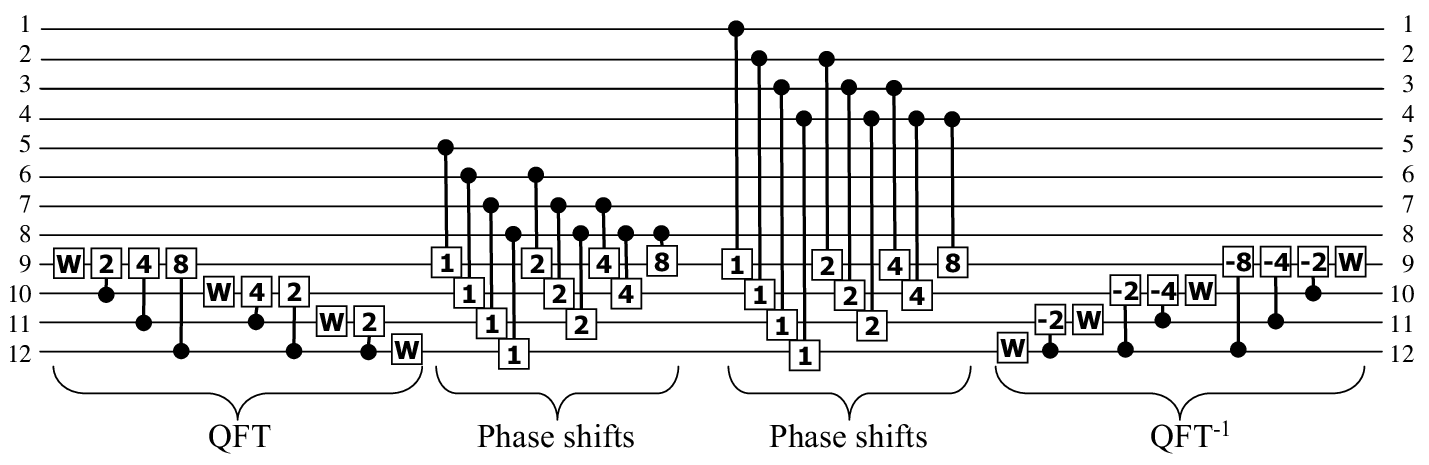}
\caption{Quantum network of a three-input adder, as described in Ref.~\cite{BETT02}.
The algorithm performs a quantum Fourier transform of register 3 (qubits 9 to 12),
adds the content of register 2 (qubits 5 to 8)
and the content of register 1 (qubits 1 to 4),
followed by a QFT on register 3 to yield the final answer
(register 1 + register 2 + register 3 $\mod 16$).
W denotes the Walsh-Hadamard transform.
The operations ``$\pm1$,'' ``$\pm2$,'' ``$\pm4$,'' and ``$\pm8$'' perform controlled phase shifts
with angles $\pm\pi$, $\pm\pi/2$, $\pm\pi/4$ and $\pm\pi/8$, respectively.
Squares that touch each other indicate operations that can be done simultaneously.}
\label{addernetwork}
\end{center}
\end{figure*}

This subsection gives another illustration of the use of the QFT: a quantum
algorithm to add the content of three four-qubit registers.
This example is taken from the Ph.D. thesis of S. Bettelli~\cite{BETT02}.
The quantum network of the complete circuit is shown in
Fig.~\ref{addernetwork}.
The modular structure of this approach is clear.
Note that with respect to the QFT network of Fig.~\ref{qftnetwork},
both the labeling of qubits and the order of operations have been reversed.
The former is merely a change of notation
and the latter allowed because quantum algorithms
are reversible (unitary transformations) by construction~\cite{NIEL00}.
The basis idea of this algorithm is to use the QFT to first transfer
the information in a register to the phase factors of the amplitudes
and then use controlled phase shifts to add information from the two other
registers and finally QFT back to the original representation.
Note that this quantum network differs considerably from
the one described in Ref.~\cite{BARE95} and is also
easier to implement.

\subsection{Number Partitioning}

\setlength{\unitlength}{1cm}
\begin{figure*}[t]
\begin{center}
\includegraphics[width=\figsize]{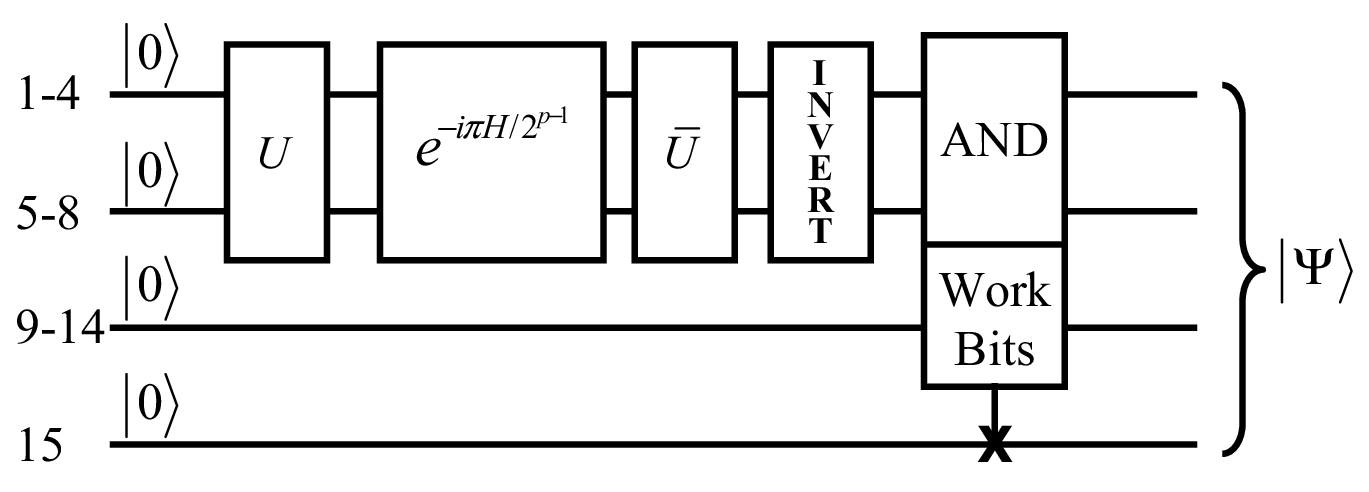}
\end{center}
\caption{
Block diagram of the quantum algorithm that solves the number
partitioning problem. The first four qubits are used to represent the sum of the four integers to be
partitioned. Qubits 5 to 8 are used to determine the number
of solutions of the number partitioning problem. The remaining seven qubits
are used to relate the number of solutions $n_s$ to a physically measurable quantity: the
expectation value of the 15th qubit.
The unitary transformation $U$ prepares the uniform superposition
of the first eight qubits, $\BAR U$ is the inverse of $U$, and
the combination of INVERT and AND gates
sets the 15th qubit to one if and only if the first eight qubits are all one.}
\label{nppnetwork}
\label{npp}
\end{figure*}

As a final example, we discuss a quantum algorithm
to count the number of solutions of the number partitioning problem (NPP)~\cite{RAED01}.
The NPP is defined as follows:
Does there exist a partitioning of the set $A=\{a_1,...,a_n\}$
of $n$ positive integers $a_j$ into two disjoint sets $A_1$ and $A_2=A-A_1$
such that the difference of the sum of the elements of $A_1$
and the sum of the elements of $A_2$ is either zero
(if the sum of all elements of $A$ is even) or one
(if the sum of all elements of $A$ is odd)?
The following simple example may be useful to understand the problem.
If $A=\{1,2,3,4\}$,
the answer to the NPP is yes because for $A_1=\{1,4\}$ and $A_2=\{2,3\}$,
the sum of the elements of $A_1$ and $A_2$ are both equal to five.
In this case there are two solutions because we can interchange $A_1$ and $A_2$.
If $A=\{1,1,1,4\}$
the answer is again yes, as there is one solution, namely $A_1=\{1,1,1\}$  and $A_2=\{4\}$.
The difference of the sum of the elements of $A_1$ and $A_2$ is equal to one,
and that is all right
because the sum of all elements of $A$ is odd. If $A=\{2,2,2,4\}$, there is no solution to the NPP.

The quantum network for the NPP quantum algorithm for the case
that the sum of four integers to be partioned is maximally equal to 16 is shown in Fig.~\ref{nppnetwork}.
The basic idea behind this quantum algorithm is a  number of transformations
that reduce the counting of the number of solutions of the NPP to
finding the number of zero eigenvalues of a spin-1/2 Hamiltonian~\cite{RAED01}.
The largest part (in terms of the number of operations) of the quantum algorithm is a combination
of gates that transforms the state of the quantum computer such that the number of solutions $n_s$
is a physically measureable quantity (the expectation values of the 15th spin).
For $n=4$, we have $n_s=16\sqrt{Q_{15}^z}$~\cite{RAED01}.

\section{Simulation of Ideal Quantum Computers}\label{idealqc}
On the most abstract level, a quantum
algorithm that runs on an ideal quantum computer
performs one unitary transformation ($D\times D$ matrix)
of the state vector (of length $D=2^L$, where $L$ is the number of qubits).
On a conventional computer, this calculation would
take \ORDER{D^2} arithmetic operations.
As in the case of programming a conventional computer,
it is extremely difficult to write down this one-step operation explicitly.
Usually an algorithm consists of many steps, each step being a simple
construct that easily translates into the elementary operations
of the processing units.

As explained in Section~\ref{universal}, for a quantum computer
the set of single-qubit rotations and the CNOT gate
or any other set of gates that can be used for universal quantum computation
will qualify as elementary operations.
Thus, a first task is to decompose a given quantum algorithm (unitary matrix)
into a sequence of unitary matrices that act on one or two qubits.
Recall that each of these matrices is very sparse (see Section~\ref{numa}).
Multiplying a very sparse matrix and a vector can be done
very efficiently on a conventional computer, typically in \ORDER{D}
arithmetic operations.
Hence the remaining important question is whether a quantum
algorithm can be broken down such that the number of elementary
operations is much less than $D$.
The general answer to this question is no~\cite{NIEL00}.
However, for all quantum algorithms reviewed in this chapter, the answer is affirmative.
Finding the shortest decomposition of a given unitary matrix
in terms of a fixed set of sparse unitary matrices is a difficult optimization problem.

Assuming that we have a representation of a quantum algorithm in terms
of the elementary set of sparse unitary matrices, simulating this
algorithm on a conventional computer is conceptually simple.
The basic techniques are reviewed in Section~\ref{numa}.
Software that performs these calculations
is called a quantum computer gate-level simulator.
It is an essential tool to validate the gate-level design
of a quantum algorithm on an ideal quantum computer.
Examples of such simulators are discussed in Section~\ref{simulators}.

\section{Simulation of Physical Models of Quantum Computers}\label{physical}

The qubits on an ideal quantum computer are ideal two-state quantum systems.
Therefore, the operation of an ideal quantum computer does not depend on
the intrinsic dynamics of its qubits.
However, a physically realizable quantum computer is a many-body system of qubits.
The quantum dynamics of these qubits are at the heart of real quantum computation.
Conceptually, the main differences between simulating quantum
computers on a quantum-gate level (see Section~\ref{idealqc}) and physical models
of quantum computers are that the dynamics of the latter are specified
in terms of a time-dependent Hamiltonian (not in terms of unitary matrices)
and that the physical quantum computer is a single physical system
(not a collection of abstract mathematical entities that can be
controlled with infinite precision).

There are many proposals for building quantum computer hardware~\cite{%
CIRA95,KANE98,MAKH99,MOLM99,SORE99,ORLA99,
DIVI00,BERM00,MAHK00,
TOTH01,FERR01,HARR01,
HU03,IMAM03},
and the criteria that a physical system should satisfy
to qualify for quantum computation have been examined~\cite{DIVI00b,DYAK03,KEYE03,BART03}.
Candidate technologies for building quantum gates include
ion traps, cavity QED, Josephson junctions, and nuclear magnetic resonance (NMR)
technology~\cite{%
CIRA95,MONR95,SLEA95,DOMO95,JONE98a,JONE98b,CHUA98a,CHUA98b,
HUGH98,KANE98,WINE98,MAKH99,NAKA99,NOGU99,MOLM99,SORE99,FAZI99,ORLA99,
BLIA00,OLIV00,TOTH01,SYPE00,DIVI00,BERM00,BERM00a,MAHK00,
MARX00,KNIL00,CORY00,JONE00,NIEL00,SYPE01a,SYPE01b,TAKE01,JONE01,SCHO01,
LUCA03,CLAR03,ARDA03,BARN03,
FERR01,HARR01,HU03,IMAM03}.
With the exception of NMR techniques, none of these technologies has demonstrated nontrivial
(that is, more than one CNOT operation on an arbitrary linear superposition)
quantum computation.
In this respect, the field of simulating quantum computers is much
more developed: it is relatively easy to simulate realistic physical
models of NMR quantum computers~\cite{RAED00,RAED02}
or a pair of Josephson junction qubits~\cite{JOHN03,STRA03}.

In some cases (for instance, electron spins, nuclear spins), there is a one-to-one
mapping from the mathematical object of a qubit to
a dynamical variable in the quantum mechanical description.
In other cases (for example, Josephson junctions), this mapping
is not self-evident and requires a very careful analysis
of the dynamics of the physical system~\cite{JOHN03,STRA03}.
Mathematically, nothing prevents us
from using three- or even many-level systems as basic units for quantum computation.
However, inventing useful quantum algorithms that use
two-state systems as qubits already seems to be a major tour de force.
Therefore, it is not clear whether working with three- or more-valued qubits
instead of two-state systems will bring more than additional complications.
Thus, in the following we take the point of view that the
quantum computation will be carried out with physical systems
that are described by (but not necessarily are) two-state quantum systems.

The physical system defined by Eq.~\Eq{hamiltonian}
is sufficiently general to serve as a physical model for
a generic quantum computer at zero temperature.
For instance, Eq.~\Eq{hamiltonian} includes the simplest (Ising) model
of a universal quantum computer~\cite{LLOY93,BERM94}

\begin{equation}
H_{Ising}=-\sum_{i,j=1}^L J^z_{i,j}(t)S_i^zS_j^z
-\sum_{j=1}^3\sum_{\alpha = x,y,z}h_j^{\alpha}(t)S_j^{\alpha}.
\label{ideal}
\end{equation}
More specific candidate hardware realizations of
Eq.~\Eq{hamiltonian} include linear arrays of
quantum dots~\cite{TOTH01}, Josephson junctions~\cite{NAKA99},
and NMR systems~\cite{JONE98a,JONE98b,CHUA98a,CHUA98b,MARX00,KNIL00,CORY00,JONE00}.
An approximate model for the linear arrays of quantum dots reads

\begin{equation}
H(t)=-\sum_{j=1}^L E_jS_j^zS_{j+1}^z
-\sum_{j=1}^L h_j^x(t)S_j^x+E_0\sum_{j=1}^L P_j(t)S_j^z,
\end{equation}
where $E_j=E_0$ ($E_j=2E_0$) when $j$ is odd (even) and $h_j^x(t)$ and $P_j(t)$ are external control
parameters~\cite{TOTH01}.

Projection of the Josephson-junction model onto a
subspace of two states per qubit yields~\cite{MAKH99,FAZI99}

\begin{equation}
H(t)=-2E_I(t)\sum_{j=1}^L S_j^yS_{j+1}^y-E_J\sum_{j=1}^L
S_j^x-\sum_{j=1}^L h_j^z(t)S_j^z,
\end{equation}
where the energy of the Josephson tunneling is represented by $E_J$ and $E_I(t)$
denotes the energy associated with
the inductive coupling between the qubits~\cite{MAKH99,FAZI99}. Here $h_j^z(t)$ and $E_I(t)$
may be controlled externally.

For nuclei with spin quantum number $S=1/2$, the Hamiltonian
that describes the interacting spin system is of the form \Eq{hamiltonian}~\cite{ERNS87}.
NMR uses radio-frequency electromagnetic pulses to rotate the spins~\cite{SLIC90,ERNS87}.
By tuning the radio frequency of the pulse to the precession (Larmor)
frequency of a particular spin, the
power of the applied pulse (= intensity times duration)
controls how much the spin will rotate.
The axis of the rotation is determined by the direction of the applied RF-field.
By selecting the appropriate radio-frequency pulses,
arbitrary single-spin rotations can be carried out.
In other words, using radio-frequency pulses we can perform any single-qubit operation.
The simplest model for the interaction of the spins with
the external magnetic fields reads

\begin{equation}
\label{eq:NMRQC}
h_j^{\alpha}(t)=\hat h_j^{\alpha}+\widetilde h_j^{\alpha}
\sin (2\pi f_j^{\alpha}t+\varphi_j^{\alpha}),
\end{equation}
where $\hat h_j^{\alpha}$ and $\widetilde h_j^{\alpha}$ represent
the static magnetic field and radio-frequency field acting
on the $j$th spin, respectively.
The frequency and phase of the periodic field are denoted
by $f_j^{\alpha}$ and $\varphi_j^{\alpha}$.
The static field $\hat h_j^{\alpha}$ fixes the computational basis.
It is convention to choose $\hat h_j^{x}=\hat h_j^{y}=0$ and
$\hat h_j^{z}\not=0$ so that the direction of the static
field corresponds to the $z$-axis of the spins.
Under fairly general conditions and to a very good approximation,
the nuclear spin Hamiltonian
(containing exchange interactions, dipole-dipole interactions, and so on)
reduces to the universal quantum computer model \Eq{Ising}
in which there are interactions only between the $z$-components of the spin operators.
All NMR quantum computer experiments to date have been interpreted with this model.
Therefore, it makes sense to use model \Eq{Ising} as
a starting point for simulating physical realizations of quantum computers.

\subsection{NMR-like Quantum Computer}

In this section, we illustrate the difference between simulating
an ideal, computer science-type quantum computer and a
more realistic, physical model of quantum computer hardware~\cite{RAED00,RAED02}.
We limit our presentation to the implementation
of the CNOT gate and Grover's algorithm on a two-qubit, NMR-like quantum computer.
A more extensive discussion, as well as many other examples, can be found
in Ref.~\cite{RAED02}.
All simulations have been carried out using the
quantum computer emulator software~\cite{QCEDOWNLOAD}.
The examples given in this section are included in the
software distribution of the quantum computer emulator software~\cite{QCEDOWNLOAD}.

As a prototype quantum computer model, we will take the Hamiltonian for
the two nuclear spins of the $^1$H and $^{13}$C
atoms in a carbon-13 labeled chloroform molecule that has been used
in NMR-quantum computer experiments~\cite{CHUA98a,CHUA98b}.
The strong external magnetic field in the $z$-direction defines the computational basis.
If the spin is aligned along the direction
of the field, the state of the qubit is $\KET{0}$; if the spin points
in the other direction, the state of the qubit is $\KET{1}$.
In the absence of interactions with other degrees of freedom,
the Hamiltonian of this spin-1/2 system reads

\begin{equation}
H_{NMR}=-J_{1,2}^zS_1^zS_2^z-h_1^zS_1^z-h_2^zS_2^z,
\label{NMRmodel}
\end{equation}
where $h_1^z/2\pi \approx 500$MHz, $h_2^z/2\pi\approx 125$MHz, and
$J\equiv J_{1,2}^z/2\pi\approx -215$Hz~\cite{CHUA98a}.
In our numerical work, we use the rescaled model parameters

\begin{equation}
J=-0.43\times10^{-6},\quad h_{1}^z=1, \quad h_{2}^z=1/4.
\label{NMRmodelparameters}
\end{equation}
The ratio $\gamma= h_{2}^z/ h_{1}^z=1/4$ expresses
the difference in the gyromagnetic ratio of the nuclear spin
of the $^1$H and $^{13}$C atom.

\subsubsection{Single-Qubit Operations}\label{NMRsqubit}

In NMR experiments, it is impossible to shield a particular spin from the sinusoidal field.
An application of a sinusoidal field not only affects the state of the resonant spin
but also changes the state of the other spins
(unless they are both perfectly aligned along the $z$-axis).
An analytical, quantitative analysis of this simple-looking
many-body problem is rather difficult.
The values of the model parameters (\ref{NMRmodelparameters}) suggest that
the interaction between the spins will have a negligible impact
on the time evolution of the spins
during application of the sinusoidal pulse if the duration of the pulse
is much shorter than $1/J$ (we use units such that $\tau J$ is dimensionless).
Thus, as far as the single-qubit operations are concerned, we may
neglect the interaction between the two spins (which is also confirmed
by numerical simulation of (\ref{NMRmodel}), see later).
In this section, we closely follow~\cite{RAED02}.

We consider
the two-spin system described by the  time-dependent Schr\"odinger equation

\begin{equation}
\label{RFTDSE1}
i{\partial\over\partial t}\KET{\Phi(t)}=
-\left[
h_{1}^z S_1^z +h_{2}^z S_2^z
+\widetilde h_{1}^x( S_1^x \sin \omega t +S_1^y \cos \omega t)
+\widetilde h_{2}^x( S_2^x \sin \omega t +S_2^y \cos \omega t)
\right]
 \KET{\Phi(t)},
\end{equation}
for two interacting spins in a static and a rotating sinusoidal field.
As usual, it is convenient to work in a rotating frame~\cite{SLIC90}.
Substituting $\KET{\Phi(t)}=e^{it\omega( S_1^z+S_2^z)}\KET{\Psi(t)}$, we obtain

\begin{equation}
\label{RFTDSE2}
i{\partial\over\partial t}\KET{\Psi(t)}=
-\left[
(h_{1}^z -\omega)S_1^z +(h_{2}^z-\omega) S_2^z
+\widetilde h_{1}^xS_1^y +\widetilde h_{2}^x S_2^y
\right]
 \KET{\Psi(t)}.
\end{equation}
Our aim is to determine the conditions under which we can
rotate spin 1 by an angle $\varphi_1$ without affecting
the state of spin 2.
First we choose

\begin{equation}
\label{CONDITION0}
\omega=h_{1}^z,
\end{equation}%
that is, the frequency of the sinusoidal field is tuned to the resonance frequency of spin 1.
Then (\ref{RFTDSE2}) can easily be integrated. The result is

\begin{equation}
\label{RFTDSE3}
\KET{\Phi(t)}=
e^{ith_{1}^z( S_1^z+S_2^z)}
e^{it\widetilde h_{1}^x S_1^y}
e^{it{\bf S}_2\cdot{\bf v}_{12}}
 \KET{\Phi(0)},
\end{equation}%
where ${\bf v}_{nm}\equiv(0,\widetilde h_{m}^x,h_{m}^z - h_{n}^z )$.
The third factor in (\ref{RFTDSE3}) rotates spin 2 about the vector ${\bf v}_{12}$.
This factor can be expressed as

\begin{equation}
\label{RFTDSE4}
e^{it{\bf S}_m\cdot{\bf v}_{nm}}
=
\left(
\begin{array}{cc}
1&0 \\
0&1 \\
\end{array}
\right)
\cos \frac{t|{\bf v}_{nm}|}{2}
+
i|{\bf v}_{nm}|^{-1}\left(
\begin{array}{cc}
h_{m}^z - h_{n}^z&-i\widetilde h_{m}^x \\
i\widetilde h_{m}^x&h_{n}^z - h_{m}^z \\
\end{array}
\right)
\sin \frac{t|{\bf v}_{nm}|}{2},
\end{equation}%
and we see that the sinusoidal field will not change the state of spin 2
if and only if the duration $t_1$ of the pulse satisfies

\begin{equation}
\label{CONDITION3}
t_1|{\bf v}_{12}|=t_1\sqrt{(h_{1}^z - h_{2}^z)^2+(\widetilde h_{2}^x)^2}=4\pi n_1,
\end{equation}%
where $n_1$ is a positive integer.
The second factor in (\ref{RFTDSE3}) is a special case of (\ref{RFTDSE4}).
Putting

\begin{equation}
\label{CONDITION2}
t_1\widetilde h_{1}^x=\varphi_1,
\end{equation}%
the second factor in (\ref{RFTDSE3})
will rotate spin 1 by $\varphi_1$ about the $y$-axis.
Therefore, if conditions
(\ref{CONDITION0}),
(\ref{CONDITION3}), and
(\ref{CONDITION2})
are satisfied, we can rotate spin 1 by $\varphi_1$
without affecting the state of spin 2,
independent of the physical realization of the quantum computer.
Combining these conditions yields the so-called $2\pi k$
method for suppressing nonresonant effects~\cite{BERM97,BERM00}:

\begin{equation}
\label{CONDITIONk}
t_1=\frac{\pi}{|h^z_1(1-\gamma)|}\sqrt{16n_1^2-(\varphi_1/\pi)^2}.
\end{equation}%

If we would use the conditions
(\ref{CONDITION0}), (\ref{CONDITION2}) and (\ref{CONDITIONk}) to determine the parameters
$\omega$, $\widetilde h_{1}^x$ and $t_1$ of the radio-frequency pulse,
the first factor in (\ref{RFTDSE3}) can still generate a phase shift.
This phase shift clearly depends on the state of the spins.
Although it drops out in the expression of the expectation value of single qubits,
for quantum computation purposes it has to be taken into account
(as is confirmed by numerical calculations~\cite{RAED02}).
Adding the condition

\begin{equation}
\label{CONDITION1}
t_1h_{1}^z=4\pi k_1,
\end{equation}%
where $k_1$ is a positive integer ($h_{i}^z>0$ by definition),
the first factor in (\ref{RFTDSE3}) is always equal to one.
A last constraint on the choice of the pulse parameters comes
from the fact that

\begin{equation}
\label{CONDTION4}
h_{2}^\alpha=\gamma h_{1}^\alpha\quad,\quad
\widetilde h_{2}^\alpha=\gamma\widetilde h_{1}^\alpha\quad;\quad  \alpha=x,y,z.
\end{equation}%
Without loss of generality, we may assume that $0<\gamma<1$.

Using conditions
(\ref{CONDITION0}),
(\ref{CONDITION3}),
(\ref{CONDITION2}),
(\ref{CONDITION1}), and
(\ref{CONDTION4})
and reversing the role of spin 1 and spin 2, we obtain

\begin{eqnarray}
(1-\gamma)^2 k_1^2+\frac{\gamma^2}{4}\left(\frac{\varphi_1}{2\pi}\right)^2=n_1^2
\quad,\quad
(1-\frac{1}{\gamma})^2 k_2^2+\frac{1}{4\gamma^2}\left(\frac{\varphi_2}{2\pi}\right)^2=n_2^2,
\label{CONDITION}
\end{eqnarray}%
where $k_1$, $k_2$, $n_1$, and $n_2$ are positive integers.
The angles of rotation about the $y$-axis can be chosen such that
$0\le \varphi_1\le 2\pi$ and $0\le \varphi_2\le 2\pi$.
Of course, similar expressions hold for rotations about the $x$-axis.

In general, (\ref{CONDITION}) has no solution, but a good approximate
solution may be obtained if
$\gamma$ is a rational number and $k_1$ and $k_2$ are large.
Let  $\gamma=N/M$ where $N$ and $M$ are integers satisfying $0<N<M$.
It follows that
the representation $k_1=kMN^2$ and $k_2=kNM^2$ will generate
sufficiently accurate solutions of (\ref{CONDITION})
if the integer $k$ is chosen such that
$2kNM(M-N)\gg1$.
%
In terms of $k$, $N$, and $M$, the relevant physical quantities are then given by

\begin{equation}
\frac{t_1h_{1}^z}{2\pi}=2kMN^2
\quad,\quad
\frac{\widetilde h_{1}^x}{h_{1}^z}=\frac{1}{2kMN^2}\frac{\varphi_1}{2\pi},
\quad,\quad
\frac{t_2h_{1}^z}{2\pi}=2kM^3
\quad,\quad
\frac{\widetilde h_{2}^x}{h_{1}^z}=\frac{1}{2kM^3}\frac{\varphi_2}{2\pi}.
\label{PARAMETERS}
\end{equation}%

We have derived conditions (\ref{PARAMETERS})
under the assumption of ideal sinusoidal RF pulses.
In an experiment, there is no such limitation: the sinusoidal fields
may be modulated by almost any waveform \cite{ERNS87,FREE97}.
However, the fact that in quantum computer applications it is necessary to use
single-spin pulses that do not change the state of the other spins remains.
For general pulses, finding the form of the pulse that rotates spin 1 such that the state
of spin 2 is not affected is a complicated nonlinear optimization problem~\cite{RAED02,PALA02}.

To summarize: If conditions
(\ref{CONDITION0}),
(\ref{CONDITION3}),
(\ref{CONDITION2}), and
(\ref{CONDITION1})
are satisfied, we can rotate spin 1 by $\varphi_1$
without affecting the state of spin 2
and without introducing unwanted phase shifts.
In numerical experiments, Eq.(\ref{PARAMETERS}) may be used to determine
the duration of the sinusoidal pulses.
These sinusoidal pulses will then be optimized in the sense
that a pulse that rotates spin 1 (2) will
change the state of spin 2 (1) only slightly
if $k$ satisfies $2kNM(M-N)\gg1$.

\subsubsection{Two-Qubit Operations}

In Section~\ref{twoqubit}, we implemented the CNOT sequence \Eq{COMM3} using
the Ising model $H=-JS_1^zS_2^z-h(S_1^z+S_2^z)$.
The implementation of the CNOT operation using Eq.~\Eq{NMRmodel}
requires additional steps to account
for the fact that the two nuclear spins feel different static fields.
The additional rotations are

\begin{equation}
\label{eq:cnotnmr}
\CNOT=
{\BAR Y_2}
e^{-i\tau(h_{1}^z-h)S_1^{z}}
e^{-i\tau(h_{2}^z-h)S_2^{z}}
e^{-i\tau H_{NMR}}{\NOBAR Y_2}
=
{\BAR Y_2}
e^{-i\tau(h_{1}^z-h)S_1^{z}}
e^{-i\tau(h_{2}^z-h)S_2^{z}}
{\NOBAR Y_2}
{\BAR Y_2}
e^{-i\tau H_{NMR}}{\NOBAR Y_2}
,
\end{equation}
where we used the fact that ${\NOBAR Y_2}{\BAR Y_2}=1$.
The extra phase shifts in (\ref{eq:cnotnmr}) can be expressed
in terms of single-qubit operations. The identities
\begin{equation}
\label{eq:cnotnmr1}
e^{-i\tau(h_{1}^z-h)S_1^{z}}
=
{\NOBAR Y_1} {\NOBAR X_1^\prime} {\BAR Y_1}
=
{\BAR X_1} {\NOBAR Y_1^\prime} {\NOBAR X_1}
,\quad
\label{eq:cnotnmr2}
e^{-i\tau(h_{2}^z-h)S_2^{z}}
=
{\NOBAR Y_2} {\NOBAR X_2^\prime} {\BAR Y_2},
\end{equation}
define the single-spin rotations
${\NOBAR X_1^\prime}$,
${\NOBAR Y_1^\prime}$,
and
${\NOBAR X_2^\prime}$.

In the case of Grover's database search algorithm,
the representation of $G$ in terms of the time evolution of (\ref{NMRmodel}) reads

\begin{equation}
\label{eq:Gnmr}
G=e^{-i\pi S_1S_2}=e^{-i\tau h_{1}^z S_1^{z}}
e^{-i\tau h_{2}^zS_2^{z}} e^{-i\tau H_{NMR}}
=
{\NOBAR Y_2} {\NOBAR X_2^{\prime\prime}} {\BAR Y_2}
{\NOBAR Y_1} {\NOBAR X_1^{\prime\prime}} {\BAR Y_1}
e^{-i\tau H_{NMR}}
,
\end{equation}
where $\tau =-\pi/J$.
This choice of $\tau$ also fixes the angles of the rotations
and all parameters of the operations
${\NOBAR X_1^{\prime\prime}}$ and ${\NOBAR X_2^{\prime\prime}}$.

Equation \Eq{eq:cnotnmr1} suggests that
there are many different, logically equivalent
sequences that implement the CNOT gate on an NMR-like quantum computer. We have chosen to limit ourselves
to the respresentations

\begin{eqnarray}
\label{eq:cnot1}
\CNOT_{1}
=
{\NOBAR Y_1}{\NOBAR X_1^\prime}{\BAR Y_1}{\NOBAR X_2^\prime}{\BAR Y_2}{I^\prime}{\NOBAR Y_2},
\quad
\label{eq:cnot2}
\CNOT_{2}
=
{\NOBAR Y_1}{\NOBAR X_1^\prime}{\NOBAR X_2^\prime}{\BAR Y_1}{\BAR Y_2}{I^\prime}{\NOBAR Y_2},
\quad
\label{eq:cnot3}
\label{eq:cnot123}
\CNOT_{3}
=
{\BAR X_1}{\NOBAR Y_1^\prime}{\NOBAR X_2^\prime}{\BAR Y_2}{\NOBAR X_1}{I^\prime}{\NOBAR Y_2},
\end{eqnarray}
where we introduced the symbol $I^\prime$ to represent the time evolution $e^{-i\tau H_{NMR}}$
with $\tau=-\pi/J$.

On an ideal quantum computer, there is no difference between
the logical and physical computer and the sequences (\ref{eq:cnot1}) give identical results.
However, on a physical quantum computer such as the NMR-like quantum computer (\ref{NMRmodel}),
this is not the case.
On a physically realizable NMR-like quantum computer $X_1X_2\not=X_2X_1$
unless $2kNM(M-N)\gg1$ and (\ref{PARAMETERS}) are satisfied {\sl exactly}.
Next we use the sequences (\ref{eq:cnot1})
to demonstrate that this unpleasant feature of physical quantum computers
may give rise to large systematic errors.

\begin{table*}[t]
\begin{center}
\caption{Model parameters of single-qubit operations on an NMR-like quantum computer using rotating sinusoidal fields
for the case ($k=1$, $N=1$, $M=4$); see (\ref{PARAMETERS}).
Parameters of model (\ref{RFTDSE1}) that do not
appear in this table are zero, except for the interaction $J=-0.43\times10^{-6}$
and the constant magnetic fields $h_{1}^z=1$ and $h_{2}^z=0.25$.
The  time-dependent Schr\"odinger equation is solved using a time step $\delta/2\pi=0.01$.}
\begin{ruledtabular}
\begin{tabular}{ccccccccc}
& $\tau/2\pi$ & $\omega$ & $\widetilde h_{1}^x$ & $\widetilde h_{2}^x$ & $\varphi_x$ & $\widetilde h_{1}^y$ & $\widetilde h_{2}^y$ & $\varphi_y$ \\
\hline
\noalign{\vskip 4pt}
$X_1$ &  8 & 1.00        & -0.0312500 & -0.0078125 & $-\pi/2$ & -0.0312500& -0.0078125 & 0 \\
$X_2$ & 128 & 0.25       & -0.0078125 & -0.0019531 & $-\pi/2$ & -0.0078125& -0.0019531 & 0 \\
$Y_1$ &  8 & 1.00        &  0.0312500 &  0.0078125 & 0        &  0.0312500&  0.0078125 & $\pi/2$ \\
$Y_2$ & 128 & 0.25       &  0.0078125 &  0.0019531 & 0        &  0.0078125&  0.0019531 & $\pi/2$ \\
$X_1^\prime$ &  8 & 1.00 &  0.0559593 &  0.0139898 & $-\pi/2$ &  0.0559593&  0.0139898 & 0 \\
$X_2^\prime$ & 128 & 0.25&  0.0445131 &  0.0111283 & $-\pi/2$ &  0.0445131&  0.0111283 & 0 \\
$Y_1^\prime$ &  8 & 1.00 & -0.0559593 & -0.0139898 & 0        & -0.0559593& -0.0139898 & $\pi/2$ \\
$X_1^{\prime\prime}$ &  8 & 1.00 &  0.0872093 &  0.0218023 & $-\pi/2$ &  0.0872093&  0.0218023 & 0 \\
$X_2^{\prime\prime}$ & 128 & 0.25&  0.0523256 &  0.0130914 & $-\pi/2$ &  0.0523256&  0.0130914 & 0 \\
\end{tabular}
\end{ruledtabular}
\label{tab:NMRRFQC}
\end{center}
\end{table*}

\begin{table*}[t]
    \caption{Expectation values of the two qubits as obtained by
    performing a sequence of five CNOT operations on an NMR-like  quantum computer.
    The initial states $\KET{10}$, $\KET{01}$, $\KET{11}$, and
    $\KET{singlet}=(\KET{01}-\KET{10})/\sqrt{2}$ have been prepared by starting from the state
    $\KET{00}$ and performing exact rotations of the spins.
    The CNOT operations on the singlet state are followed by a $\pi/2$ rotation of spin 1 to yield a nonzero
    value of qubit 1.
    The time $s=\tau/2\pi=2kMN^2$ determines the duration and strength of the
    sinusoidal pulses through relations (\ref{PARAMETERS}), see
    Table \ref{tab:NMRRFQC} for the example of the case $s=8$.
    The CNOT operation itself was implemented by applying the CNOT sequence given by (\ref{eq:cnot1}).
    On an ideal quantum computer, CNOT$^4$ is the identity operation.
    For $s=256$, all results are exact within an error of 0.01.}
  \begin{center}
\begin{ruledtabular}
\begin{tabular}{ccccccccccc}
\multicolumn{1}{c}{ }&\multicolumn{2}{c}{Ideal quantum computer}&\multicolumn{2}{c}{$s=8$} &
\multicolumn{2}{c}{$s=16$}&
\multicolumn{2}{c}{$s=32$}&
\multicolumn{2}{c}{$s=64$}\\
Operation&
$Q_1^z$&$Q_2^z$&
$Q_1^z$&$Q_2^z$&
$Q_1^z$&$Q_2^z$&
$Q_1^z$&$Q_2^z$&
$Q_1^z$&$Q_2^z$\\
\hline
\noalign{\vskip 4pt}
$(\CNOT_1)^5\KET{00}$          &0.00&0.00&0.00&0.00&0.00&0.00&0.00&0.00&0.00&0.00\\
$(\CNOT_2)^5\KET{00}$          &0.00&0.00&0.24&0.76&0.50&0.26&0.20&0.07&0.06&0.02\\
$(\CNOT_3)^5\KET{00}$          &0.00&0.00&0.23&0.76&0.50&0.26&0.20&0.07&0.06&0.02\\
\hline
\noalign{\vskip 4pt}
$(\CNOT_1)^5\KET{01}$          &1.00&1.00&1.00&1.00&1.00&1.00&1.00&1.00&1.00&1.00\\
$(\CNOT_2)^5\KET{01}$          &1.00&1.00&0.76&0.24&0.50&0.74&0.80&0.93&0.95&0.98\\
$(\CNOT_3)^5\KET{01}$          &1.00&1.00&0.77&0.24&0.50&0.74&0.80&0.93&0.95&0.98\\
\hline
\noalign{\vskip 4pt}
$(\CNOT_1)^5\KET{10}$          &0.00&1.00&0.00&1.00&0.00&1.00&0.00&1.00&0.00&1.00\\
$(\CNOT_2)^5\KET{10}$          &0.00&1.00&0.24&0.24&0.51&0.74&0.20&0.93&0.06&0.98\\
$(\CNOT_3)^5\KET{10}$          &0.00&1.00&0.23&0.24&0.51&0.74&0.20&0.93&0.06&0.98\\
\hline
\noalign{\vskip 4pt}
$(\CNOT_1)^5\KET{11}$          &1.00&0.00&1.00&0.00&1.00&0.00&1.00&0.00&1.00&0.00\\
$(\CNOT_2)^5\KET{11}$          &1.00&0.00&0.76&0.76&0.50&0.26&0.80&0.07&0.95&0.02\\
$(\CNOT_3)^5\KET{11}$          &1.00&0.00&0.77&0.76&0.50&0.26&0.80&0.07&0.95&0.02\\
\hline
\noalign{\vskip 4pt}
$Y_1 (\CNOT_1)^5\KET{singlet}$ &1.00&1.00&0.90&1.00&0.03&1.00&0.58&1.00&0.88&1.00\\
$Y_1 (\CNOT_2)^5\KET{singlet}$ &1.00&1.00&0.98&0.24&0.95&0.74&0.98&0.93&0.99&0.98\\
$Y_1 (\CNOT_3)^5\KET{singlet}$ &1.00&1.00&0.79&0.24&0.55&0.74&0.82&0.93&0.95&0.98\\
\end{tabular}
\label{tab:NMRRFRESULT1}
\end{ruledtabular}
  \end{center}
\end{table*}

\begin{table*}[t]
    \caption{Expectation values of the two qubits as obtained by
    running Grover's database search algorithm on an NMR-like quantum computer.
    The time $s=\tau/2\pi=2kMN^2$ determines the duration and strength of the
    sinusoidal pulses through relations (\ref{PARAMETERS}), see
    Table \ref{tab:NMRRFQC} for the example of the case $s=8$.
    Within two-digit accuracy, all results for $s=256$ are exact.}
  \begin{center}
\begin{ruledtabular}
\begin{tabular}{ccccccccccc}
\multicolumn{1}{c}{ }&\multicolumn{2}{c}{Ideal quantum computer}&\multicolumn{2}{c}{$s=8$} &
\multicolumn{2}{c}{$s=16$}&
\multicolumn{2}{c}{$s=32$}&
\multicolumn{2}{c}{$s=64$}\\
Item position&
$Q_1^z$&$Q_2^z$&
$Q_1^z$&$Q_2^z$&
$Q_1^z$&$Q_2^z$&
$Q_1^z$&$Q_2^z$&
$Q_1^z$&$Q_2^z$\\
\hline
\noalign{\vskip 4pt}
0&0.00&0.00&0.48&0.53&0.15&0.16&0.04&0.04&0.01&0.01\\
1&1.00&0.00&0.52&0.50&0.85&0.15&0.96&0.04&0.99&0.01\\
2&0.00&1.00&0.55&0.48&0.15&0.84&0.04&0.96&0.01&0.99\\
3&1.00&1.00&0.45&0.50&0.85&0.85&0.96&0.96&0.99&0.99\\
\end{tabular}
\end{ruledtabular}
    \label{tab:GROVRESULT1}
  \end{center}
\end{table*}

\subsubsection{Simulation Results}\label{nmrresults}

The model parameters for the rotating sinusoidal fields are determined according
to the theory outlined previously.
We use the integer $k$ to compute all free parameters and
label the results of the quantum computer calculation by the subscript $s=2kMN^2$.
For reference we present the set of parameters corresponding to $s=8$ ($k=1$) in
Table \ref{tab:NMRRFQC}.
Multiplying $s$ (the duration of the sinusoidal pulse) with the unit of time
(2 ns for the case at hand) shows that in our simulations,
single-qubit operations are implemented by using short
pulses that are, in NMR terminology, nonselective and hard.
In contrast to the analytical treatment given in Section~\ref{NMRsqubit}, in all our
simulations the interaction $J$ is nonzero.

The two-qubit operation $I^\prime$ can be implemented by letting the system
evolve in time according to Hamiltonian $H_{NMR}$, given by (\ref{NMRmodel}).
$I^\prime$ is the same for both an ideal or NMR-like quantum computer.
Note that the condition $\tau J=-\pi$ yields
$\tau/2\pi=1162790.6977$, a fairly large number
[compared to $h_1^z=1$, see (\ref{NMRmodel})].
Also note the digits after the decimal point: this accuracy is necessary
for correct operation of the quantum computer~\cite{RAED02}.

As a first check we execute all sequences on an implementation of the ideal quantum computer
and confirm that they give the exact answers (results not shown).
It is also necessary to rule out that the numerical results depend on the
time step $\delta$ used to solve the  time-dependent Schr\"odinger equation.
The numerical error of the product formula used by QCE is proportional to $\delta^2$
\cite{RAED87}.
It goes down by a factor of about one 100 if we reduce the time step by a factor of 10.
Within the two-digit accuracy used to present our data,
there is no difference between the results for $\delta=0.01$ and $\delta=0.001$.

In Table \ref{tab:NMRRFRESULT1} we present simulation results
for $\CNOT^5$ acting on one of the basis states and $Y_1 \CNOT^5$ acting on a singlet state,
using the three logically equivalent but physically different
implementations $\CNOT_1$, $\CNOT_2$, and $\CNOT_3$ [see Eq.(\ref{eq:cnot123})].
It is clear that some of the least accurate implementations ($s=8$)
do not reproduce the correct answers if the input corresponds to
one of the four basis states.
Moreover, if the operations act on the exact singlet state,
the results strongly depend on the CNOT implementation if $s\le32$.
In agreement with the theoretical analysis given previously,
the exact results are recovered if $s$ is sufficiently large.
On the time scale set by $J$, the pulses used in these simulations are so short
that the presence of a nonzero $J$ has a negligible effect on the single-qubit pulses.
These simulations also demonstrate that in order for a quantum algorithm to work
properly, it is not sufficient to show that it correctly operates on the basis states.

In contrast to computation in the classical framework, quantum
computation can make use of entangled states.
At the point where the quantum algorithm actually uses an
entangled state, the quantum algorithm is most sensitive to (accumulated) phase errors.
As another illustration of this phenomenon, we present
in Table~\ref{tab:GROVRESULT1}
some typical results obtained by executing Grover's database search algorithm.
We used the same NMR-like quantum computer as for the CNOT calculations.
We conclude that reasonably good answers are obtained if $s\ge32$,
in concert with our observations for the CNOT operation.

\subsection{Decoherence}

An important topic that we have not discussed so far is the effect of the interaction
of the quantum computer with its environment (dissipation, decoherence).
The general belief seems to be that engineers will be able
to cope with systematic errors (see Section~\ref{nmrresults} for examples)
due to the physics of the qubits
and that problems due to decoherence will be the main stumbling block
for building useful quantum computers~\cite{NIEL00}.
As most theoretical work~\cite{LUMI98,
ZANA98,SUN98,BARN99,BEIG00,SMIR00,TSEN00,CARV01,MANC01,THOR01,
WU02,BARR03,DALT03,YU03,STOR03,PROT03,LI03,TEKL03}
is based on approximations, the
validity of which needs to be established,
computer simulation of physical models that include decoherence
may be of great value to gain insight into this challenging problem.
Dissipation cannot be treated within the context of the simulation models
that we have covered in this review.
Instead of solving the time-dependent Schr\"odinger equation,
we have to solve the equations of motion of the full density matrix
of an interacting many-body system.
Although still feasible for a small number of qubits $L$,
the computation time now scales with $2^{2L}$ instead of with $2^L$.

In the absence of dissipation, it is straightforward to incorporate
into the class of simulation
models physical processes that lead to loss of phase information during the time evolution.
In Section~\ref{comments} we have given some hints as how this can be done.
Needless to say, performing such simulations is costly in terms of CPU time
but the pay-off may be significant: the simple, highly idealized uncorrelated
random processes that are being used to analyze error correction and fault-tolerant
quantum computing are very far from being physically realistic~\cite{PRES98}.
Quantum error correction schemes that work well on an ideal quantum computer
require many extra qubits and many additional operations
to detect and correct errors~\cite{PRES98}.
The systematic errors discussed previously are not included
in the current model~\cite{NIEL00} of quantum error correction and fault tolerant computing.
Our simulation results for a most simple NMR-like quantum computer
demonstrate that systematic errors are hard to avoid, in particular
if the number of qubits increases (which is a basic
requirement for fault-tolerant quantum computation).
It would be interesting to simulate a two-qubit quantum computer
that interacts with other spins and analyze the effect of decoherence
on, for example, the CNOT operation. Such simulations are definitely within
reach but, to our knowledge, no results have been reported.

\section{Quantum Computer Simulators}{\label{simulators}}
\subsection{Review}

In this section, we review software tools that run on conventional computers and
can be used to study various aspects of quantum computation.
The term ``quantum computer simulator'' is used in a broad sense:
not all the software discussed in this section actually simulates a quantum computer.
In recent years many quantum computer simulators have been developed.
The level of complexity of the simulations they can perform varies considerably:
some deal with both the quantum hardware and software while others
focus on quantum computer algorithms and software.
An early detailed survey is given in Ref.~\cite{WALL01}.
In the survey that follows, we confine ourselves to software that was
accessible via the Web at the time of writing.

Because quantum computer hardware is not readily available,
quantum computer simulators are valuable tools to develop and test quantum algorithms
and to simulate physical models for the hardware implementation of a quantum processor.
Simulation is an essential part of the design of conventional digital circuits
and microprocessors in particular.
It is hardly conceivable that it will be possible to construct a real quantum computer
without using simulation tools to validate its design and operation.

There is a very important difference in simulating
conventional microprocessors and quantum processors.
In conventional digital circuits, the internal working of each logical circuit
is irrelevant for the logical operation of the processor.
However, in a quantum computer the internal quantum dynamics
of each elementary building block is a key ingredient of the quantum computer itself.
Therefore, in the end the physics of the elementary units that make up the quantum computer
have to be incorporated into the simulation model of the quantum computer.
It is not sufficient to model a quantum computer in terms of hypothetical
qubits that evolve according to the rules of a mathematical model
of a hypothetically ideal quantum computer.
Including models for noise and errors also is no substitute for the
physics of the qubits that comes into play when one wants to make
contact to a real physical system.

Quantum computer simulators come in different flavors and have different levels of complexity.
A first group of software deals with programming languages for quantum computers.
Programming languages express the semantics of a computation in an abstract manner
and automatically generate a sequence of elementary operations to control the computer.
An overview of quantum programming languages is given in Table~\ref{tabsim0}.
A second group of simulators comprises the quantum compilers,
as summarized in Table~\ref{tabsim1}.
A quantum compiler takes as input a unitary transformation and returns
a sequence of elementary one-qubit and two-quibit operations
that performs the desired quantum operation.
Quantum circuit simulators, or gate-level simulators, form the third group of simulators.
On an abstract level, quantum computation on an ideal quantum computer
amounts to performing unitary transformations on a complex-valued (state) vector.
A conventional computer can perform these operations equally well, provided
there is enough memory to store all the numbers of the vector.
That is exactly what quantum circuit simulators do: they provide
a software environment to simulate ideal quantum computers.
A summary of quantum circuit simulators is given in Table~\ref{tabsim2}.
The fourth group, collected in Table~\ref{tabsim3}, consists of software
that uses time-dependent Hamiltonians to implement the unitary transformations on the qubits.
These simulators make it possible to emulate various hardware designs of quantum computers,
that is, they simulate models for physical realizations of quantum computers.
Simulators of this group can also be used as gate-level simulators.
Finally, Table~\ref{tabsim4} describes a number of purely pedagogical software products.

\subsection{Example: Quantum Computer Emulator}

In this section, we briefly discuss some features of QCE (see Table~\ref{tabsim3}),
a software tool that simulates ideal quantum computers
and also emulates physical realizations of quantum computers.
The QCE is freely distributed as a self-installing executable, containing the program, documentation and
many examples of quantum algorithms, including all the quantum algorithms discussed in this review.
The file help.htm~\cite{HELPHTM} contains information about how to install and how to start the QCE.
The QCE runs in a Windows environment.
It consists of a simulator of a generic, general-purpose quantum computer,
a graphical user interface, and a real-time visualization module.

QCE's simulation engine is built on the principles reviewed in this chapter.
The engine simulates the physical processes
that govern the operation of the hardware quantum processor,
strictly according to the laws of quantum mechanics.
It solves the time-dependent Schr\"odinger equation (\ref{TDSE0})
by a Suzuki product-formula (see Section~\ref{SUZTROT})
in terms of elementary unitary operations.
For all practical purposes, the results obtained by this technique are indistinguishable
from the exact solution of the time-dependent Schr\"odinger equation.
The graphical user interface is used to control the simulator,
to define the hardware of the quantum computer, and
to debug and execute quantum algorithms.
Using the graphical user interface requires no skills
other than the basic ones needed to run a standard MS-Windows applications.
The current version of QCE (8.1.1) simulates quantum computers
with a maximum of 16 qubits.

QCE can be used to validate designs of physically realizable quantum processors.
QCE is also an interactive educational tool to learn about quantum computers and
quantum algorithms.
As an illustration, we give a short exposition of the implementation
of the three-input adder (see Fig.~\ref{addernetwork}) on
an ideal quantum computer and of Grover's database
search algorithm on an NMR-like quantum computer.
Other examples are given in Refs.\cite{RAED00,RAED01,RAED02,MICH02}.

\subsubsection{General Aspects}
As described in Section~\ref{qas}, a quantum algorithm for a quantum computer
model (\ref{hamiltonian}) consists of a sequence of
elementary operations that change the state $\KET{\Phi}$ of the quantum
processor according to the  time-dependent Schr\"odinger equation (\ref{TDSE0}),
namely, by (a product of) unitary transformations.
We call these elementary operations microinstructions in the sequel.
They do not play exactly the same role as microinstructions in digital processors.
They merely represent the smallest units of operation of the quantum processor.
The action of a microinstruction on the state $\KET{\Phi}$
of the quantum processor is defined by specifying
how long it acts (that is, the time interval it is active)
and the values of all the $J$s and $h$s appearing in $H(t)$ (\ref{hamiltonian}),
the model Hamiltonian of the quantum computer.
A microinstruction transforms the input state $\KET{\Phi(t)}$
into the output state $\KET{\Phi(t+\tau)}$, where $\tau$ denotes the
time interval during which the microinstruction is active.
During this time interval the only time-dependence of $H(t)$
is through the time-dependence of the external fields on the spins.
Microinstructions completely specify
the particular (ideal or physical) realization of the quantum computer
that one wants to emulate.
Quantum algorithms are translated into quantum programs
that are written as a sequence of microinstructions.
The graphical user interface of the QCE has been developed to
facilitate the specification of the microinstructions
(to model the quantum computer hardware)
and the execution of quantum programs.
A detailed exposition of how to use the QCE can be found in Ref.~\cite{MICH03}.

\begin{figure*}[t]
\begin{center}
\includegraphics[width=12cm]{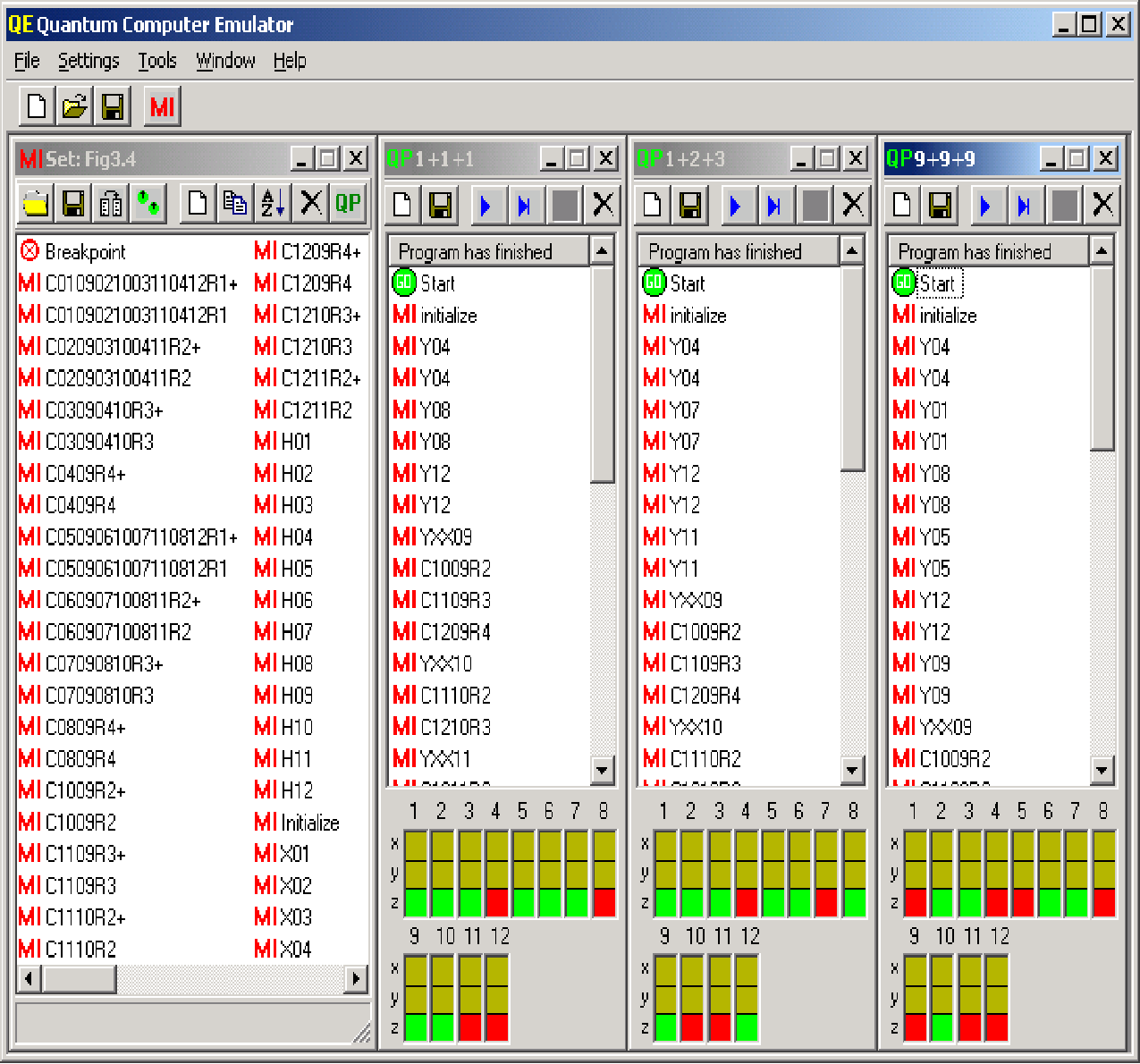}
\caption{QCE implementation of the quantum network of Fig.~\ref{addernetwork} for adding three quantum
registers of four qubits each on an ideal 12-qubit quantum computer.
Quantum programs (from left to right) compute 1+2+3, 1+1+1, and 9+9+9 mod 16.
Not shown are the microinstructions that set the initial values of the
three four-qubit registers.}
\label{adderQCE}
\end{center}
\end{figure*}

\subsubsection{Three-Input Adder on an Ideal Quantum Computer}
A QCE implementation of the three-input adder described
in Section~\ref{adder} is shown in Fig.~\ref{adderQCE}.
The left panel shows the microinstruction set ``adder'' and
the three panels on the right show the quantum programs
``1+1+1,'' ``1+2+3,'' and ``9+9+9.''
These can be found in the quantum program directory ``adder.''
The microinstruction set contains all microinstructions that are needed
to execute the quantum programs ``1+1+1,'' ``1+2+3,'' and ``9+9+9''
on an ideal 12-qubit quantum computer.
The results (in binary notation) of the examples (1+1+1, 1+2+3, and 9+9+9)
can be read off from the values of qubits 9 to 12 at
the bottom of the quantum programs.
Qubit 12 corresponds to the least significant bit.
The numerical values of the qubits appear when the mouse moves over the bottom region
of the quantum program window
(red or dark gray corresponds to 1, green or light gray to 0,
and greenish brown or middle gray to 0.5).
The graphics area in Fig.~\ref{adderQCE} is too small to show all microinstructions
(on the computer the scroll bars allow the user to open/edit all microinstructions).
The same holds for the three quantum programs.
This example suggests that programming more complicated quantum algorithms
like this one should not be done by hand; in fact, the microinstructions
and quantum programs for these examples have been generated by another computer program.
Some of the microinstructions may look rather complicated,
but that is a little misleading: Whenever it is logically allowed
to perform operations simultaneously (see Fig.~\ref{adderQCE}),
these operations have been put into one microinstruction.

\begin{figure*}[t]
\begin{center}
\includegraphics[width=14cm]{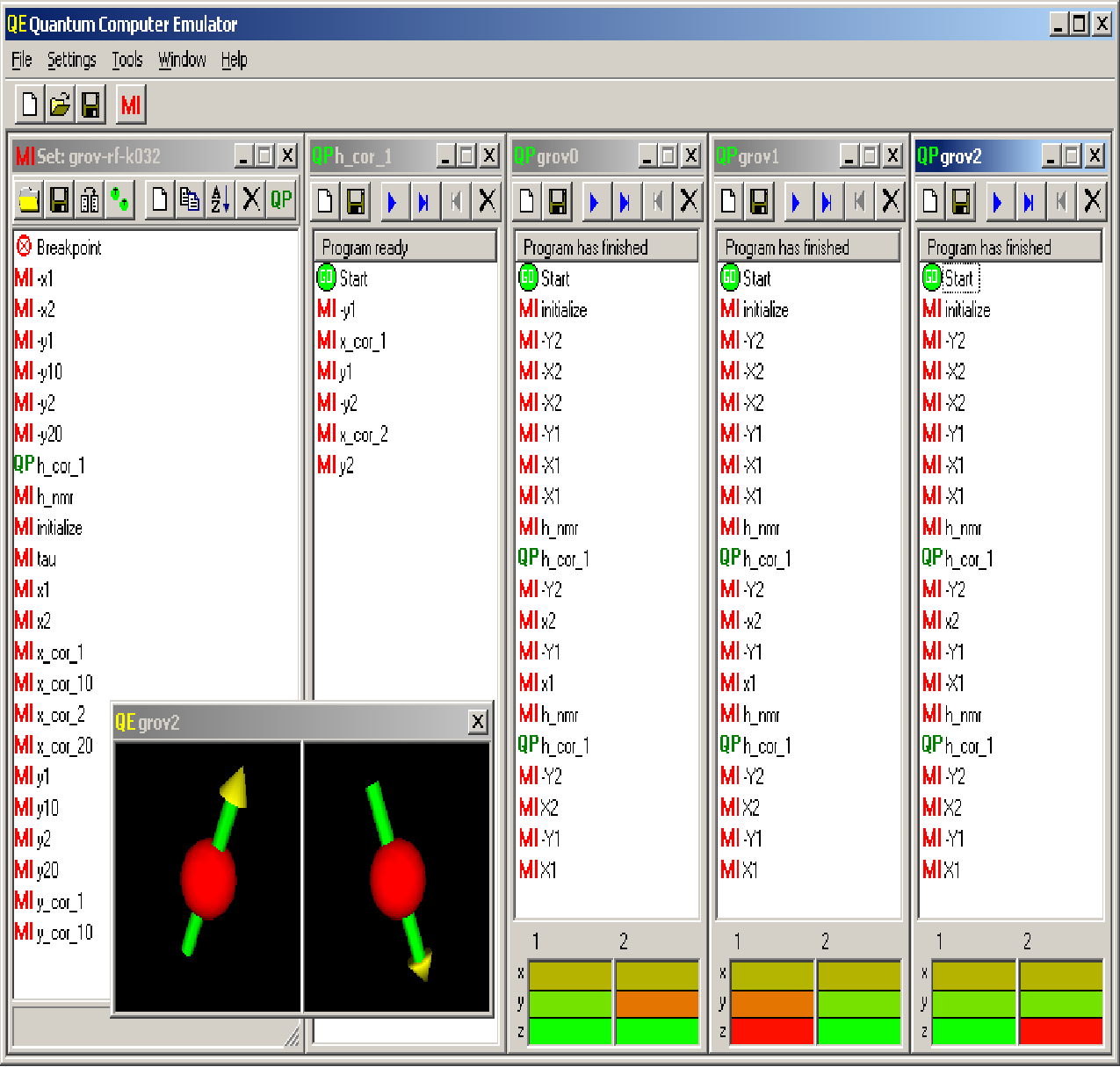}
\caption{Microinstruction set window ``grov-rf-k032''
together with the quantum program windows
``h\_cor\_1,'' ``grov0,'' ``grov1,'' and ``grov2''
that implement Grover's database search algorithm on a
two-qubit NMR-like quantum computer
for the cases $g_0(x)$, $g_1(x)$, and $g_2(x)$ [see Eq.~\Eq{GROV0}].}
\label{grover_NMR}
\end{center}
\end{figure*}

\subsubsection{Grover's Algorithm on an NMR-like Quantum Computer}
The QCE software comes with examples of
quantum programs that perform Grover's database search algorithm with four items
on ideal and NMR-like quantum computers (\ref{NMRmodel}).
In this application, the quantum computer has two qubits.
In Fig.~\ref{grover_NMR}, we show the QCE window after
loading the microinstruction set ``grov-rf-k032'' and
the quantum programs ``grov0'', $\ldots$, ``grov3''
from the QCE program directory ``grover2.''
The quantum programs ``grov0'', $\ldots$, ``grov3'' implement
the Grover algorithms for which
the searched-for item corresponds to item 0, 1, 2 or 3, respectively.
For printing purposes, quantum program ``grov3'' has been omitted from Fig.~\ref{grover_NMR}.
The microinstruction set ``grov-rf-k032'' corresponds to the
parameter set for the rotating sinusoidal fields
labeled with $s=32$ in Table~\ref{tab:GROVRESULT1}.
In the microinstruction set ``grov-rf-k032,''
quantum program ``h\_cor\_1'' and micro instruction ``h\_nmr''
are used to construct operation $G$ [see (\ref{eq:Gnmr})].
Running the four quantum programs yields
the values for $Q_1^z$ and $Q_2^z$ given in colums eight and nine of
Table~\ref{tab:GROVRESULT1}.
In contrast to the example of the three-input adder, we use the
inverse binary notation: qubit 1 corresponds to the least significant bit.

QCE has an option to visualize the time evolution of the state of the quantum computer
in terms of arrows representing the expectation values of the qubits.
In Fig.~\ref{grover_NMR} this option was used to visualize the outcome of the program ``grov2.''
Microinstruction sets corresponding to the parameter sets labeled with $s=8$, $s=16$, $s=64$,
and $s=256$ are also provided with the QCE software.
Running the programs ``grov0'', $\ldots$, ``grov3'' with these instruction
sets demonstrates that the exact results are recovered if $s$ is sufficiently
large (see colums two and three in Table~\ref{tab:GROVRESULT1}).

\section{Summary and Outlook}

The simulation methods reviewed in this chapter
have the potential to faithfully simulate
quantum computers of 20 -- 30 qubits,
considerably more than what is
projected to be realizable in the laboratory
within the next decade.
We say ``potential'' because whether
such simulations can be performed in a reasonable
(from the perspective of the researcher or funding organization)
amount of real time depends on how the implementation
(for example, the vectorization, parallelization, memory usage)
efficiently uses conventional computer resources.

An important topic that is virtually unexplored is
the implementation of fault-tolerant quantum computation on
quantum computers other than the ideal one.
The implementation of quantum error correction codes requires additional qubits.
It is not known whether fault-tolerant quantum computation
is robust with respect to systematic errors that
inevitably affect the operation of physically realistic models of quantum computers.
It most likely is not, and then it is good to know how to
circumvent this problem.
The complexity of this problem is such that it is
not accessible to theoretical analysis (unless
drastic, unrealistic simplifications are introduced),
whereas computer simulation may give valuable insight.

Decoherence is another topic that, albeit of much broader scope,
is also very important for quantum computation.
Although related to the issue of fault-tolerant quantum computation,
the effect of the coupling between the quantum computer
and its environment on the operation of the quantum computer
is a problem that is hardly accessible to
theoretical analysis, except perhaps for extremely idealized cases.
Also here computer simulation may help.

Finally, we believe that the software approach to quantum
computation can help students gain insight into
quantum physics during projects in which they improve their computational skills.
The algorithms used to simulate quantum computers
are fairly generic, that is, they solve differential
equations of the parabolic type, and the concepts and techniques learned
find applications in many other fields of computational science.

\section*{Acknowledgment}
We thank V.V. Dobrovitski, A.H. Hams, S. Miyashita, K. De Raedt, and K. Saito
for stimulating discussions.
Support from the Nederlandse Stichting Nationale Computer Faciliteiten (NCF)
is gratefully acknowledged.

\clearpage
\newpage
\begin{table}
\begin{center}
\caption{Programming languages for quantum computers. URLs last accessed on July 29, 2004.}
\label{tabsim0}
\begin{ruledtabular}
\begin{tabular}{p{4cm}p{13.5cm}}
Name and&Description\\
latest release&\\
\noalign{\vskip 2pt\hrule\vskip 4pt}
QCL
(Beta) Version 0.5.1 (0.6.1), March 30, 2004&
QCL~\cite{OEME98,OEME00,OEME03b,OEME03} (Quantum Computation Language) is a high-level, architecture-independent programming
language for quantum computers, with a syntax derived from classical procedural languages like C or Pascal.
Examples such as the quantum Fourier transform, Shor's algorithm, and Grover's algorithm are included.
QCL has been developed under Linux; version 0.5.0 compiles with the GNU C++ compiler 3.3.
The current version of QCL (sources and i386 Linux binaries) can be downloaded freely from:
http://tph.tuwien.ac.at/\~{}oemer/qcl.html
\\
\noalign{\vskip 2pt\hrule\vskip 4pt}
Q language Version 0.5.8,

February 18, 2002&
Q language~\cite{BETT02,BETT03} is a C++ implementation of a quantum programming language.
The source code can be downloaded freely from:
http://sra.itc.it/people/serafini/qlang
\\
\noalign{\vskip 2pt\hrule\vskip 4pt}
Quantum Superpositions

Version 2.02, April 22, 2003&
Quantum Superpositions is a PERL library that enables programmers to use variables that can hold more than
one value at the same time.
The module can be downloaded freely from:
http://search.cpan.org/\~{}lembark/Quantum-Superpositions/lib/Quantum/Superpositions.pm
\\
\noalign{\vskip 2pt\hrule\vskip 4pt}
QuBit,
July 24, 2001&
QuBit is a C++ library that supports Quantum Superpositions.
The library is rewritten starting from the
Quantum Superpositions library in PERL.
A complete implementation of the QuBit code can be downloaded from:
http://www.bluedust.com/qubit/default.asp
\\
\noalign{\vskip 2pt\hrule\vskip 4pt}
Quantum Entanglement

Version 0.32, June 5, 2002&
Quantum Entanglement is a PERL library that allows the user to put variables in a superposition of states,
have them interact with each other, and then observe them.
The module can be downloaded freely from:
http://search.cpan.org/dist/Quantum-Entanglement
\\
\noalign{\vskip 2pt\hrule\vskip 4pt}
Q-gol Version 3,
September 11, 1998&
Q-gol is a visual programming language.
The fundamental symbols of the system are gates (represented by raised blocks),
and directed wires.
The graphical editor lets the user select gates, place
them on sheets, and wire them together.
An implementation of Shor's algorithm is included.
The source code can be downloaded freely from:
http://www.ifost.org.au/\~{}gregb/q-gol/index.html
\\
\noalign{\vskip 2pt\hrule\vskip 4pt}
Quantum Fog Version 1.6,
June 25, 2003&
Quantum Fog is a Macintosh application for modeling physical situations that exhibit quantum mechanical behavior.
It is a tool for investigating and discussing quantum measurement problems graphically,
in terms of quantum Bayesian nets.
It simulates a general-purpose quantum computer.
The software can be downloaded freely from:
http://www.macupdate.com/info.php/id/12181 , http://www.ar-tiste.com
\\
\noalign{\vskip 2pt\hrule\vskip 4pt}
QDD
Version 0.2, February 4, 2003&
QDD is a C++ library that provides a relatively intuitive set of quantum computing constructs within the
context of the C++ programming environment. The emulation of quantum computing is based upon a
Binary Decision Diagram representation of the quantum state.
Shor's factoring algorithm is included in the QDD library, SHORNUF.
The software can be downloaded freely from:
http://thegreves.com/david/software
\\
\noalign{\vskip 2pt\hrule\vskip 4pt}
Quantum Lambda
Calculus 2003&
Functional language based on Scheme for expressing and simulating quantum algorithms.
Access at: http://www.het.brown.edu/people/andre/qlambda/index.htmls
\\
\end{tabular}
\end{ruledtabular}
\end{center}
\end{table}

\begin{table}
\begin{center}
\caption{Quantum compilers. URLs last accessed on July 29, 2004.}
\label{tabsim1}
\begin{ruledtabular}
\begin{tabular}{p{4cm}p{13.5cm}}
Name and&Description\\
latest release&\\
\noalign{\vskip 2pt\hrule\vskip 4pt}
Qubiter
Version 1.1, April 6, 1999&
Qubiter is a quantum compiler written in C++.
Qubiter takes as input an arbitrary unitary matrix
and returns as output an equivalent sequence of elementary quantum operations.
The software can be downloaded freely from:
http://www.ar-tiste.com/qubiter.html
\\
\noalign{\vskip 2pt\hrule\vskip 4pt}
GQC,
2002&
GQC\cite{BREM02} is an online quantum compiler
that returns a circuit for the CNOT in terms
of a user-specified unitary transformations.
Access at:
http://www.physics.uq.edu.au/gqc/
\\
\end{tabular}
\end{ruledtabular}
\end{center}
\end{table}

\begin{table}
\begin{center}
\caption{Quantum circuit simulators. URLs last accessed on July 29, 2004.}
\label{tabsim2}
\begin{ruledtabular}
\begin{tabular}{p{4cm}p{13.5cm}}
Name and &Description\\
latest release&\\
\noalign{\vskip 2pt\hrule\vskip 4pt}
QCAD Version 1.80,
May 8, 2003&
QCAD is a graphical Windows 98/2000 environment to design quantum circuits.
It can export the circuits as BMP or EPS files,
simulate the circuits, and show the states of the qubits.
QCAD can be downloaded freely from:
http://acolyte.t.u-tokyo.ac.jp/\~{}kaityo/qcad
\\
\noalign{\vskip 2pt\hrule\vskip 4pt}
QuaSi(2),
March 18, 2002&
QuaSi(2) is a general-purpose quantum circuit simulator.
It enables the user to build and simulate quantum circuits
in a graphical user interface.
Demo circuits for Shor's, Grover's and the Deutsch-Josza algorithm are included.
QuaSi simulates up to 20 qubits.
A Java applet is provided for use over the internet.
The full version of QuaSi2 can be downloaded freely from:
http://iaks-www.ira.uka.de/QIV/QuaSi/aboutquasi.html
\\
\noalign{\vskip 2pt\hrule\vskip 4pt}
JaQuzzi
Version 0.1, January 14, 2001&
JaQuzzi~\cite{SCHU00} is an interactive quantum computer simulator to design, test, and
visualize quantum algorithms with up to 20 qubits.
The program can either run standalone or as a Web-based applet.
To run jaQuzzi, a JAVA virtual machine of version 1.3 or higher is required.
The software can be downloaded freely from:

http://www.eng.buffalo.edu/\~{}phygons/jaQuzzi/jaQuzzi.html
\\
\noalign{\vskip 2pt\hrule\vskip 4pt}
QCSimulator

Version 1.1, March 8, 2000&
QCSimulator is a quantum computer simulator for Macintosh and Windows machines.
A graphical user interface is used to build a circuit representation
of a quantum algorithm and to simulate
quantum algorithms by exchanging unitary elements with Mathematica~\cite{MATH}.
Shor's factorization algorithm, Grover's database search algorithm,
the discrete Fourier transform, and an adder for two numbers are included.
The complete software package can be ordered from:
http://www.senko-corp.co.jp/qcs/
\\
\noalign{\vskip 2pt\hrule\vskip 4pt}
Libquantum Version 0.2.2,
November 3, 2003&
Libquantum is a C library for the simulation of an ideal quantum computer.
Basic operations for register manipulation such as the Hadamard gate
or the Controlled-NOT gate are available.
Measurements can be performed on either single qubits or the whole quantum register.
Implementations of Shor's factoring algorithm, and Grover's search algorithm are included.
Libquantum contains features to study decoherence and quantum error correction.
Libquantum is developed on a GNU/Linux platform and requires the installation of a C compiler with
complex number support. It can be downloaded freely from:
http://www.enyo.de/libquantum
\\
\noalign{\vskip 2pt\hrule\vskip 4pt}
OpenQUACS,
May 22, 2000&
OpenQUACS~\cite{MCCU00} (Open-Source QUAntum Computer Simulator) is a library written in Maple
that simulates the capabilities of an ideal quantum computer.
The simulator comes with a full tutorial. Several quantum algorithms such as Deutsch's
algorithm, quantum teleportation, Grover's search algorithm and a quantum adder are included.
The software can be downloaded freely from:
http://userpages.umbc.edu/\~{}cmccub1/quacs/quacs.html
\\
\noalign{\vskip 2pt\hrule\vskip 4pt}
QuCalc Version 2.13,
November 8, 2001&
QuCalc is a library of Mathematica functions~\cite{MATH} to simulate quantum circuits and
solve problems of quantum computation.
The Mathematica package can be downloaded freely from:
http://crypto.cs.mcgill.ca/QuCalc
\\
\noalign{\vskip 2pt\hrule\vskip 4pt}
QGAME
Version 1, July 7, 2002&
QGAME~\cite{SPEC99a,SPEC99b} (Quantum Gate And Measurement Emulator) is a system, written in
Common Lisp, that allows a user to run quantum computing algorithms
on a digital computer.
QGAME's graphical user interface (GUI) is a quick hack intended to allow people with no
knowledge of Lisp to experiment with QGAME. It uses Macintosh Common Lisp (MCL) interface
code and will work only under MacOS with MCL.
Not all features of QGAME are available from the GUI.
QGAME itself is platform independent (it will run on any platform for which a Common
Lisp environment is available); only the GUI requires Macintosh Common Lisp.
The full QGAME source code can be downloaded freely from:
http://hampshire.edu/lspector/qgame.html
\\
\noalign{\vskip 2pt\hrule\vskip 4pt}
QCompute,
July 1997&
QCompute is a quantum computer simulator written in Pascal that performs quantum gate
operations on an arbitrary number of qubits. The source code and sample gate-networks for a 1-bit and
a 2-bit adder can be found in the appendices to a thesis that can be downloaded from:
http://w3.physics.uiuc.edu/\~{}menscher/quantum.html
\\
\noalign{\vskip 2pt\hrule\vskip 4pt}
Quantum,
July 1997&
Quantum is a quantum circuit simulator written in C++ for a Windows environment.
It includes an implementation of Shor's algorithm.
The software can be downloaded freely from:
http://www.themilkyway.com/quantum
\\
\noalign{\vskip 2pt\hrule\vskip 4pt}
Eqcs
Version 0.0.5, March 19, 1999&
Eqcs is a library allowing clients to simulate a quantum computer. It includes a program showing the creation of
a CNOT gate. The software can be downloaded freely from:
http://home.snafu.de/pbelkner/eqcs/index.html
\\
\noalign{\vskip 2pt\hrule\vskip 4pt}
QCS,
January 11, 2001&
QCS (Quantum Computer Simulator) is a quantum computer library in C++.
The software can be downloaded freely from:
http://www-imai.is.s.u-tokyo.ac.jp/\~{}tokunaga/QCS/simulator.html
\\
\end{tabular}
\end{ruledtabular}
\end{center}
\end{table}

\begin{table}
\begin{center}
\caption{Quantum Computer Emulators. URLs last accessed on July 29, 2004.}
\label{tabsim3}
\begin{ruledtabular}
\begin{tabular}{p{4cm}p{13.5cm}}
Name and &Description\\
latest release&\\
\noalign{\vskip 2pt\hrule\vskip 4pt}
QCE
Version 8.1.1, June 27, 2004&
QCE~\cite{RAED00,MICH02} (Quantum Computer Emulator) is a software tool that
emulates ideal quantum computers as well as physical implementations of quantum computer
hardware.
QCE uses time-dependent Hamiltonians and unitary time evolutions
to simulate the physical processes that govern the operation of a
hardware quantum processor.
QCE provides an environment to debug and execute
quantum algorithms under realistic experimental conditions.
The QCE package includes many examples such as Shor's algorithm for factoring integers, order finding,
number partitioning~\cite{RAED01}, the quantum Fourier transform, various
implementations of the Deutsch-Josza algorithm,
and Grover's database search on ideal and more realistic quantum computers, such as
those used in the 2-qubit NMR quantum computer.
The software consists of a Graphical User Interface and the simulator itself.
It runs under Windows 98/NT4/2000/ME/(XP with SP1)
and Linux+VMware on Intel/AMD processor machines.
The software can be downloaded freely from:
http://www.compphys.org/qce.htm
\\
\noalign{\vskip 2pt\hrule\vskip 4pt}
QSS,
June 2000&
QSS~\cite{SCHN00} (Quantum System Simulator) is a software tool 
that simulates quantum computations with time-dependent Hamiltonians.
It provides a simple and convenient graphical user interface that enables users to specify
complex Hamiltonians as sums of smaller ones.
The simulator engine is designed as a self-contained module, so that
it is independent of the user interface, and can be easily enhanced.
The QSS runs on a Windows operating system and can be downloaded freely from:
http://web.mit.edu/scottsch/www/qc
\\
\end{tabular}
\end{ruledtabular}
\end{center}
\end{table}

\begin{table*}
\begin{center}
\caption{Pedagogical software. URLs last accessed on July 29, 2004.}
\label{tabsim4}
\begin{ruledtabular}
\begin{tabular}{p{4cm}p{13.5cm}}
Name and &Description\\
latest release&\\
\noalign{\vskip 2pt\hrule\vskip 4pt}
Quantum Turing machine

simulator 2002\hfil&
Mathematica toolkit to construct, run, and research quantum Turing machines.
Subscribers to {\sl The Mathematica Journal} can download the toolkit from:
http://www.mathematica-journal.com/issue/v8i3/features/hertel
\\
\noalign{\vskip 2pt\hrule\vskip 4pt}
QTM simulator,
1995&
Classical simulator of a quantum Turing machine written in C++.
The machine has one one-sided infinite tape and one read/write-head and
simulates a Fourier transform on a parity function.
It can be downloaded freely from:
http://www.lri.fr/\~{}durr/Attic/qtm
\\
\noalign{\vskip 2pt\hrule\vskip 4pt}
Quantum Search Simulator,
October 10, 2002&
Quantum Search Simulator is a Java applet that demonstrates the
operation of Grover's quantum search algorithm on a database of
four items on a quantum computer based on optical interference.
Access at:
http://strc.herts.ac.uk/tp/info/qucomp/qucompApplet.html
\\
\noalign{\vskip 2pt\hrule\vskip 4pt}
Grover's algorithm,

June 2001&
Mathematica-compatible notebook demonstrating Grover's algorithm.
It can be downloaded freely from:
http://www.cs.caltech.edu/\~\ chenyang/cs20proj-grover6.nb
\\
\noalign{\vskip 2pt\hrule\vskip 4pt}
CS 596 Quantum Computing,
Spring 1999&
Matlab programs that demonstrate some features of quantum computing.
Demonstrations of the RSA algorithm and
Shor's algorithm on a conventional computer are included.
It can be downloaded freely from:
http://www.sci.sdsu.edu/Faculty/Don.Short/QuantumC/cs662.htm
\\
\noalign{\vskip 2pt\hrule\vskip 4pt}
Shor's algorithm,
June 2001&
QCL code demonstrating Shor's algorithm.
It can be downloaded freely from:

http://www.cs.caltech.edu/\~\ chenyang/myshor.qcl
\\
\noalign{\vskip 2pt\hrule\vskip 4pt}
Shor's algorithm simulator,
January 23, 2002&
C++ program that simulates the operation of a quantum computer performing Shor's algorithm.
It can be downloaded freely from:
http://alumni.imsa.edu/\~{}matth/quant
\\
\end{tabular}
\end{ruledtabular}
\end{center}
\end{table*}

\clearpage
\newpage
\raggedright

\end{document}